\documentclass[preprint]{aastex}

\newcommand{\rShock}{R_{\mbox{\tiny{Sh}}}}
\newcommand{\rPNS}{R_{\mbox{\tiny{PNS}}}}
\newcommand{\genasis}{GenASiS}
\newcommand{\f}[2]{\frac{#1}{#2}}
\newcommand{\vect}[1]{\mathbf{#1}}
\newcommand{\pderiv}[2]{\frac{\partial #1}{\partial #2}}
\newcommand{\divergence}{\mathbf{\nabla} \cdot}
\newcommand{\gradient}{\mathbf{\nabla}}
\newcommand{\curl}{\mathbf{\nabla} \times}
\newcommand{\rateCompression}{\sigma_{\divergence\vect{u}}}
\newcommand{\rateAdvection}{\sigma_{\vect{u}\cdot\nabla}}
\newcommand{\rateStretching}{\sigma_{\nabla\vect{u}}}
\newcommand{\rateMonopoles}{\sigma_{\divergence\vect{B}}}
\newcommand{\rateLorentzWork}{\sigma_{\vect{J}\times\vect{B}}}
\newcommand{\lambdacurvature}{\lambda_{\mbox{\tiny c}}}
\newcommand{\lambdarms}{\lambda_{\mbox{\tiny rms}}}
\newcommand{\fEmag}[1]{f_{\mbox{\tiny #1 G}}}

\defcitealias{blondin_etal_2003}{BMD03}
\defcitealias{blondin_mezzacappa_2007}{BM07}

\shorttitle{}
\shortauthors{Endeve et al.}

\begin{document}

\title{Generation of Magnetic Fields by the Stationary Accretion Shock Instability}

\author{Eirik Endeve\altaffilmark{1,2,3}, Christian Y. Cardall\altaffilmark{1,2}, Reuben D. Budiardja\altaffilmark{1,2,3}, and Anthony Mezzacappa\altaffilmark{1}}
\email{endevee@ornl.gov}

\altaffiltext{1}{Physics Division, Oak Ridge National Laboratory, Oak Ridge, TN 37831-6354, USA}
\altaffiltext{2}{Department of Physics and Astronomy, University of Tennessee, Knoxville, TN 37996-1200, USA}
\altaffiltext{3}{Joint Institute for Heavy Ion Research, Oak Ridge National Laboratory, Oak Ridge, TN 37831-6374, USA}

\begin{abstract}

We begin an exploration of the capacity of the stationary accretion shock instability (SASI) to generate magnetic fields by adding a weak, stationary, and radial (but bipolar) magnetic field, and in some cases rotation, to an initially spherically symmetric fluid configuration that models a stalled shock in the post-bounce supernova environment.  In axisymmetric simulations we find that cycles of latitudinal flows into and radial flows out of the polar regions amplify the field parallel to the symmetry axis, typically increasing the total magnetic energy by about two orders of magnitude.  Nonaxisymmetric calculations result in fundamentally different flows and a larger magnetic energy increase: shearing associated with the SASI spiral mode contributes to a widespread and turbulent field amplification mechanism, boosting the magnetic energy by almost four orders of magnitude (a result which remains very sensitive to the spatial resolution of the numerical simulations).  While the SASI may contribute to neutron star magnetization, these simulations do not show qualitatively new features in the global evolution of the shock as a result of SASI-induced magnetic field amplification.  

\end{abstract}


\section{Introduction}

Key aspects of the core-collapse supernova explosion mechanism remain unknown. It is known that a shock wave forms when the central density of a collapsing massive star ($\gtrsim 8 M_\odot$) exceeds nuclear density: at this point the repulsive short-range nuclear force stiffens the equation of state (EoS), and the resulting core bounce gives rise to a compression wave that steepens into a shock when it reaches the sonic point that separates the subsonically collapsing inner stellar core from the supersonically infalling outer core.  As the roughly spherical shock wave propagates radially outward through infalling material  it loses energy through dissociation of heavy nuclei and neutrino emission, eventually stalling to form an accretion shock at a radial distance of $100-200$~km from the center of the star. It is expected that the shock is revived within a second or so, enabling it to disrupt the star's outer layers and give rise to the supernova.  The details of this shock revival is still uncertain, but recent progress has been made \citep[e.g.,][]{marek_janka_2009, bruenn_etal_2009}. Although a meager 1\% of the gravitational energy released during collapse (most of which is carried away by neutrinos) is needed to account for the kinetic energy of the explosion, sophisticated spherically symmetric models fail to reproduce one of nature's most energetic events \citep{ramppJanka_2000, liebendorfer_etal_2001, thompson_etal_2003}, except in the case of the lightest supernova progenitors with O-Ne-Mg cores \citep{kitaura_etal_2006}.  The solution to the core-collapse supernova problem will require multidimensional multiphysics simulations, including multi-frequency (or multi-frequency {\it and} multi-angle) neutrino transport, nuclear EoSs and reaction networks, rotation, and (magneto)hydrodynamic instabilities \citep[e.g.,][]{mezzacappa_2005, woosley_janka_2005}.  

Investigation of the role of magnetic fields in core-collapse supernovae began in the 1970s and 1980s \citep{leblanc_wilson_1970, bisnovati-kogan_etal_1976, meier_etal_1976, symbalisty_1984}.  These studies seemed to show that both a strong magnetic field and rapid rotation were needed at the pre-collapse stage for the magnetic field to have any significant effect on the ensuing dynamics.  The mechanisms responsible for magnetic field amplification in these studies were (1) compression during collapse, and (2) winding during the post-bounce phase. Magnetic field enhancement through compression occurs because the stellar core may be regarded as a perfect electrical conductor; hence the magnetic field is `frozen-in' to the fluid and the magnetic field strength grows with the mass density roughly as $B\propto\rho^{2/3}$, resulting in a boost of three orders of magnitude due to a five orders of magnitude density increase during core collapse.  Differential rotation---an inevitable outcome of the collapse of a stellar core with any initial rotation---winds the magnetic field up, amplifying the azimuthal component linearly with time.  The inefficiency of these classic mechanisms in producing dynamically significant magnetic fields from typical progenitor stars, expected to have modest initial rotation and field strength, is one reason magnetohydrodynamic (MHD) effects were not seriously reconsidered for almost two decades.  

While the relevance of magnetic fields to the supernova explosion mechanism remains unclear, the discovery of magnetars (neutron stars with magnetic fields in the range $10^{14}-10^{15}$~G) \citep{duncan_thompson_1992}, the generically observed asphericity of core-collapse supernova explosions \citep{wang_etal_2001}, the observation of collimated jets in supernovae associated with gamma ray bursts (GRBs) \citep{woosley_bloom_2006}, and the theoretical discovery of the magnetorotational instability (MRI) \citep{balbus_hawley_1991} have significantly increased recent interest in the role of magnetic fields in core-collapse supernovae \citep{kotake_etal_2004, ardeljan_etal_2005, obergaulinger_etal_2005, thompson_etal_2005, shibata_etal_2006, burrows_etal_2007, cerda_etal_2007, endeve_etal_2007, suzuki_etal_2008, mikami_etal_2008, takiwaki_etal_2009}.  For example, the work of \citet{burrows_etal_2007} illustrates that a strong magnetic field, organized by rapid differential rotation into `magnetic towers' along the rotation axis, can drive bipolar explosions when the evolution is followed for several hundred milliseconds after bounce.  

In particular, the MRI, a weak-field instability, operates quite generically in differentially rotating fluids with a negative radial gradient in the angular velocity and may lead to exponential field growth on the rotation time scale.  \citet{akiyama_etal_2003} showed, in proof-of-principle calculations, that the MRI may operate in the region between the  proto-neutron star (PNS) and the supernova shock to generate magnetic fields exceeding $10^{15}$~G within a few 100s of milliseconds.  The extent of operation and ultimate impact of the MRI are still uncertain, in large part due to the requirement of high spatial resolution to numerically resolve the MRI:  simple estimates suggest the wavelength of the fastest growing unstable mode to be roughly $100$~m \citep[e.g.,][]{obergaulinger_etal_2005}, and must be resolved by $\sim10$ mesh points \citep{etienne_etal_2006}.  \citet{shibata_etal_2006} employed simplified high resolution magnetorotational simulations of stellar collapse and observed episodes of exponential field growth due to the MRI in localized regions shortly after bounce, but also found that the field generated by the MRI had little influence on the growth of the global magnetic energy, which was dominated by the field generated by the winding mechanism.  \citet{obergaulinger_etal_2009} demonstrated, through what they coined semi-global simulations, that the MRI may lead to dynamically significant magnetic fields in the supernova environment, but were unable to draw definitive conclusions on the magnetic field's impact on the global dynamics.  Despite existing uncertainties, it is frequently argued that the MRI may operate efficiently in the post-bounce phase, and be responsible for generating the strong large-scale field that is needed to drive bipolar outflows \citep[e.g.,][]{burrows_etal_2007}.  

While rotation was long assumed to be essential to the observed asphericity of supernova explosions (perhaps in connection with magnetic fields), recent simulations of an idealized model constructed to mimic the stalled supernova shock wave have shown that the accretion shock is unstable to non-radial perturbations \citep[][hereafter \citetalias{blondin_etal_2003}]{blondin_etal_2003}, and that this `stationary accretion shock instability' (SASI) may give rise to asymmetric explosion dynamics even without initial rotation.  Axisymmetric, two-dimensional (2D) simulations show that the instability is dominated by $\ell=1$ (dipole) and $\ell=2$ (quadrupole) modes, leading to an asymmetric expansion of the supernova shock wave.  Three-dimensional (3D) simulations of the same configuration \citep[][hereafter \citetalias{blondin_mezzacappa_2007}]{blondin_mezzacappa_2007} show that the non-axisymmetric $m=1$ (spiral) mode may eventually dominate, and even result in significant spin-up of the PNS.  Axisymmetric multiphysics supernova simulations have confirmed the existence of the SASI \citep{bruenn_etal_2006, buras_etal_2006, burrows_etal_2006, scheck_etal_2008, marek_janka_2009} and underscored its importance:  the SASI may improve the conditions for energy deposition by neutrinos in the post shock gas \citep{scheck_etal_2008,marek_janka_2009}, and result in natal pulsar kicks that explain the velocity distribution of young pulsars \citep{scheck_etal_2006}.  

The fact that the SASI can generate phenomena previously attributed to progenitor rotation---in particular, asymmetric shock expansion and pulsar spin---raises the question of whether the SASI can also generate strong magnetic fields in the absence of initial rotation. The above-referenced MHD studies of stellar core collapse have taken rapid progenitor rotation for granted as a necessary ingredient, both as the energy source and mechanical agent of field amplification.  The requirement of rapid rotation is an important constraint since the rotational free energy available in a millisecond PNS is sufficient to account for the kinetic energy of the `canonical' supernova, but requires the magnetic field to grow strong enough to facilitate the conversion of rotational kinetic energy to kinetic energy of outward flow.  However, an even smaller energy reservoir (i.e., that provided by the SASI, which may reach several $\times10^{49}$~erg) may result in interesting and perhaps important, although less spectacular, MHD developments prior to the onset of explosion.  In many of the past MHD studies the computational domain has been limited to one quadrant, and some of them have only been able to follow the simulations for a few 10s of milliseconds post-bounce, thereby excluding development of the SASI.  Most studies are also carried out with axial symmetry imposed.  In this study we extend the models of \citetalias{blondin_etal_2003} and \citetalias{blondin_mezzacappa_2007} to include an initially weak magnetic field.  We then perturb the stationary initial condition and study the evolution of the magnetic field as the SASI develops.  We present results from axisymmetric 2D models, and non-rotating and slowly rotating 3D models.  

Our simulations illustrate that SASI-induced flows may be able to significantly amplify the magnetic field beneath the shock, even in the absence of progenitor rotation.  The amplification occurs on a time scale that is determined by the SASI-driven flows.  The magnetic field in axisymmetric models is amplified by compression and stretching due to the flows driven by the $\ell=1$ (sloshing) mode. The magnetic field then becomes concentrated in the polar regions.  Away from the polar regions, turbulent flows stretch the magnetic field into thin flux tubes.  All our 3D models eventually become dominated by the $m=1$ (spiral) mode.  The focused, high-speed accretion flow associated with the SASI spiral mode drives turbulence beneath the shock, which amplifies the magnetic field through flux tube stretching and results in a highly intermittent flux rope structure.  The amplification does not lead to dynamically significant magnetic fields in our simulations, but the magnetic energy growth is ultimately suppressed by (numerical) resistivity due to finite spatial resolution.  We are therefore unable to make precise statements about the final strength of the magnetic field generated by the SASI, but our simulations suggest that SASI-generated magnetic fields are not likely to have a dynamical impact on global supernova dynamics.  Nevertheless, our results suggest a mechanism for PNS magnetization that does not rely on progenitor rotation.  

\section{Model Setup and Numerical Solution}
\label{sec:model}

In this section we describe our simplified MHD model of the post-bounce supernova environment, briefly describe our numerical scheme, and demonstrate that our code can maintain stationary unperturbed initial conditions.

\subsection{An Idealized Model}

In this study of magnetic field generation by the SASI we adopt an idealized description of the post-bounce supernova environment.  We do not explicitly include neutrino transport, general relativity, or a nuclear equation of state---crucial components of a realistic supernova model---and let the magnetized fluid be described by the non-relativistic ideal MHD equations:
\begin{eqnarray}
  \pderiv{\rho}{t} + \divergence \left(\rho \vect{u} \right) &=& 0,
  \label{eq:mass} \\
  \pderiv{\rho \vect{u}}{t} 
  + \divergence \left( \rho\vect{u}\vect{u} - \f{1}{\mu_{0}} \vect{B}\vect{B} + \mathbf{I} P^{\star}\right)
  &=& - \rho \gradient \Phi,
  \label{eq:momentum} \\
  \pderiv{E}{t} + \divergence \left[ \left( E+P^{\star} \right)\vect{u} 
                                    - \f{1}{\mu_{0}} \vect{B}\left( \vect{B}\cdot\vect{u} \right) \right]
  &=& - \rho \vect{u} \cdot \gradient \Phi, 
  \label{eq:energy} \\
  \pderiv{\vect{B}}{t} &=& \curl \left( \vect{u} \times \vect{B} \right).  
  \label{eq:induction}
\end{eqnarray}
In Eqs. (\ref{eq:mass})-(\ref{eq:induction}) $\rho$, $\vect{u}$, $\Phi$, and $\vect{B}$ represent mass density, fluid velocity, gravitational potential, and magnetic field respectively.  The total (fluid plus magnetic) pressure is $P^{\star}=P+\vect{B}\cdot\vect{B} / (2\mu_{0})$, where $\mu_0$ is the vacuum permeability; we adopt a polytropic equation of state, $P \propto \rho^{\gamma}$, with the ratio of specific heats $\gamma$ set to $4/3$. The fluid energy density is $E=e+e_{\mbox{\tiny kin}}+e_{\mbox{\tiny mag}}$, where $e=P/\left(\gamma-1\right)$ is the internal energy density, $e_{\mbox{\tiny kin}}= \rho\vect{u}\cdot\vect{u} / 2$ is the kinetic energy density, and $e_{\mbox{\tiny mag}}=\vect{B}\cdot\vect{B} / (2\mu_{0})$ is the magnetic energy density (also referred to as the magnetic pressure).  The unit tensor is $\mathbf{I}$.  The evolution of the magnetic field is also constrained to satisfy the divergence-free condition,  $\divergence\vect{B}=0$.  In the problem studied here we take the gravitational potential to be given by the point-mass formula $\Phi=-GM/r$, where $G$ is Newton's constant, $M$ is the mass of the central object, and $r$ is the radial distance from the center of the star.  

Our initial setup resembles a post-bounce stalled supernova shock configuration \citepalias[e.g.,][and references therein]{blondin_etal_2003,blondin_mezzacappa_2007}.  We place a steady-state, spherically symmetric accretion shock at $r=\rShock=200$ km from the center of the star.  At larger radii matter falls into the shock at the free-fall speed $\left( 2GM/r \right)^{1/2}$, and a constant, highly supersonic Mach number of 300 is used to set the pressure in the pre-shock gas.  The mass of the central object is set to $M=1.2 M_{\odot}$ and is not allowed to change with time.  We do not include the self-gravity of matter on the grid (which excludes the central object).  The accretion rate is set to 0.36 M$_{\odot}$ s$^{-1}$, which is kept constant throughout the simulations.  This accretion rate is large enough to increase the mass of the central object considerably over the time scales considered in this study ($\sim$30 percent in one second), but in order to construct a steady state for the unperturbed initial condition, and compare against it, we set the rate of mass change of the central object to $\partial M / \partial t=0$ in all the models presented here.  The Rankine-Hugoniot conditions \citep[cf.][]{landau_lifshitz_1959} determine the hydrodynamic state just inside the shock, and the Bernoulli equation is solved for the fluid structure from the shock down to $r=\rPNS=40$ km, which serves as an inner `cutout' boundary of our grid, and may be loosely interpreted as the surface of the proto-neutron star (PNS).  Thus, we have $\rShock/\rPNS=5$. We allow fluid flow through the inner boundary in a manner that, in our experience, allows maintenance of a steady state in simulations with unperturbed initial conditions. First, the fluid velocity just inside the inner boundary is held fixed to its initial value. Second, power laws for mass density ($\rho \propto r^{-3}$) and pressure ($P \propto r^{-4}$) obtained from the Bernoulli equation near the PNS are used to dynamically interpolate values from zones just outside the inner boundary radius to ghost zones just inside the boundary radius.  

Neither the strength nor the topology of the magnetic field in supernova progenitors is known with confidence.  One-dimensional stellar evolution calculations indicate that the magnetic field at the onset of stellar core collapse may be dominated by the azimuthal component $B_{\phi} \sim 10^{9}$ G, with a radial component about three orders of magnitude lower \citep{heger_etal_2005}.  In this study we will only consider models with a purely radial initial field, disregarding any initial azimuthal component.  This choice is consistent with a steady-state initial condition, and in any case collapse will tend to drag any initial higher order multipole moments of the field (e.g., dipole, quadrupole, etc.) into a more radial configuration.  Our rotating models are also initialized with a purely radial magnetic field, but the initial (differential) rotation profile quickly winds up the magnetic field and results in a configuration with a significant azimuthal magnetic field inside the shock.  During collapse the magnetic field is `frozen-in' to the fluid, and its strength increases with density roughly as $B\propto\rho^{2/3}$.  With an initial poloidal field of about $10^{6}$ G---which is comparable to the magnitude of poloidal field expected from the stellar evolution calculations mentioned above---and five orders of magnitude density increase during collapse, the field strength in the collapsed core is a few times $10^{9}$ G.  In the 2D base model presented in this study we set $B_{0}=10^{10}$ G, where $B_{0}$ is the strength of the magnetic field at $r=\rPNS$.  This poloidal field strength is not expected to have any influence on the dynamics in the early stages of the development of the SASI:  the ratio of the magnetic energy density to the fluid pressure, $\beta^{-1}=e_{\mbox{\tiny mag}}/P$, is less than $2\times10^{-11}$ everywhere inside the shock.  The ratio of the magnetic energy density to the kinetic energy density, $e_{\mbox{\tiny mag}}/e_{\mbox{\tiny kin}}$, is about $2\times10^{-8}$ near the inner boundary, and falls off to about $10^{-10}$ just inside the shock.  We will also present results from models where we have varied the strength of the initial magnetic field, $B_{0}$.  

Accordingly, to the fluid setup described previously we add a `split monopole' magnetic field which, being purely radial, is consistent with a steady state.  This is given by $\vect{B}=B_{r} \vect{e}_{r}$, with $B_{r}=\mbox{sign} \left(\cos\theta\right)\times B_{0}\left(\rPNS/r\right)^{2}$, where  $\theta$ is the polar angle in a spherical coordinate system.  The magnetic field has positive polarity in the northern hemisphere and negative polarity in the southern hemisphere (the opposite polarities in the two hemispheres imply the existence of a thin current sheet in the equatorial plane).  The magnetic field is held fixed at the outer boundary throughout the simulations.  At the inner boundary we copy magnetic field components parallel to the boundary face to ghost zones inside $\rPNS$, so that any non-radial magnetic field that develops is advected with the fluid through the inner boundary. Because magnetic field components perpendicular to zone faces `live' on the respective faces in our scheme, the component perpendicular to an inner boundary face is allowed to evolve freely.  

Note that $B_{0}$ refers to the magnetic field strength at the surface of the PNS.  The RMS magnetic field inside the shock is initially about an order of magnitude below $B_{0}$:  $B_{\mbox{\tiny RMS}}\approx\sqrt{3\left(\rPNS/\rShock\right)^{3}}\times B_{0}\approx0.15 \times B_{0}$.  

\subsection{Numerical Scheme}

The simulations presented here were performed with our astrophysical simulation code \genasis, in which we have implemented a time-explicit, second-order, semi-discrete central-upwind scheme \citep{kurganov_etal_2001, londrillo_delZanna_2004} for the integration of the time-dependent ideal MHD equations in the form they are presented in Eqs. (\ref{eq:mass})-(\ref{eq:induction}). This finite-volume approach handles shocks, and preserves the divergence-free condition on the magnetic field via the method of constrained transport \citep{evans_hawley_1988}. 
We use the HLL formulae given in \citet{londrillo_delZanna_2004} to compute the fluid fluxes on zone faces and electric fields, $\vect{E}=-\vect{u}\times\vect{B}$, on zone edges. 
Second-order temporal accuracy is obtained with a two-step Runge-Kutta algorithm \citep[e.g.,][]{kurganov_tadmor_2000}. In order to achieve second-order spatial accuracy in smooth regions of the flow, while maintaining non-oscillatory behavior near shocks and discontinuities, we use slope-limited linear interpolation inside computational zones to slide variables to the appropriate faces and edges for the flux and electric field computations.  
In particular, we use a one-parameter family of minmod limiters to evaluate the slopes inside computational zones \citep[see for example][]{kurganov_tadmor_2000}: the slope of an arbitrary variable $f$ in (for example) the $x$-direction, in a zone whose center is indexed by $(i,j,k)$, is
\begin{eqnarray}
  \left. \pderiv{f}{x} \right|_{ijk}
   &=& \mathrm{minmod} \left[ \vartheta \left(\f{f_{ijk}-f_{i-1jk}}{x_{i}-x_{i-1}}\right), \right. \nonumber \\
   & &                 \left.   \left(\f{f_{i+1jk}-f_{i-1jk}}{x_{i+1}-x_{i-1}}\right), 
                                            \vartheta \left(\f{f_{i+1jk}-f_{ijk}}{x_{i+1}-x_{i}}\right) \right].  
  \label{eq:slopeLimiter}
\end{eqnarray}
Here $\vartheta\in [1,2]$, and the multivariable minmod function selects the least steep slope, provided all of the arguments have the same sign; otherwise it is zero. Smaller values of the slope weighting parameter $\vartheta$ make the scheme more dissipative.  We have found that setting $\vartheta=1.4$ gives satisfactory results in most situations, with a combination of reduced oscillations behind the stationary accretion shock and low numerical diffusion.  Setting $\vartheta=1$---which is equivalent to the traditional two-variable minmod that selects between only the left- and right-sided slopes---yields unacceptable results, as a significantly higher resolution (compared to $\vartheta>1$) is required in order to maintain the near-hydrostatic equilibrium inside the shock, near the PNS. 

The 3D simulations presented in this study are carried out on a uniform Cartesian mesh ($x$,$y$,$z$), while the axisymmetric simulations are carried out in cylindrical coordinates ($r_{\perp}$,$z$).  The latter choice avoids complications associated with using spherical coordinate systems to solve axisymmetric problems, such as the Courant-Friedrich-Lewy (CFL) condition---which governs the maximum stable time step---becoming severely restrictive in curvilinear coordinates due to the unavoidably small zone widths near the origin.  It is true that in the models presented in this paper we use an inner boundary at finite radius that renders the origin irrelevant and that this inner boundary is in fact more cumbersome to handle in Cartesian/cylidrical coordinates.  But the cylindrical coordinate system, with axial symmetry imposed, is similar to the 3D Cartesian version of this model.  We also plan to use this same code for other physical problems that do include the origin; therefore, in terms of code development, it is more efficient for us to use cylindrical coordinates to compute the axisymmetric models presented here.  Our implementation conserves mass (by construction), total energy (for the time-independent point-mass gravitational potential), and, for axisymmetric problems, angular momentum about the axis of symmetry.  

The resolution we used for most of the models presented here was chosen to be high enough to give acceptable results in maintaining steady-state unperturbed initial conditions (see the next subsection), as well as convergence in the early stage of the non-linear evolution for the perturbed models.  This was a zone width of $\Delta l \approx 1.56$~km, or $\Delta l/\rPNS \approx 0.039$.  We do not, as expected for the highly nonlinear flow with structures on all grid scales, find convergence at late times in the calculations for the chosen resolution.  However, global values (total kinetic energy inside the shock, etc.) are qualitatively the same for models where the resolution has been varied.  We will also include results from models with $\Delta l\approx 2.34$~km and $\Delta l\approx 1.17$~km.  The axisymmetric models are computed on a fixed computational domain, with $x\in\left[0,1200\right]$~km and $z\in\left[-1200,1200\right]$~km.  The 3D models are, in order to save computational resources, computed on a growing computational domain.  Initially, the box boundary is placed at a distance of $1.5~\rShock$ along all coordinate directions:  $x(,y,z)\in\left[-300,300\right]$~km.  As the initial perturbations grow and the SASI evolves into the nonlinear stage, we  expand the computational domain to accommodate the growing volume occupied by the shock.  In the models presented here, the computational domain has been allowed to grow until its sides has doubled; i.e., $x(,y,z)\in\left[-600,600\right]$~km.  The computational domain is then covered by $1024^{3}$ zones for the model with the highest resolution.  

\subsection{Maintenance of a Steady-state Initial Condition}

We will now demonstrate \genasis's ability to maintain steady state initial conditions in the absence of initial perturbations.  We run our models to an end time that is comparable to the time between bounce and shock revival in a core collapse supernova---about 1~s.  Thus, our code must be able to integrate the MHD equations accurately on this time scale.  Because the analytic initial condition is not an exact solution to the discretized MHD equations, initial transients are expected; but the system should eventually settle to a stationary configuration.  

One minor deviation from the initial steady state is the appearance of post-shock entropy oscillations, which commonly appear in numerical solutions of the MHD equations. Adjusting the slope weighting parameter $\vartheta$ reduces these oscillations in our simulations, but does not entirely suppress them.  To further reduce these post-shock oscillations we have implemented the flattening procedure described in \citet{fryxell_etal_2000}.  The oscillations are strongest at the beginning, and then damp out; they are barely noticeable at the end of an unperturbed run (see for example Figure \ref{fig:magneticAndKineticEnergy_medRes_2D_B0_1e10_l0_0e00_perturbed}).  In particular, they are not sufficient to initiate the SASI, at least on a time scale of about $1$~s.  

\subsubsection{Non-rotating Spherically Symmetric Initial Condition}

\begin{figure}
  \epsscale{1.0}
  \plotone{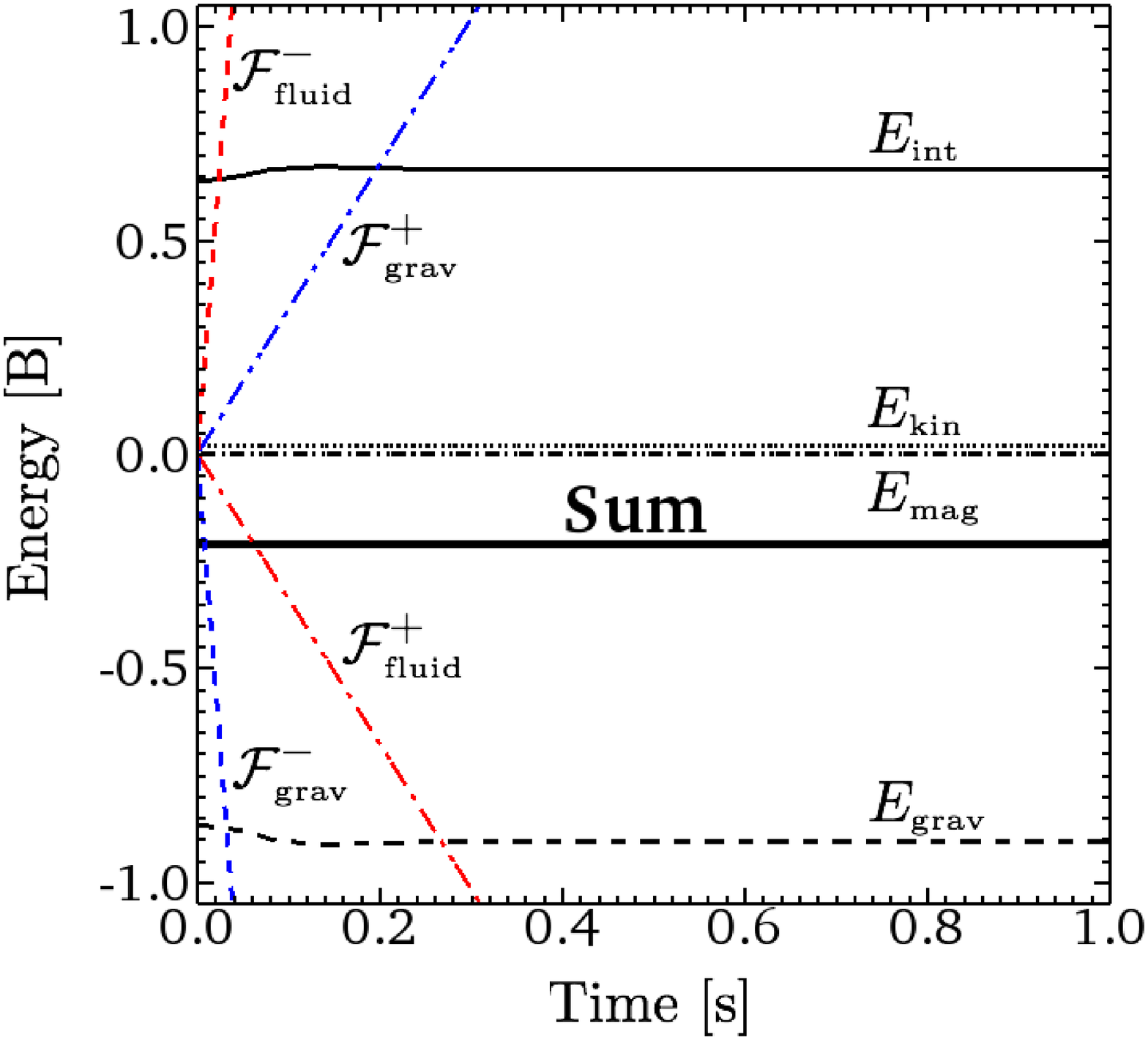}
  \caption{Energy conservation in an unperturbed model ($1 \ {\rm B} = 10^{51}$~erg).  Plotted versus time are the internal energy (black solid line), kinetic energy (black dotted line), magnetic energy (black dash-dot line), and gravitational energy (black dashed line) on the grid; the magneto-fluid energy (internal plus kinetic plus magnetic) and gravitational energy lost from the grid through the inner boundary (red and blue dashed lines respectively); the magneto-fuid energy and gravitational energy lost from the grid through the outer boundary (red and blue dot-dashed lines respectively); and the sum of all these (thick black solid line), which remains constant within numerical precision.  \label{fig:energyConservationUnperturbed}}
\end{figure}

Figure \ref{fig:energyConservationUnperturbed} illustrates how the unperturbed model evolves and settles to a stationary state with our `standard' resolution, $\Delta l\approx 1.56$~km.  We plot $E_{\mbox{\tiny{int}}}$, $E_{\mbox{\tiny{kin}}}$, $E_{\mbox{\tiny{mag}}}$, and $E_{\mbox{\tiny{grav}}}$, which are respectively the volume integrals, extending over the entire computational domain, of internal energy density $e$, kinetic energy density $e_{\mbox{\tiny kin}}$, magnetic energy density $e_{\mbox{\tiny mag}}$, and gravitational energy density $e_{\mbox{\tiny grav}}=\rho\Phi$.  Using the magneto-fluid energy flux density $\vect{F}_{\mbox{\tiny fluid}}=\left( E+P^{\star} \right)\vect{u} - \vect{B}\left( \vect{B}\cdot\vect{u} \right) / \mu_0$ appearing in equation (\ref{eq:energy}) and the gravitational energy flux density $\vect{F}_{\mbox{\tiny grav}}=\rho \Phi \vect{u}$, we also keep track of the total magneto-fluid (internal + kinetic + magnetic) energy and gravitational energy that have been lost from the computational domain by fluid flow across its boundaries:  In Figure \ref{fig:energyConservationUnperturbed} we plot the accumulated energies $\mathcal{F}_{\mbox{\tiny{fluid}}}^{-}$ and $\mathcal{F}_{\mbox{\tiny{grav}}}^{-}$ lost through the inner boundary enclosing the surface of the PNS, and also the accumulated energies $\mathcal{F}_{\mbox{\tiny{fluid}}}^{+}$ and $\mathcal{F}_{\mbox{\tiny{grav}}}^{+}$ `lost' through inflow through the outer boundary at large radius.

The energies in the computational domain remain remarkably constant after a transient adjustment period.  From $t=0$ to 200~ms the shock adjusts to a steady state consistent with the discrete MHD equations; it starts off at $\rShock = 200$~km and expands a bit before settling at $\rShock \approx 207$~km. Because of this net initial expansion, the volume of shocked, high-density, subsonic material increases slightly; this corresponds to a small net increase in $E_{\mbox{\tiny{int}}}$ and small net decrease in $E_{\mbox{\tiny{kin}}}$ between $t=0$ and 200 ms. The magnetic energy is many orders of magnitude below the scale of Figure \ref{fig:energyConservationUnperturbed}; it is visible on a logarithmic scale in the left panel of Figure \ref{fig:magneticAndKineticEnergy_medRes_2D_B0_1e10_l0_0e00_perturbed}. For $t>200$~ms the shock radius stays roughly constant, with relative variations smaller than 0.1\%. During this period $E_{\mbox{\tiny{int}}}$ also remains constant to about $0.1\%$, while the kinetic energy varies at about the 0.2\% level.  (The kinetic energy of the subsonic flow inside the shock is plotted on a logarithmic scale in the right panel of Figure \ref{fig:magneticAndKineticEnergy_medRes_2D_B0_1e10_l0_0e00_perturbed}). These variations are caused by the numerically-induced small amplitude entropy oscillations behind the stationary accretion shock, which introduce perturbations to the radial flow inside the shock. These variations also induce small lateral motions, whose kinetic energy remains about two orders of magnitude smaller than the total kinetic energy in the shocked region, with no overall growth trend (which would be an indication that the SASI is developing \citepalias[cf.][]{blondin_etal_2003}). These lateral motions couple to the magnetic field and induce variations in the magnetic energy, which are at the $\sim 3$\% level near the end of the calculation.

The sum of all the curves in Figure \ref{fig:energyConservationUnperturbed}, represented by the thick solid line, is constant over time to numerical precision.  The change in the total energy on the grid (about $0.01$~B) is offset by a difference in the fluxes through the surface enclosing the PNS.  The relation  $\mathcal{F}_{\mbox{\tiny fluid}}^{-} + \mathcal{F}_{\mbox{\tiny grav}}^{-} = 0$ is analytically expected\footnote{For our steady state setup the total energy fluxes $\vect{F}_{\mathcal{E}} = \vect{F}_{\mbox{\tiny{fluid}}} + \vect{F}_{\mbox{\tiny{grav}}}$ through the inner and outer bounding surfaces are individually zero:  at any point outside the shock the fluid is assumed to be in highly supersonic free-fall, and we have $\vect{F}_{\mathcal{E}} \approx \left( \f{1}{2} \vect{u} \cdot \vect{u} + \Phi \right) \rho \vect{u} = 0$,  while inside the shock we have $\vect{F}_{\mathcal{E}}=\left( \f{1}{2} \vect{u}\cdot\vect{u}+\f{e+P}{\rho}+\Phi\right) \rho \vect{u} = 0$, by virtue of the Bernoulli equation.}, but at $t = 1$~s we find $\mathcal{F}_{\mbox{\tiny fluid}}^{-} = 28.45$~B and $\mathcal{F}_{\mbox{\tiny grav}}^{-}=-28.44$~B ($1 \ {\rm B} = 10^{51}$~erg).  Analytically we also expect $\mathcal{F}_{\mbox{\tiny grav}}^{-}=-\left({G M}/{\rPNS}\right)\dot{M}t=-28.51$~B for our simulation at $t=1$~s, which deviates from our numerical value by about 0.25\%.  This difference is generated during the initial relaxation.  For $t>200$~ms we have $\mathcal{F}_{\mbox{\tiny fluid}}^{-} + \mathcal{F}_{\mbox{\tiny grav}}^{-} \approx 0$.  We conclude from this that our numerical scheme preserves the initial condition at an acceptable level.  

\subsubsection{Adding Rotation to the Initial Condition}

An analytic initial condition is not readily available in the case where the accreting material has non-zero angular momentum.  We initialize the rotating model presented in this study by setting the pre-shock gas in the spherically symmetric initial condition into rotation about the $z$-axis.  In particular, the azimuthal velocity is initialized as $u_{\phi}=l \sin\theta/r$, where $l$ is the (constant) specific angular momentum.  This method of initializing the rotating SASI model is similar to the method used by \citet{iwakami_etal_2009}, except that we include rotation from the onset of the simulation; we do not introduce rotation into the simulation in the nonlinear evolution \citep[e.g.,][]{iwakami_etal_2009}.  

\begin{figure}
  \epsscale{1.0}
  \plotone{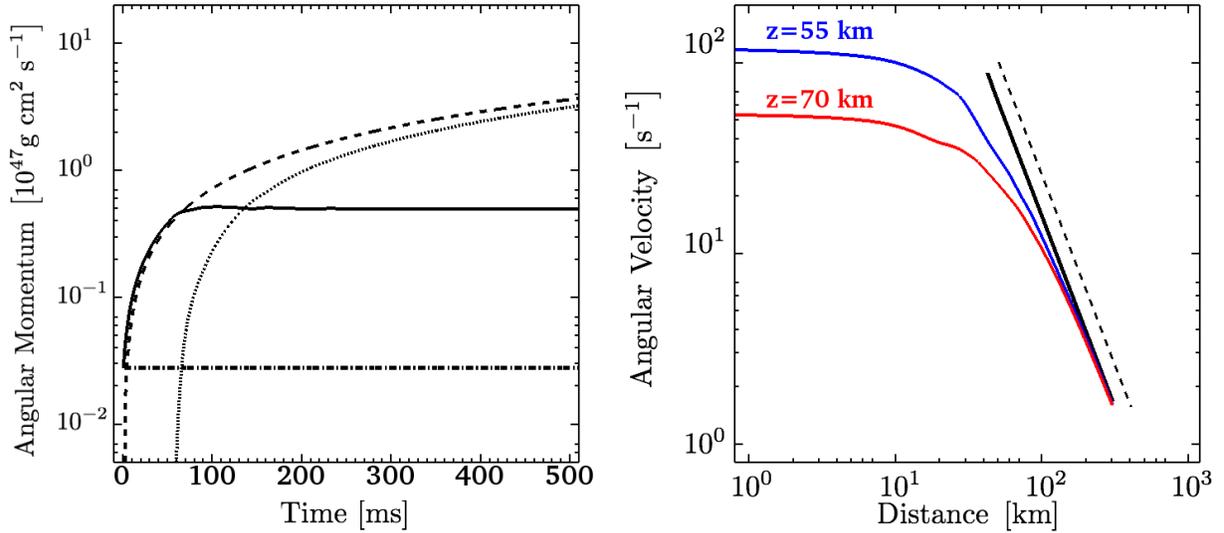}
  \caption{Results from an unperturbed model with rotation:  In the left panel we plot (versus time) the angular momentum in the computational domain (solid line), the total angular momentum entering through the outer boundaries of the computational box (dashed line), the angular momentum accreted on to the PNS through the cutout boundary (dotted line), and the total (conserved) angular momentum (dash-dot line).  In the right panel we plot angular velocity profiles, $\Omega_{z}=u_{\phi}/r_{\perp}$, at $t=300$~ms versus distance from the rotation axis, $r_{\perp}$, for selected values of the vertical coordinate $z$:  0~km (black), 55~km (blue), and 70~km (red).  The dashed reference line is proportional to $r_{\perp}^{-2}$.   \label{fig:angularMomentumAndAngularVelocityProfiles}}
\end{figure}

Due to the spherical shape of the inner `cutout' boundary our study is limited to include only models with rotation rates that do not result in a significant deviation from spherical symmetry.  Specifically, we will present results from a model where the specific angular momentum is set to $1.5\times10^{15}$~cm$^{2}$ s$^{-1}$.  This rotation rate is relatively slow, but similar to the level suggested by the stellar evolution calculations presented by \citet{heger_etal_2005}.  As demonstrated in Figure \ref{fig:angularMomentumAndAngularVelocityProfiles}, this model settles (in the absence of any explicit initial perturbations) to a quasi-steady state in a manner similar to the unperturbed, non-rotating model presented above.  For $t>0$ rotating material falls through the shock, propagates downstream, and flows through the inner boundary, beginning around $60$~ms.  The system settles rather quickly into a new equilibrium:  The angular momentum in the computational domain settles to a nearly constant value of about $0.5\times10^{47}$~g cm$^{2}$ s$^{-1}$ around $100$~ms, when the angular momentum accreting into the computational box is balanced by the angular momentum flowing through the inner cutout boundary.  At $500$~ms we find that about $3\times10^{47}$~g cm$^{2}$ s$^{-1}$ has accreted onto the PNS.  In the right panel of Figure \ref{fig:angularMomentumAndAngularVelocityProfiles} (snapshot taken at $t=300$~ms) we see that the angular velocity profile, $\Omega$, falls  off roughly as $r^{-2}$ away from the PNS.   The angular velocity is nearly constant near the rotation axis.  Near the surface of the PNS we find $\Omega\approx100$~s$^{-1}$ (or $P=2\pi/\Omega\approx 63$~ms).  The rate of change of the PNS's angular velocity due to accretion is
\begin{equation}
  \dot{\Omega}=\dot{L}/I-\left(\dot{M}/M-2\dot{R}_{\mbox{\tiny PNS}}/\rPNS\right)\Omega, 
  \label{eq:omegaDot}
\end{equation}
where $\dot{L}$, $I(\propto M R^{2})$, $\dot{M}$, and $\dot{R}_{\mbox{\tiny PNS}}$ are the rate of change of angular momentum due to accretion of rotating matter, the PNS's moment of inertia, the accretion rate, and the rate at which the PNS contracts during the accretion process, respectively.  The change in angular velocity of the PNS is inconsistent with the steady state assumption, but the inconsistency is of similar magnitude to that arising from ignoring the change in gravitational mass due to accretion.  

The small amount of rotation results in a continuous generation of magnetic field due to winding.  When the rotating model settles to a new equilibrium configuration there is a balance between magnetic energy generation from winding and loss due to advection through the inner boundary so that the total magnetic energy between the PNS and the shock stays nearly constant.  

The negative gradient (with respect to cylindrical radius) in the angular velocity profile should make the initial configuration unstable to the magnetorotational instability.  Ignoring the effects of buoyancy, the wavelength of the fastest growing unstable mode $\lambda_{\mbox{\tiny MRI}}^{\mbox{\tiny max}}$ is roughly equal to $2\pi v_{\mbox{\tiny A}}/\Omega$, where $v_{\mbox{\tiny A}}=B/\sqrt{\mu_{0}\rho}$ is the Alfv{\'e}n speed.  Using profiles from our unperturbed rotating model at $t=300$~ms we find $\lambda_{\mbox{\tiny MRI}}^{\mbox{\tiny max}}$ to be in the range of 1-10~km.  Since this spatial scale must be resolved by several ($\sim 10$) grid points we do not expect to capture the MRI in our simulations because of limited spatial resolution.  Also, the fastest growing unstable mode grows on a time scale that is proportional to the local rotation period, which, in our model, may be too long to be relevant on a time scale of 1~s.  Therefore, we do not consider the MRI in any further detail in the simulations presented here.  

Having shown that our numerical scheme is able to maintain a steady state in the absence of perturbations, we will now discuss the outcome of perturbed initial conditions in which the SASI is initiated and the magnetic field evolution is nontrivial.  

\section{Magnetic Field Amplification}

In this section we describe and explain the magnetic field amplification and saturation that result from SASI-induced flows.  

Initiation of the SASI requires a non-radial perturbation. \citetalias{blondin_etal_2003} showed that the qualitative features of the axisymmetric SASI do not depend on the details of the perturbation. Our experience confirms this result: While the details of early evolution may vary, we find that adding a small-amplitude asymmetric perturbation to the radial velocity field, applying random pressure perturbations inside the shock, and introducing overdense regions outside the shock all result in qualitatively similar outcomes.  In all the axisymmetric---as well as some of the 3D---simulations examined in this section, the initial condition is perturbed by the introduction of two regions in the pre-shock flow, one in the northern hemisphere and one in the southern hemisphere, whose density is increased by 20\% compared to the unperturbed model. In axisymmetry these overdense regions constitute rings of two different radii around the symmetry axis, but with the same circular cross section and placed at the same radial distance from the PNS \citepalias[similar to][as illustrated in the leftmost image of their Figure 6 after having fallen through the shock]{blondin_etal_2003}.  We term this the `axisymmetric perturbation'.  As was found by \citetalias{blondin_mezzacappa_2007}, relaxing the constraint imposed by axial symmetry allows for the development of the spiral SASI mode.  To avoid favoring the $\ell=1$ sloshing mode (which may occur when the axisymmetric perturbation is used to initiate the SASI in 3D models), we perturb some of the 3D simulations presented in this paper by introducing small-amplitude ($1\%$) random pressure perturbations to the initial configuration inside the shock.  

\subsection{Overview of Magnetic Field Evolution in Non-rotating Axisymmetric Models}

\begin{figure}
  \epsscale{1.0}
  \plotone{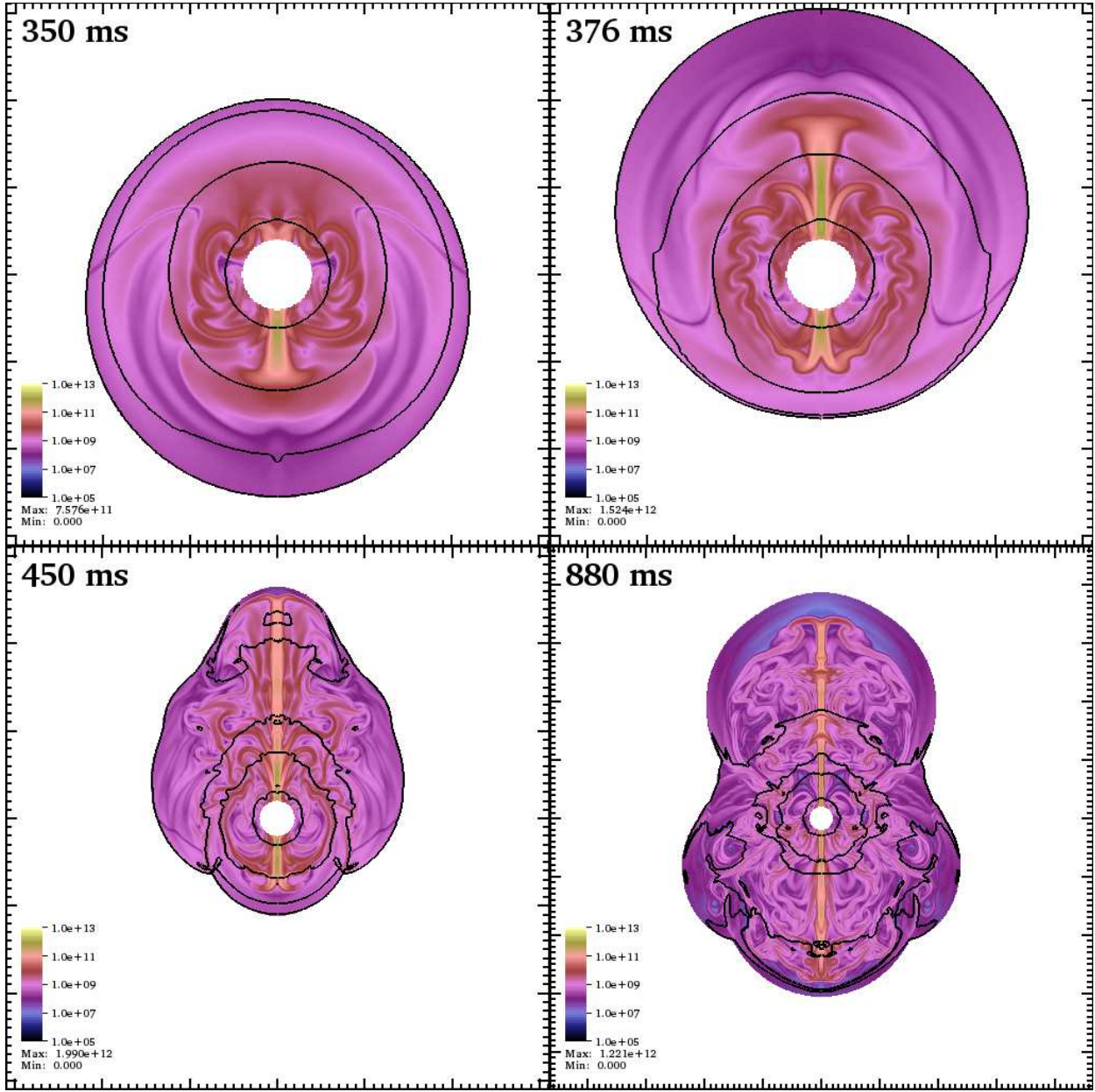}
  \caption{Selected snapshots of the evolution of the non-rotating axisymmetric SASI in the $r_{\perp}z$-plane, at times indicated in the upper left corner of each panel. The color scales give the magnetic field magnitude (in G). The black lines are contours of constant density; starting with the innermost, they denote $\rho=10^{10}$, $10^{9}$, $3 \times 10^{8}$, and $6 \times 10^{7}$ g cm$^{-3}$ (the fourth contour appearing only in the lower two panels). The sides of the two upper panels are 620~km, while the sides of the lower left and lower right panels are 1240~km and 1860~km, respectively.  We plot the solution on both sides of the symmetry axis to clearly illustrate the evolution in the polar regions .   \label{fig:magneticFieldOverview_medRes_2D_B0_1e10_l0_0e00_perturbed}}
\end{figure}

\begin{figure}
  \epsscale{1.0}
  \plottwo{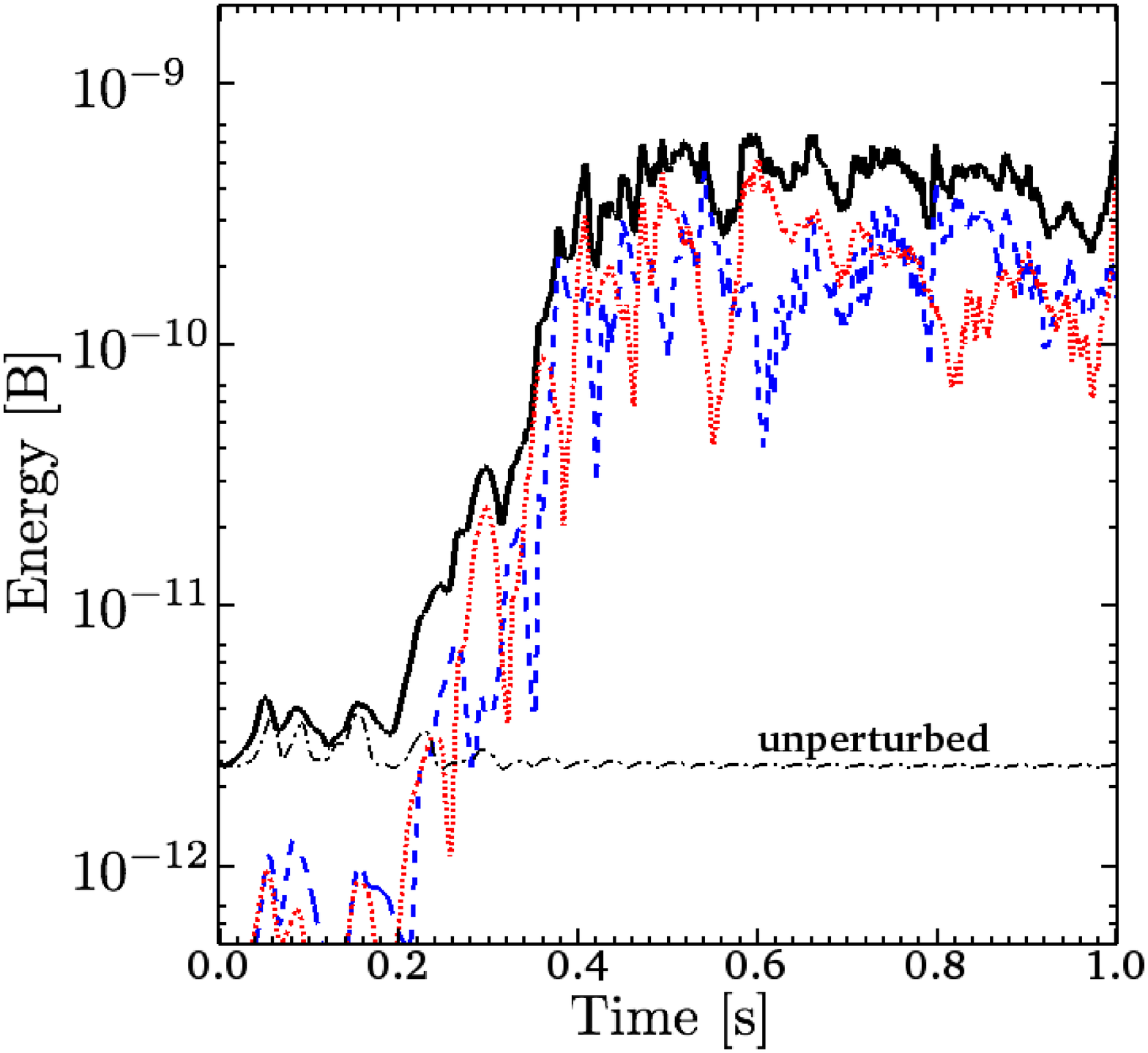}{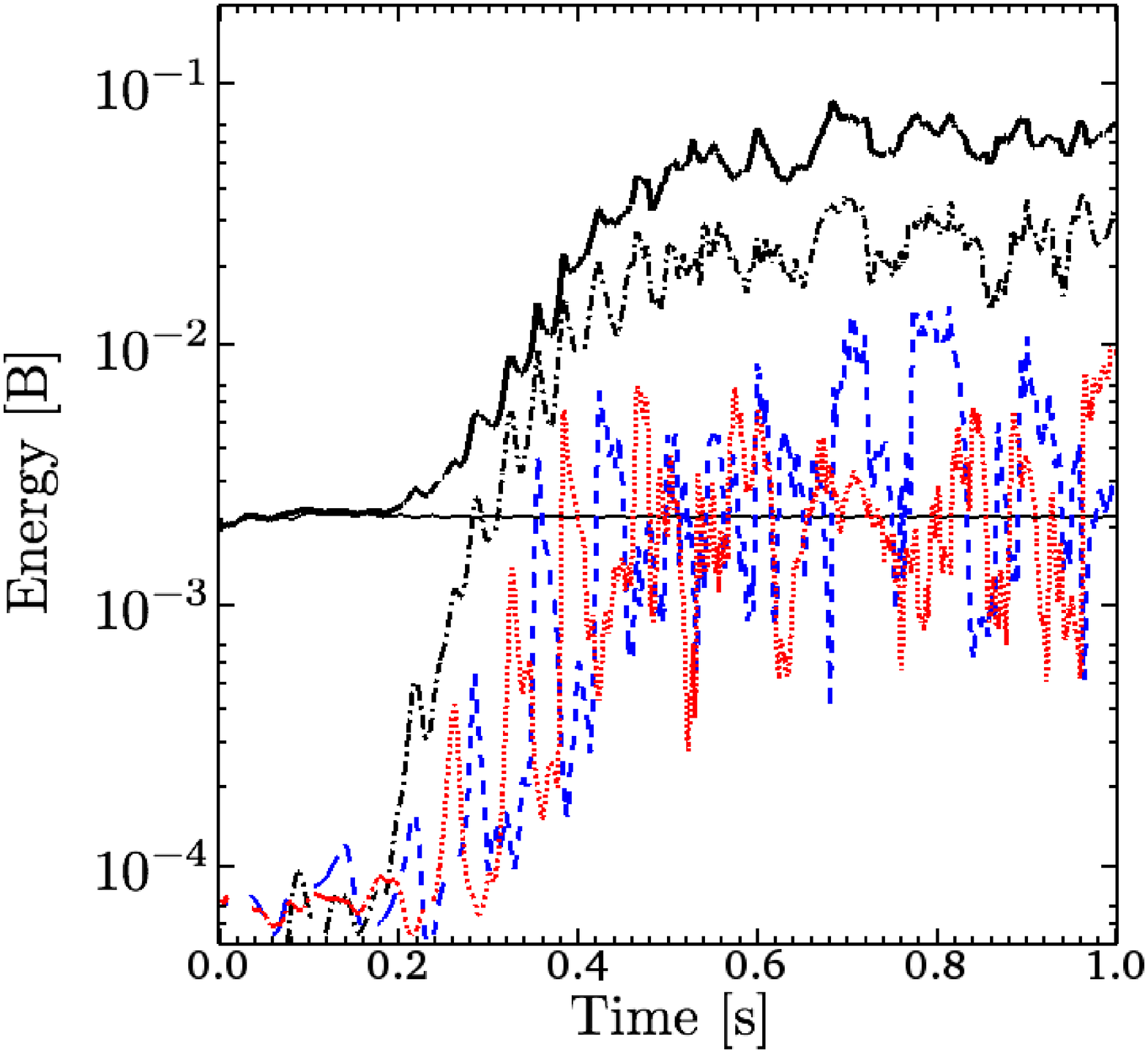}
  \caption{Magnetic (left panel) and kinetic (right panel) energies ($1 \ {\rm B} = 10^{51}$~erg) as a function of time in a non-rotating axisymmetric perturbed model.  The thick, solid lines represent energies integrated over the entire volume enclosed by the shock. Dashed (blue) and dotted (red) lines represent energies integrated over cylindrical regions centered on the symmetry axis in the northern and southern hemispheres respectively, with radius equal to $\rPNS$ and extending from the PNS surface (cutout boundary) to the shock. The dot-dashed line in the right panel represents the lateral kinetic energy in the volume enclosed by the shock. Also plotted for comparison is the magnetic and kinetic energy in the volume enclosed by the shock in an unperturbed model (dot-dashed line in the left panel and thin solid line in the right panel, respectively), which remain close to their initial values as expected.  \label{fig:magneticAndKineticEnergy_medRes_2D_B0_1e10_l0_0e00_perturbed}}
\end{figure}

Figures \ref{fig:magneticFieldOverview_medRes_2D_B0_1e10_l0_0e00_perturbed} and \ref{fig:magneticAndKineticEnergy_medRes_2D_B0_1e10_l0_0e00_perturbed} provide orienting overviews of the magnetic field evolution in an axisymmetric, non-rotating model with $B_{0}=10^{10}$~G (model 2DB10Am; all our axisymmetric models are listed in Table \ref{tbl:2Dmodels}).  In Figure \ref{fig:magneticFieldOverview_medRes_2D_B0_1e10_l0_0e00_perturbed} we plot four snapshots of the magnitude of the magnetic field inside the accretion shock, overlaid with selected density contours.  In Figure \ref{fig:magneticAndKineticEnergy_medRes_2D_B0_1e10_l0_0e00_perturbed} we plot the time evolution of subsets of magnetic (left panel) and kinetic (right panel) energies inside the accretion shock.  (See figure captions for further details.) Three epochs can be identified: An early oscillatory phase with relic transients from the initial perturbation that lasts, in this particular simulation, until about 200~ms; a period of growth in magnetic energy from about 200~ms to about 400~ms; and a final phase in which the magnetic energy remains at a more or less steady level with the strongest magnetic fields concentrated in the polar regions.  

As in the simulations of \citetalias{blondin_etal_2003}, an $\ell = 1$ `sloshing' mode heralds the onset of the SASI. In this early phase the shock remains quasi-spherical in shape, but its overall position with respect to the PNS shifts up and down along the symmetry axis with a period on the order of tens of milliseconds. The two extremes of one such oscillation are pictured in the two upper panels of Figure \ref{fig:magneticFieldOverview_medRes_2D_B0_1e10_l0_0e00_perturbed}, and the overall sloshing pattern is visibly traced by the alternating peaks in the dashed blue and dotted red curves in the right panel of Figure \ref{fig:magneticAndKineticEnergy_medRes_2D_B0_1e10_l0_0e00_perturbed} representing the kinetic energy in cylinders extending above and below the PNS, respectively.  Infalling matter strikes the shock at an oblique angle as a result of the displacement of the shock, and this introduces lateral flows inside the shocked cavity, whose share of kinetic energy steadily increases (note the growing trend of the dot-dashed curve in the right panel of Figure \ref{fig:magneticAndKineticEnergy_medRes_2D_B0_1e10_l0_0e00_perturbed} for $t\le500$~ms).  Lateral flows alternating towards and away from the symmetry axis in the northern and southern hemisphere compress and expand the magnetic field the polar regions; this results in a tendency for the maximum magnetic field to grow and decline alternately near the north and south poles, superimposed on a gradual trend of overall increase in magnetic field strength.  The total magnetic energy beneath the shock together with the magnetic energy in cylinders above and below the PNS are represented in the left panel of Figure \ref{fig:magneticAndKineticEnergy_medRes_2D_B0_1e10_l0_0e00_perturbed} by the black solid, blue dashed, and red dotted lines, respectively.  During this oscillatory phase, as exemplified in the two upper panels of Figure \ref{fig:magneticFieldOverview_medRes_2D_B0_1e10_l0_0e00_perturbed}, we note that the magnetic energy below the PNS ($z<-40$~km), at $t=350$~ms, is approaching a local maximum, while the kinetic energy in the same region is close to a local minimum.  This pattern is repeated above the PNS at $t=376$~ms:  The magnetic energy is near a local maximum, while the kinetic energy is rapidly decreasing.  The maximum magnetic field strengths reached at these instances are about $7.5\times10^{11}$~G (350~ms) and $1.5\times10^{12}$~G (376~ms), a noticeable increase from the initial condition's maximum magnitude of $10^{10}$~G.  

The increasing amplitude of the SASI's early oscillations eventually leads to vigorous fluid flows and a notably aspherical shock morphology, but without any dramatic consequences for the magnetic field growth.  As the SASI wave develops nonlinearly and the flows beneath the shock become nearly tangential to the shock surface an internal shock forms in the post-shock gas around 390~ms.  The internal shock is connected to the accretion shock in a triple point \citepalias[cf.][]{blondin_mezzacappa_2007}, which is visible as a kink in the shock surface, located south of the equatorial region ($z<0$) in the lower left panel of Figure \ref{fig:magneticFieldOverview_medRes_2D_B0_1e10_l0_0e00_perturbed}.  Ahead of the internal shock, which is traveling latitudinally southward at $t=450$~ms, a fast stream of less-severly-shocked (low-entropy) material penetrates down towards the PNS.  Such streams alternately hit the symmetry axis in the northern and southern hemisphere and result in magnetic field compression, but the magnetic energy does not grow beyond the roughly two orders of magnitude growth seen for $t<400$~ms.  (The saturation level of magnetic energy is sensitive to the spatial resolution of the numerical simulation.  We will discuss this issue further in later sections.)  For $t>400$~ms the separate curves of magnetic energy in the northern and southern polar cylinders reveal that there are intermittent episodes of rise and decline in each polar region.  From $t=550$~ms to about $600$~ms there is an episode of magnetic field growth in the southern polar region.  This growth coincides with the presence of a high-speed stream plunging into the symmetry axis south of the PNS.  The head of this stream wanders to the northern hemisphere and causes the growth seen between 600~ms and 660~ms in the dashed blue curve in the left panel of Figure \ref{fig:magneticAndKineticEnergy_medRes_2D_B0_1e10_l0_0e00_perturbed}.  The magnitude of the magnetic field is about $2\times10^{12}$~G in both the northern and southern polar regions, and the magnetic field does not have any significant impact on the fluid flows.  

The SASI is dominated by a significant $\ell=2$ mode at the latest time displayed in Figure \ref{fig:magneticFieldOverview_medRes_2D_B0_1e10_l0_0e00_perturbed} with a focused accretion funnel in the equatorial region and outflows along the pole \citepalias[cf.][]{blondin_etal_2003}.  At this time the total magnetic energy in the shocked volume (solid line in the left panel of Figure \ref{fig:magneticAndKineticEnergy_medRes_2D_B0_1e10_l0_0e00_perturbed}) is still many orders of magnitude smaller than the total kinetic energy in the shocked volume (solid line in the right panel of Figure \ref{fig:magneticAndKineticEnergy_medRes_2D_B0_1e10_l0_0e00_perturbed})---even in the polar regions, within which the overwhelming majority of the magnetic energy is concentrated.  We find that the magnetic energy near the symmetry axis may intermittently reach about 10\% of the kinetic energy in localized regions, but stays mostly below 0.01\%.  Moreover the ratio of magnetic pressure to fluid pressure, $\beta^{-1}$, reaches only a few times $10^{-4}$, and this only in localized regions.  Outside the polar regions the magnetic field develops a highly intermittent flux tube structure---or rather, in axisymmetry, a flux sheet structure.  

\begin{figure}
  \epsscale{1.0}
  \plotone{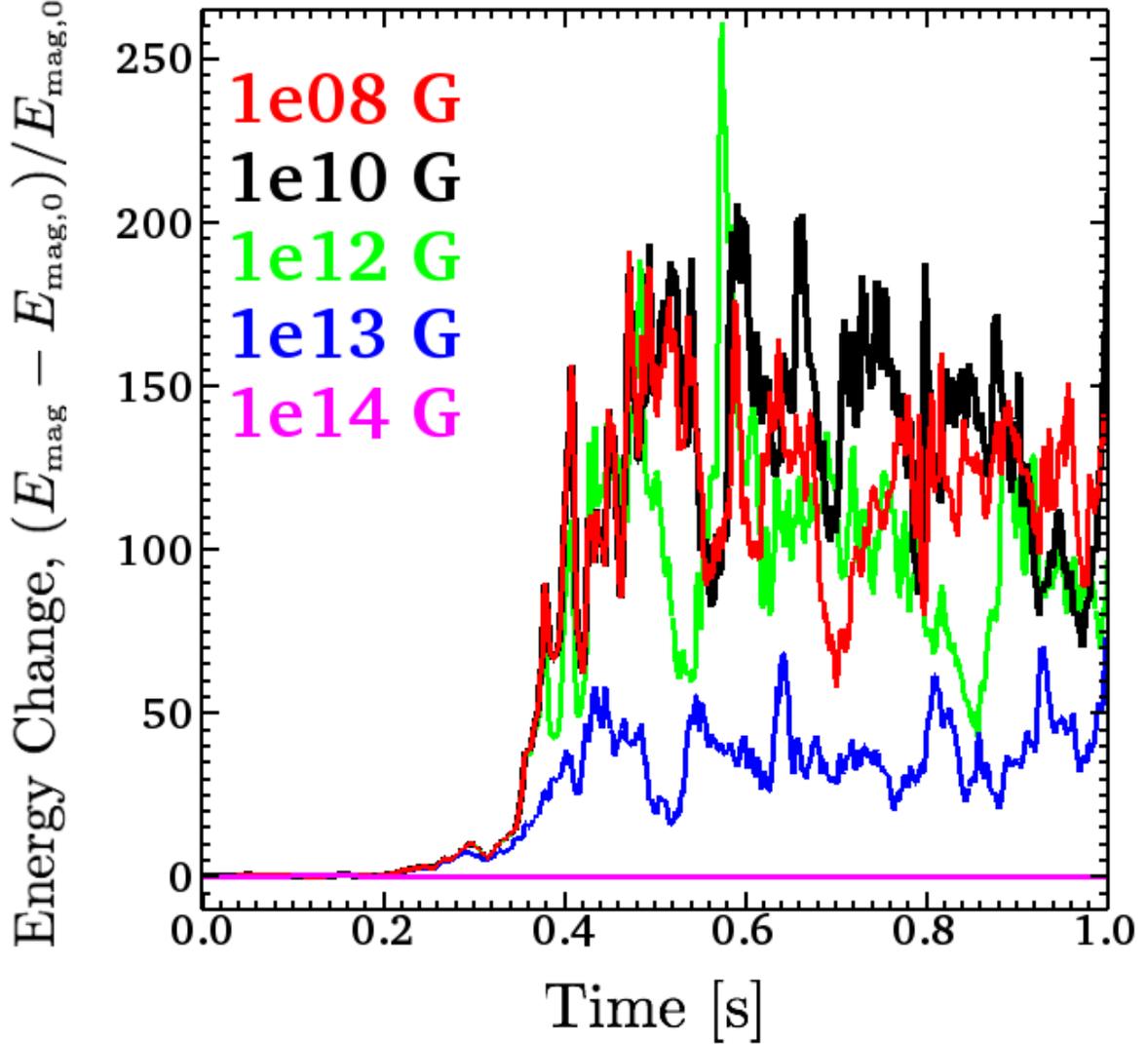}
  \caption{Relative change in magnetic energy for non-rotating axisymmetric models with different initial magnetic field strength, $B_{0}$, at the surface of the PNS:  $10^{8}$~G (red), $10^{10}$~G (black), $10^{12}$~G (green), $10^{13}$~G (blue), and $10^{14}$~G (magenta).  The initial magnetic energy, $E_{\mbox{\tiny mag,0}}$, is $2.3\times\left[10^{-16},10^{-12},10^{-8},10^{-6},10^{-4}\right]$~B, for the respective models.  \label{fig:varyingMagneticFieldStrength_2D_nonRotating_perturbed}}
\end{figure}

We have also performed calculations with different initial magnetic field strengths:  Figure \ref{fig:varyingMagneticFieldStrength_2D_nonRotating_perturbed} shows the relative change in the magnetic energy inside the accretion shock for five models in which the initial strength of the magnetic field at the surface of the PNS, $B_{0}$, ranges from $10^{8}$~G to $10^{14}$~G.  The early evolution of the magnetic field is identical in all models with $B_{0}\le10^{13}$~G up to $t=280$~ms, where the model with $B_{0}=10^{13}$~G begins to separate.  Around $t=380$~ms the model with $B_{0}=10^{12}$~G separates from the two models with the weaker initial field, which follow each other closely until $t\approx480$~ms.  The development of the SASI is entirely suppressed in the model with $B_{0}=10^{14}$~G:  the initial magnetic field is dynamically significant in this model and the axisymmetric perturbation is damped out.  We also note that the model with $B_{0}=10^{13}$~G receives a notably smaller relative boost in magnetic energy.  

We have now presented results from calculations of the axisymmetric SASI.  Dominated by the $\ell=1$ sloshing mode, it is capable of amplifying the magnetic energy beneath the shock by about two orders of magnitude (for our standard zone size of 1.56 km).  The amplified magnetic field is concentrated near the symmetry axis and is not dynamically significant for any of the models with a modest initial magnetic field (e.g., $B_{0}\le10^{12}$~G). 

\subsection{Overview of Magnetic Field Evolution in 3D Models}

The final magnetic field configuration of the axisymmetric models of the previous subsection is probably unrealistic, at least for progenitors in which rotation is not important and an axis of symmetry cannot be defined.  Thus we do not in general expect the magnetic field structure seen at late times in axisymmetric 2D models (i.e., the accumulation around the symmetry axis) to carry over to 3D models. Indeed, recent calculations by \citetalias{blondin_mezzacappa_2007} and \citet{blondin_shaw_2007} have shown that the non-axisymmetric spiral mode, which is closely related to the axisymmetric sloshing mode, may dominate the post shock flow at late times. To investigate the magnetic field evolution in non-axisymmetric flows and thereby increase the realism of our simulations, we have performed 3D calculations with a moderate initial magnetic field.  (A tabular overview of all our 3D models is given in Table \ref{tbl:3Dmodels}.) As the axisymmetric SASI is not noticeably influenced by the presence of a moderate initial magnetic field, we likewise expect the hydrodynamical aspects of our 3D MHD SASI calculations to evolve in a manner similar to 3D simulations without magnetic fields. 
  
\subsubsection{A Non-rotating Reference Model}

We begin our exploration of magnetic field evolution in 3D models by presenting a model (3DB12Am) that is perturbed by the same procedure as the axisymmetric calculations (i.e., two dense tori centered on the $z$-axis are placed outside the shock, one in the northern and one in the southern hemisphere).  There is no physical motivation for applying this particular initial perturbation other than to initiate the sloshing mode which dominates the 2D simulations.  As a reference, this model naturally ties the axisymmetric 2D models to the 3D models with random perturbations we will present below.  

The initial magnetic field strength is $B_{0}=10^{12}$~G in this model, while in axisymmetry we emphasized evolution with $B_{0}=10^{10}$~G.  We have chosen to compute most of our 3D models with $B_{0}=10^{12}$~G, but we have also computed one 3D model with $B_{0}=10^{10}$~G.  

\begin{figure}
  \epsscale{1.0}
  \plotone{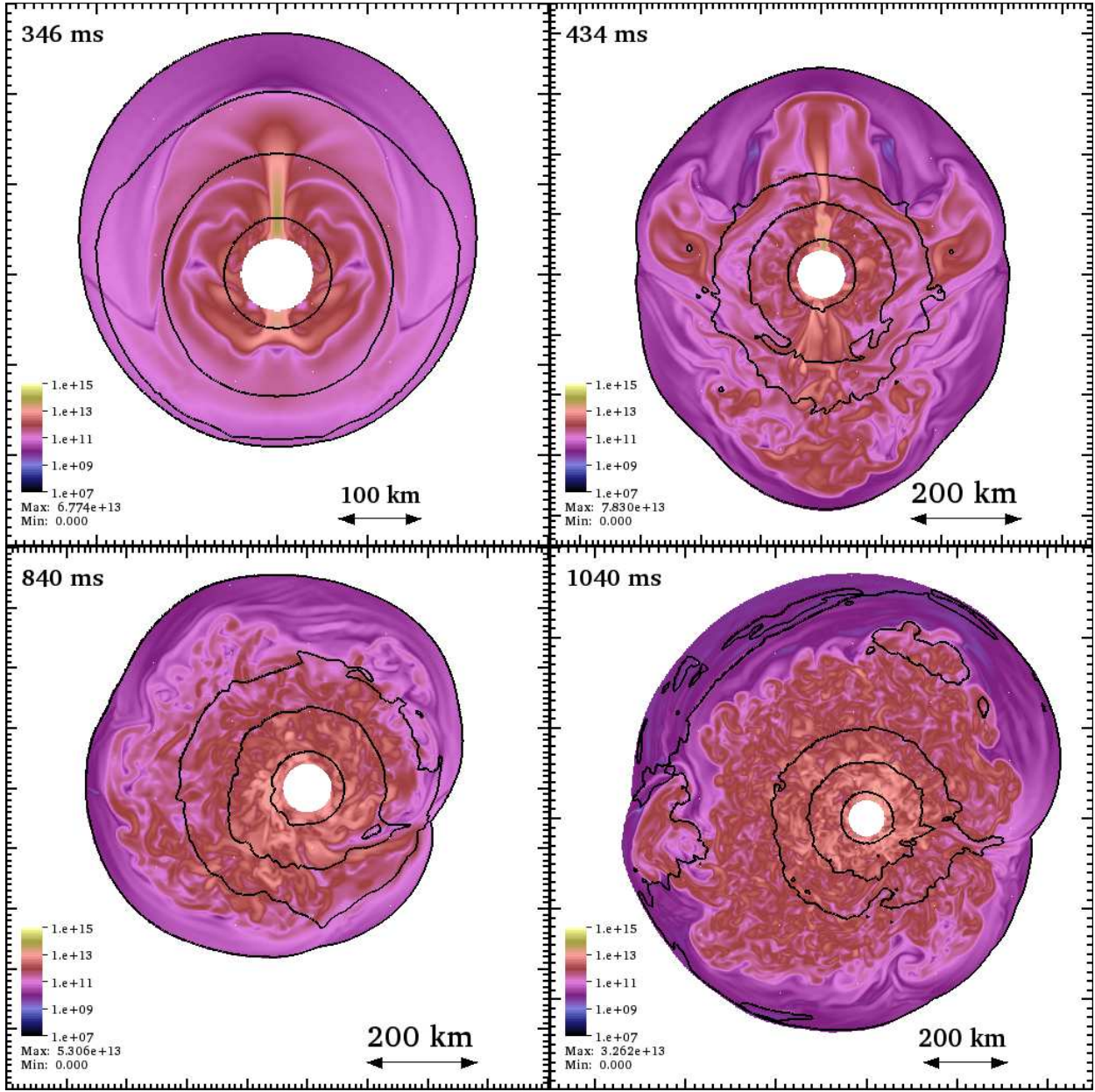}
  \caption{Selected snapshots showing the evolution and distribution of the magnitude of the magnetic field in a slice through an initially non-rotating 3D SASI model (3DB12Am), at times indicated in the upper left corner of each panel.  The two upper panels show the evolution in a slice through the $xz$-plane, while the two lower panels show the evolution in a plane whose normal vector is parallel to the total angular momentum vector of the flow between the PNS and the shock surface. Black lines are contours of constant density; starting with the innermost, they denote $\rho=10^{10}$, $10^{9}$, $3 \times 10^{8}$, and $6 \times 10^{7}$ g cm$^{-3}$.  (As in Figure \ref{fig:magneticFieldOverview_medRes_2D_B0_1e10_l0_0e00_perturbed}, the last contour is only visible in the lower right panel).  \label{fig:magneticFieldOverview_medRes_3D_B0_1e12_l0_0e00_perturbed}}
\end{figure}

\begin{figure}
  \epsscale{1.0}
  \plotone{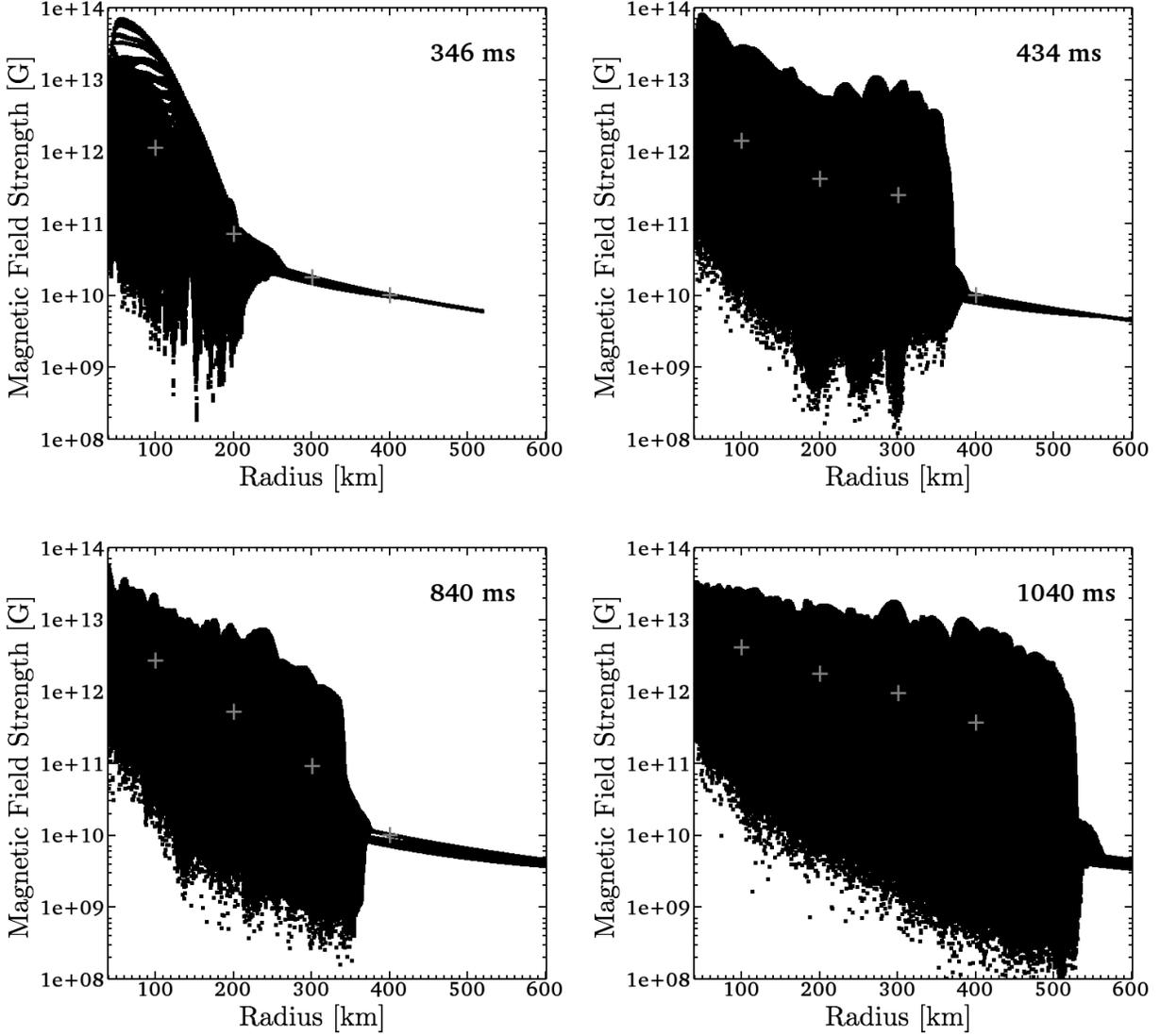}
  \caption{Scatter plots of magnetic field magnitude versus spherical radius taken from model (3DB12Am), taken at the times shown in Figure \ref{fig:magneticFieldOverview_medRes_3D_B0_1e12_l0_0e00_perturbed}.  Plus signs denote the rms value of the magnetic field, computed in spherical shells bounded by $r^{\pm}=r\pm 25$~km, with $r=100$, $200$, $300$, and $400$~km.   \label{fig:magneticFieldOverview_medRes_3D_B0_1e12_l0_0e00_perturbed_scatterPlot}}
\end{figure}

\begin{figure}
  \epsscale{1.0}
  \plottwo{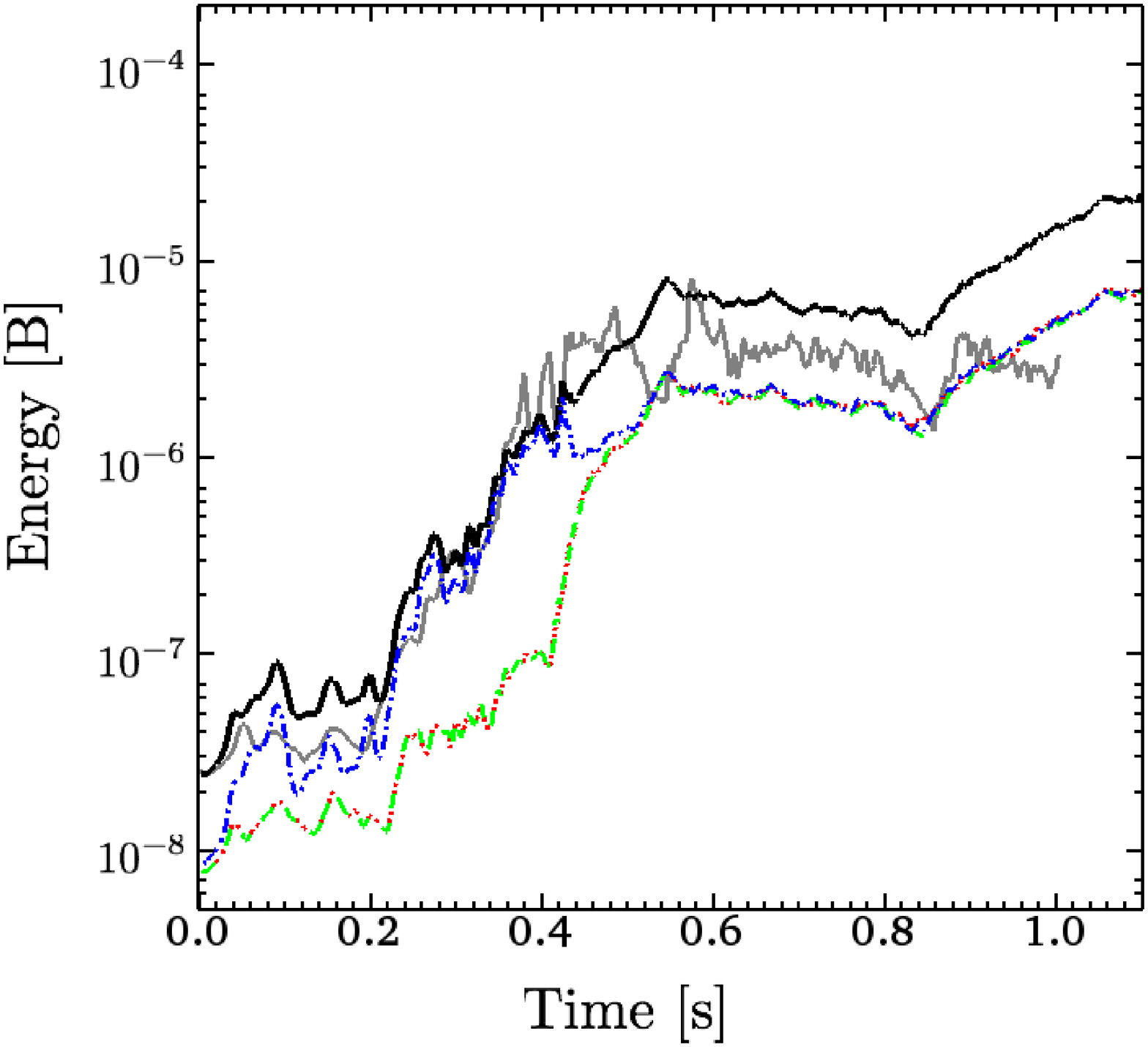}{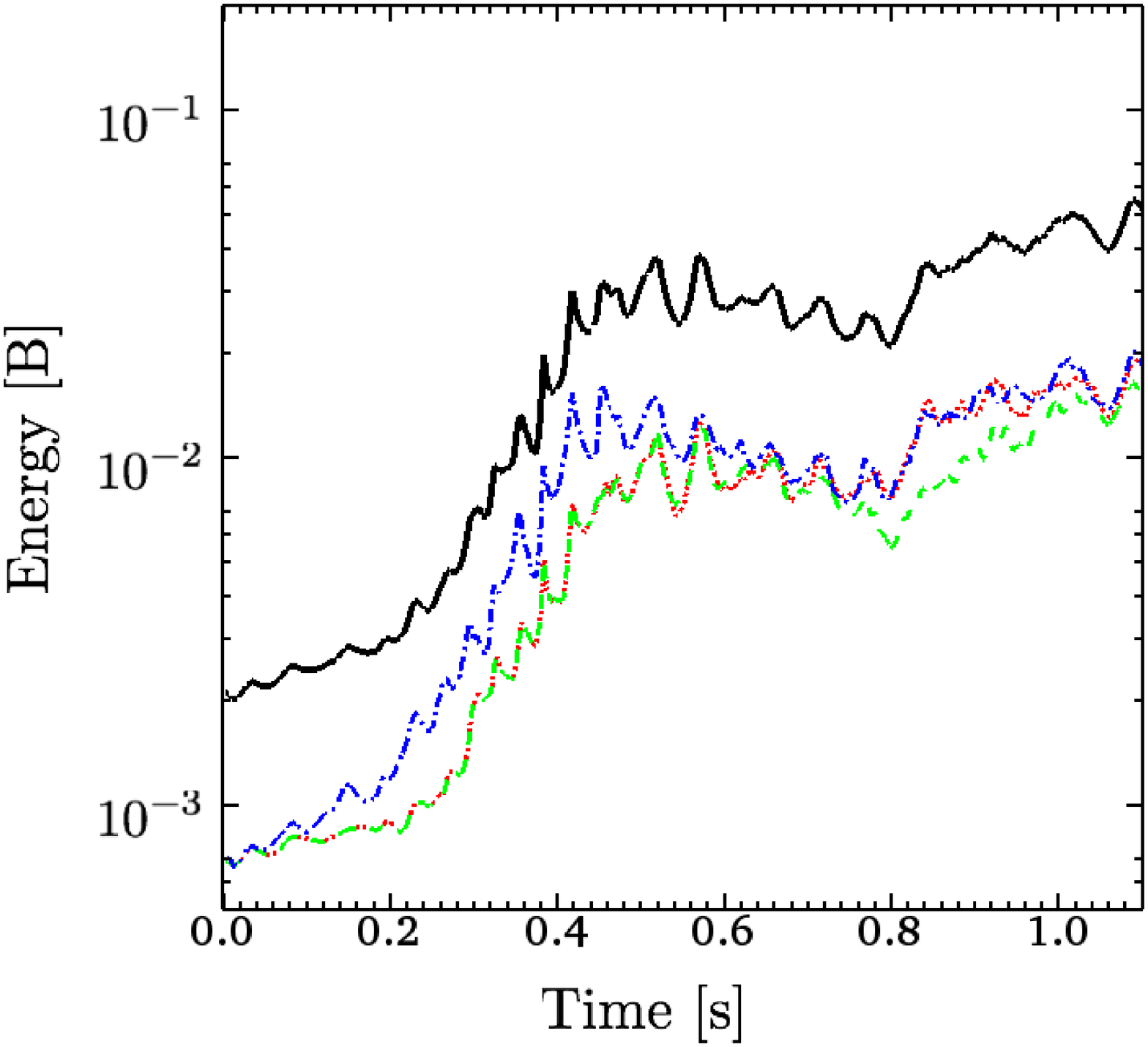}
  \caption{Evolution of magnetic and kinetic energies in a 3D model perturbed with the axisymmetric perturbation (model 3DB12Am).  In the left panel we plot the total magnetic energy inside the shock $E_{\mbox{\tiny mag}}$ (solid, thick black line),  the individual components $E_{\mbox{\tiny mag,}x}$, $E_{\mbox{\tiny mag,}y}$, and $E_{\mbox{\tiny mag,}z}$ (dotted red, dashed green, and dash-dot blue lines, respectively).  We have also plotted the total magnetic energy inside the shock for a 2D axisymmetric model (model 2DB12Am; solid grey line).  In the right panel we plot the evolution of the kinetic energy inside the accretion shock:  Total (solid black line), and the individual components $E_{\mbox{\tiny kin,}x}$ (dotted red line), $E_{\mbox{\tiny kin,}y}$ (dashed greed line), and $E_{\mbox{\tiny kin,}z}$ (dash-dot blue line).   \label{fig:magneticAndKineticEnergy_medRes_3D_B0_1e12_l0_0e00_axiSymmetricPerturbation}}
\end{figure}

Figures \ref{fig:magneticFieldOverview_medRes_3D_B0_1e12_l0_0e00_perturbed}, \ref{fig:magneticFieldOverview_medRes_3D_B0_1e12_l0_0e00_perturbed_scatterPlot}, and \ref{fig:magneticAndKineticEnergy_medRes_3D_B0_1e12_l0_0e00_axiSymmetricPerturbation} provide orienting overviews of the evolution:  in Figure \ref{fig:magneticFieldOverview_medRes_3D_B0_1e12_l0_0e00_perturbed} we show color plots of the magnitude of the magnetic field at selected times supplied with density contours; in Figure \ref{fig:magneticFieldOverview_medRes_3D_B0_1e12_l0_0e00_perturbed_scatterPlot} we show scatter plots of the magnetic field magnitude versus radius; and in Figure \ref{fig:magneticAndKineticEnergy_medRes_3D_B0_1e12_l0_0e00_axiSymmetricPerturbation} we plot the integrated magnetic (left panel) and kinetic (right panel) energy below the shock versus time.  (Movie 1\footnote{The animations submitted with this work was produced by Ross Toedte at the Oak Ridge Leadership Computing Facility, ORNL.  } in the online material shows the full evolution of the magnetic field magnitude in this model.)  

Due to the particular applied perturbation, this model's early evolution is very similar to the evolution of the axisymmetric 2D models with moderate initial magnetic fields:  the sloshing mode is clearly evident in the upper left panel of Figure \ref{fig:magneticFieldOverview_medRes_3D_B0_1e12_l0_0e00_perturbed}.  The SASI-induced flows then result in magnetic field amplification around the $z$-axis where most of the magnetic energy is concentrated.  The bulk of the magnetic energy as well as the kinetic energy of the flow beneath the shock are stored in the $z$-components, $E_{\mbox{\tiny mag,}z}=B_{z}^{2}/(2\mu_{0})$ and $E_{\mbox{\tiny kin,}z}=\rho u_{z}^{2}/2$, respectively.  Also note that the $x$- and $y$-components of both the magnetic and kinetic energies fall on top of each other, indicating good preservation of axial symmetry.  The evolution of the total magnetic energy follows the corresponding 2D model (also plotted in Figure \ref{fig:magneticAndKineticEnergy_medRes_3D_B0_1e12_l0_0e00_axiSymmetricPerturbation}) closely up to about 420~ms.  At this point non-axisymmetric modes\footnote{Our use of Cartesian coordinates and numerical roundoff errors are likely to break the initial symmetry in this model.} become significant and result in a rapid disruption of the magnetic structure which has formed along the initial symmetry axis:   Around 400~ms we find $E_{\mbox{\tiny mag}}\approx E_{\mbox{\tiny mag,}z}$, while for $t\ge480$~ms the magnetic energy is distributed equally among the components with $E_{\mbox{\tiny mag,}x}\approx E_{\mbox{\tiny mag,}y}\approx E_{\mbox{\tiny mag,}z}\approx E_{\mbox{\tiny mag}}/3$.  Remnants of the magnetic field structure along the symmetry axis is still visible in the upper right panel of Figure \ref{fig:magneticFieldOverview_medRes_3D_B0_1e12_l0_0e00_perturbed}.  

\begin{figure}
  \epsscale{1.0}
  \plotone{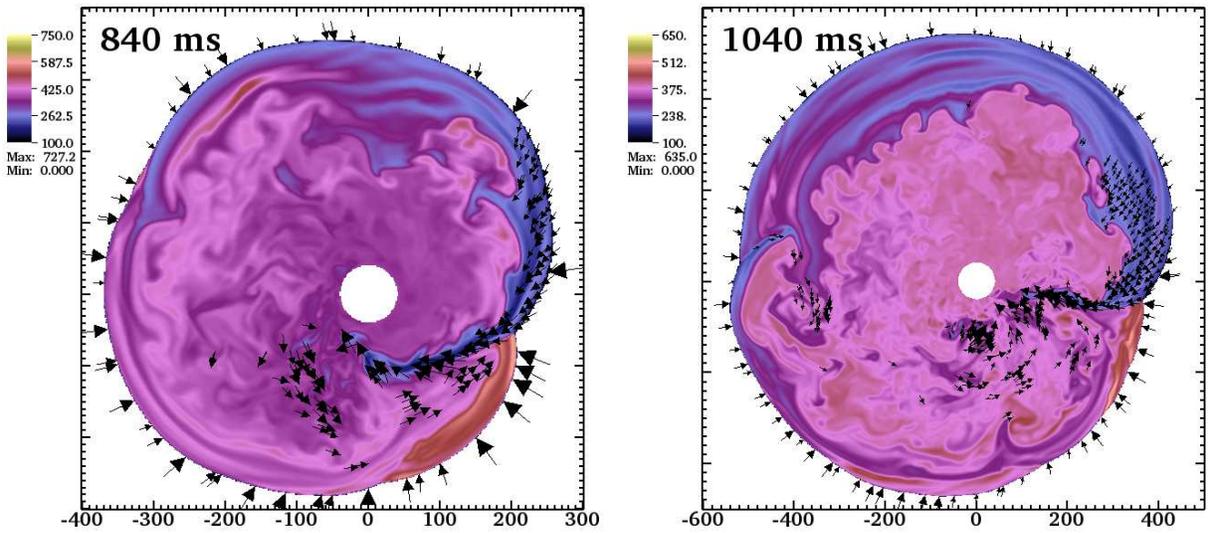}
  \caption{Color plot  of the polytropic constant ($\kappa=P/\rho^{\gamma}$) at selected times during the operation of the SASI spiral mode.  These panels correspond to the two lower panels of the magnetic field in Figure \ref{fig:magneticFieldOverview_medRes_3D_B0_1e12_l0_0e00_perturbed}.  Velocity vectors where $|\vect{u}|\ge c_{S}=\sqrt{\gamma P/\rho}$ are overlaid both plots.  \label{fig:sasi_MHD_3D_medRes_B0_1e12_polytropicConstant_twoPanel_spiralMode}}
\end{figure}

\begin{figure}
  \epsscale{1.0}
  \plotone{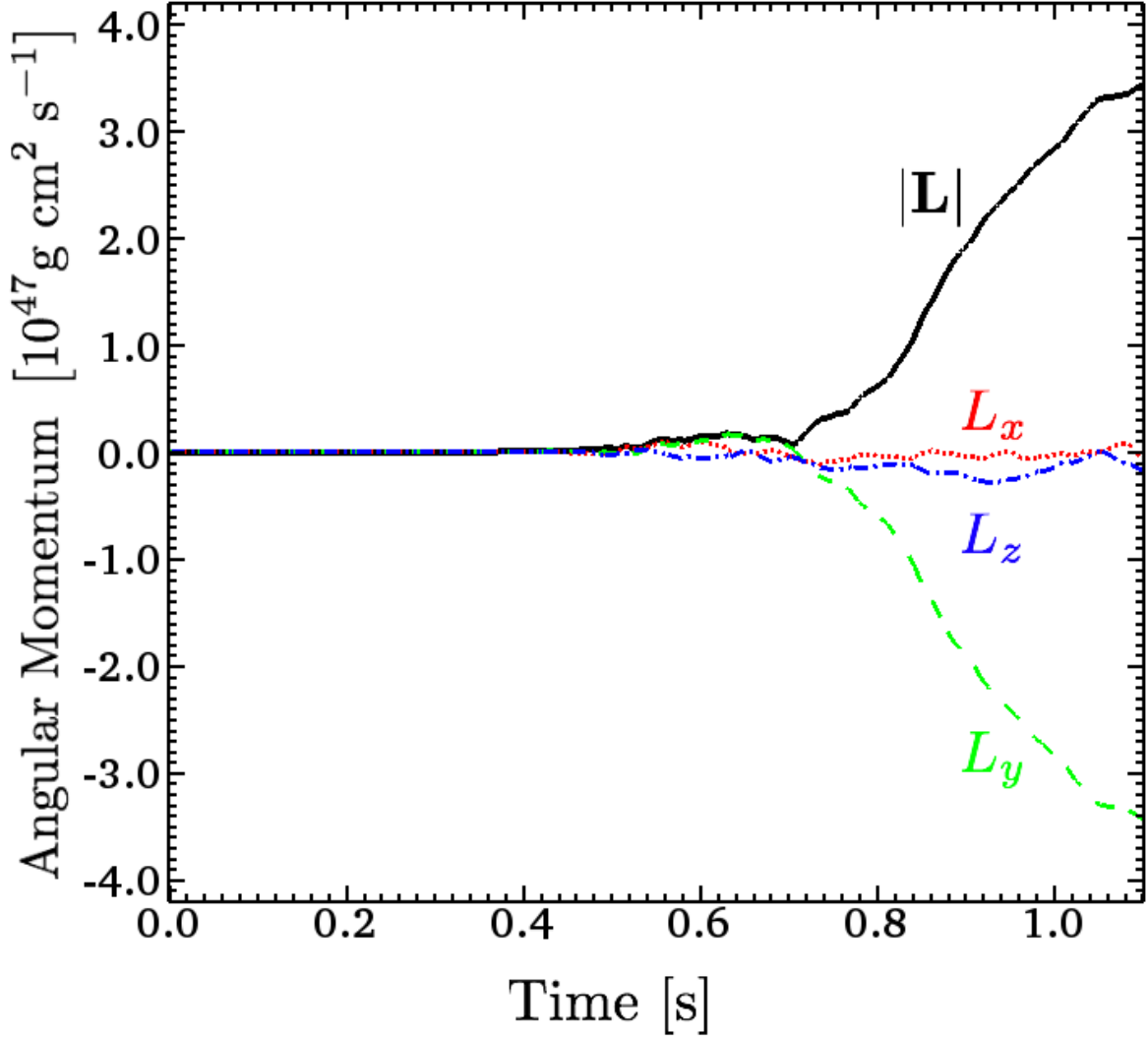}
  \caption{Angular momentum of the matter interior to the shock and exterior to the PNS for model 3DB12Am.  We plot the total angular momentum $|\vect{L}|=\sqrt{L_{x}^{2}+L_{y}^{2}+L_{z}^{2}}$ (solid black line), and the individual components $L_{x}$ (dotted red line), $L_{y}$ (dashed greed line), and $L_{z}$ (dash-dot blue line).  \label{fig:angularMomentum_3D_medRes_B0_1e12_l0_0.0e00}}
\end{figure}

The magnetic energy continues to grow as the SASI develops nonlinearly until about $t=540$~ms (see Figure \ref{fig:magneticAndKineticEnergy_medRes_3D_B0_1e12_l0_0e00_axiSymmetricPerturbation}).  Then there is a period extending almost $300$~ms, associated with the disruption of the axisymmetric polar structure, in which the magnetic energy between the PNS and the shock surface shows a declining trend before it begins to grow again for $t>830$~ms.  (The decline in magnetic energy is a combined result of a shrinking shock volume, $V_{\mbox{\tiny Sh}}$; the loss of the polar concentration of magnetic energy through the surface of the PNS; and a slowed magnetic energy generation rate while the post-shock flow rearranges.  The magnetic energy density, $E_{\mbox{\tiny mag}}/V_{\mbox{\tiny Sh}}$, continues to grow slowly during this phase.)  At the end of the calculation ($t=1.1$~s) the magnetic energy appears to be leveling off at a value of about $2\times 10^{-5}$~B, almost three orders of magnitude higher than the initial value.  After the initial increase associated with the development of the sloshing mode, the total kinetic energy beneath the shock evolves in a manner similar to the magnetic energy, exhibiting a declining trend during the `pausing' phase from $540$~ms to $830$~ms followed by a gradual increase.  The large-scale fluid flows are composed of a significant $\ell=1$ component up to about $t=615$~ms, but becomes less organized during the `pause', when the components of the kinetic energy become roughly equal (e.g., $E_{\mbox{\tiny kin,}x}\approx E_{\mbox{\tiny kin,}y}\approx E_{\mbox{\tiny kin,}z}$).  Then, around $t=800$~ms a clear spiral mode pattern emerges in the flow.  This is illustrated in Figure \ref{fig:sasi_MHD_3D_medRes_B0_1e12_polytropicConstant_twoPanel_spiralMode}, where we plot the polytropic constant (a proxy for the fluid entropy) at two instances separated by two full revolutions about the PNS, and Figure \ref{fig:angularMomentum_3D_medRes_B0_1e12_l0_0.0e00}, where the angular momentum of the fluid between the PNS and the shock is plotted versus time.  At both times displayed in Figure \ref{fig:sasi_MHD_3D_medRes_B0_1e12_polytropicConstant_twoPanel_spiralMode}, $t=840$~ms (left panel) and $t=1040$~ms (right panel), a shock triple point, moving in the counterclockwise direction, is clearly visible in the right portion.  This `cleft' extends over most of a hemisphere on the shock surface.  The development of the spiral mode results in a significant amount of angular momentum about the PNS:  the angular momentum of the flow between the PNS and the shock is $3.4\times 10^{47}$ g cm$^{2}$ s$^{-1}$ at $t=1.1$~s, which is consistent with what was reported by \citetalias{blondin_mezzacappa_2007}.  (This angular momentum is balanced by angular momentum advected through the cutout and deposited onto the PNS so that the total angular momentum is conserved.  The presence of a sufficiently strong magnetic field in this region might enable angular momentum transport between these counter-rotating flows, but the exclusion of the PNS from the computational domain prohibits meaningful treatment of this interaction.)  For this particular model we see in Figure \ref{fig:angularMomentum_3D_medRes_B0_1e12_l0_0.0e00} that the total angular momentum is roughly aligned with the $y$-axis, which is also consistent with the $x$- and $z$-components of the kinetic energy being roughly equal and slightly larger than the $y$-component as is seen in the right panel of Figure \ref{fig:magneticAndKineticEnergy_medRes_3D_B0_1e12_l0_0e00_axiSymmetricPerturbation}.  

\begin{figure}
  \epsscale{1.0}
  \plotone{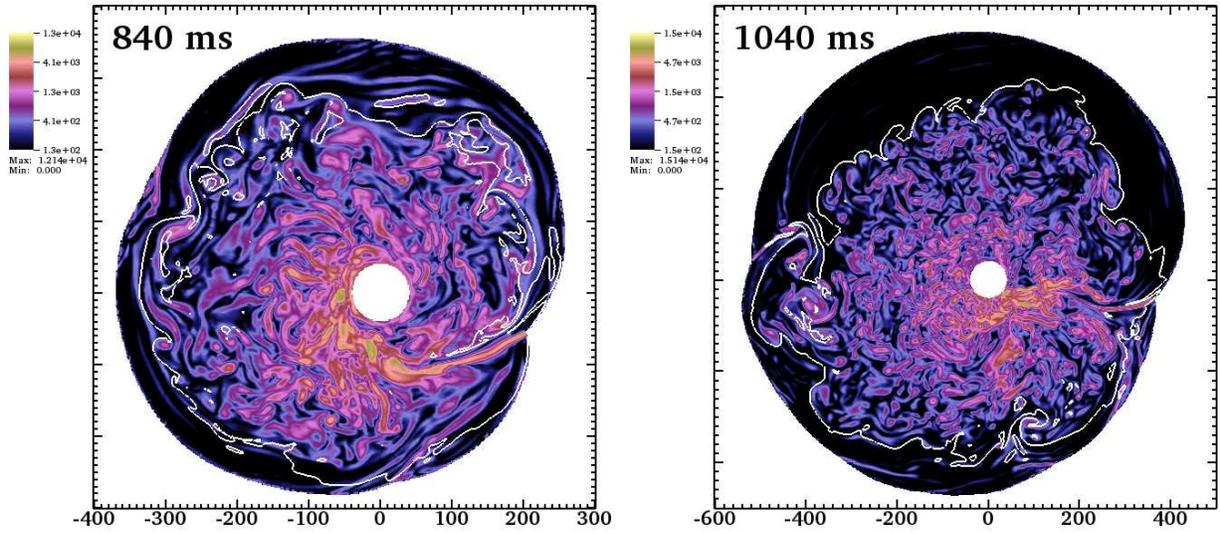}
  \caption{Color plot showing the distribution of fluid vorticity $\left|\vect{\omega}\right|\left(=\left|\nabla\times\vect{u}\right|\right)$.  Vorticity is plotted in units of s$^{-1}$.  These panels correspond to the two lower panels of the magnetic field in Figure \ref{fig:magneticFieldOverview_medRes_3D_B0_1e12_l0_0e00_perturbed}.  White contours are drawn where the magnitude of the magnetic field equals $6\times10^{10}$~G (left panel) and $4\times10^{10}$~G (right panel).  \label{fig:sasi_MHD_3D_medRes_B0_1e12_vorticityMagnitude_twoPanel_spiralMode}}
\end{figure}

Ahead of the triple point a supersonic stream (indicated by the black velocity vectors in Figure \ref{fig:sasi_MHD_3D_medRes_B0_1e12_polytropicConstant_twoPanel_spiralMode}) penetrates down toward the PNS.   This plunging stream introduces a region of persistent shear flow inside the supernova shock wave.  This flow may be susceptible to the Kelvin-Helmholtz instability and other fluid instabilities associated with velocity shear which often result in turbulence.  We find that in our calculations the emergence of the triple point and the associated plunging stream results in a highly turbulent flow beneath the shock which appears to be important for subsequent magnetic field evolution.  In Figure \ref{fig:sasi_MHD_3D_medRes_B0_1e12_vorticityMagnitude_twoPanel_spiralMode} we plot the fluid vorticity (a local measure of the rate of rotation of the fluid) at the same times selected for the two lower panels of Figure \ref{fig:magneticFieldOverview_medRes_3D_B0_1e12_l0_0e00_perturbed} (also the times selected for the polytropic constant in Figure \ref{fig:sasi_MHD_3D_medRes_B0_1e12_polytropicConstant_twoPanel_spiralMode}).  The vorticity is generated in the shearing region connected to the shock triple point and distributed in the post-shock flow.  Note that the main body of the dominant magnetic field becomes somewhat separated from the larger shock volume and there is clearly a similarity in the distribution of magnetic field and fluid vorticity at these times in the simulation.  It has been pointed out \citep[e.g.,][]{mee_brandenburg_2006} that the presence of fluid vorticity is helpful for magnetic field generation.  

At the end of the simulation the magnetic field has evolved into a highly intermittent `flux rope' structure (this 'flux rope' structure is illustrated in Movie 2 and Movie 3, which show the magnitude of the magnetic field during the operation of the SASI spiral mode from $t=832$~ms to $t=1094$~ms and a full revolution of a still frame at $t=1040$~ms, respectively).  Most of the magnetic energy beneath the shock is stored in fields with strength around $10^{12}$~G, but there are extended regions where the magnetic field strength exceeds $10^{13}$~G (cf. Figure \ref{fig:magneticFieldOverview_medRes_3D_B0_1e12_l0_0e00_perturbed_scatterPlot}, but also discussion in Section \ref{sec:varyingResolution3D}) that contributes in a non-negligible way to the total magnetic energy.  However, the magnetic energy is still far below both the internal and kinetic energy of the post shock flow.  In localized regions the magnetic energy density reaches up to about $10\%$ of the kinetic energy density, and $\beta^{-1}$ does (intermittently) reach values of a few times $10^{-2}$, but overall the magnetic field does not impact the shock dynamics in any significant way.  

We note that this model quickly developed the $\ell=1$ sloshing mode due to the somewhat artificial nature of the imposed perturbation.  The flow resulted in magnetic field amplification similar to the 2D models, but the breaking of axial symmetry quickly disrupted the magnetic structure along the $z$-axis.  As the sloshing mode developed into the spiral mode a more widespread turbulent flow emerged, which resulted in a different, and perhaps more robust, magnetic field amplification mechanism.  We continue our presentation with a number of models where the initial condition is perturbed by adding random fluctuations to the initial pressure profile.  

\subsubsection{Models with Random Pressure Perturbations}

To complement the reference model presented above and probe the dependence on perturbation method, initial field strength, and the effect of initial rotation, we have computed three additional 3D models at the same spatial resolution, all perturbed by random pressure perturbations in the post shock gas:  one non-rotating model with initial magnetic field $B_{0}=10^{12}$~G (3DB12Rm); one non-rotating model with a weaker initial field $B_{0}=10^{10}$~G (3DB10Rm); and one rotating model with $B_{0}=10^{12}$~G and specific angular momentum $l=1.5\times10^{15}$~cm$^{2}$ s$^{-1}$ about the $z$-axis (3DB12$\Omega$Rm).  

\begin{figure}
  \epsscale{1.0}
  \plotone{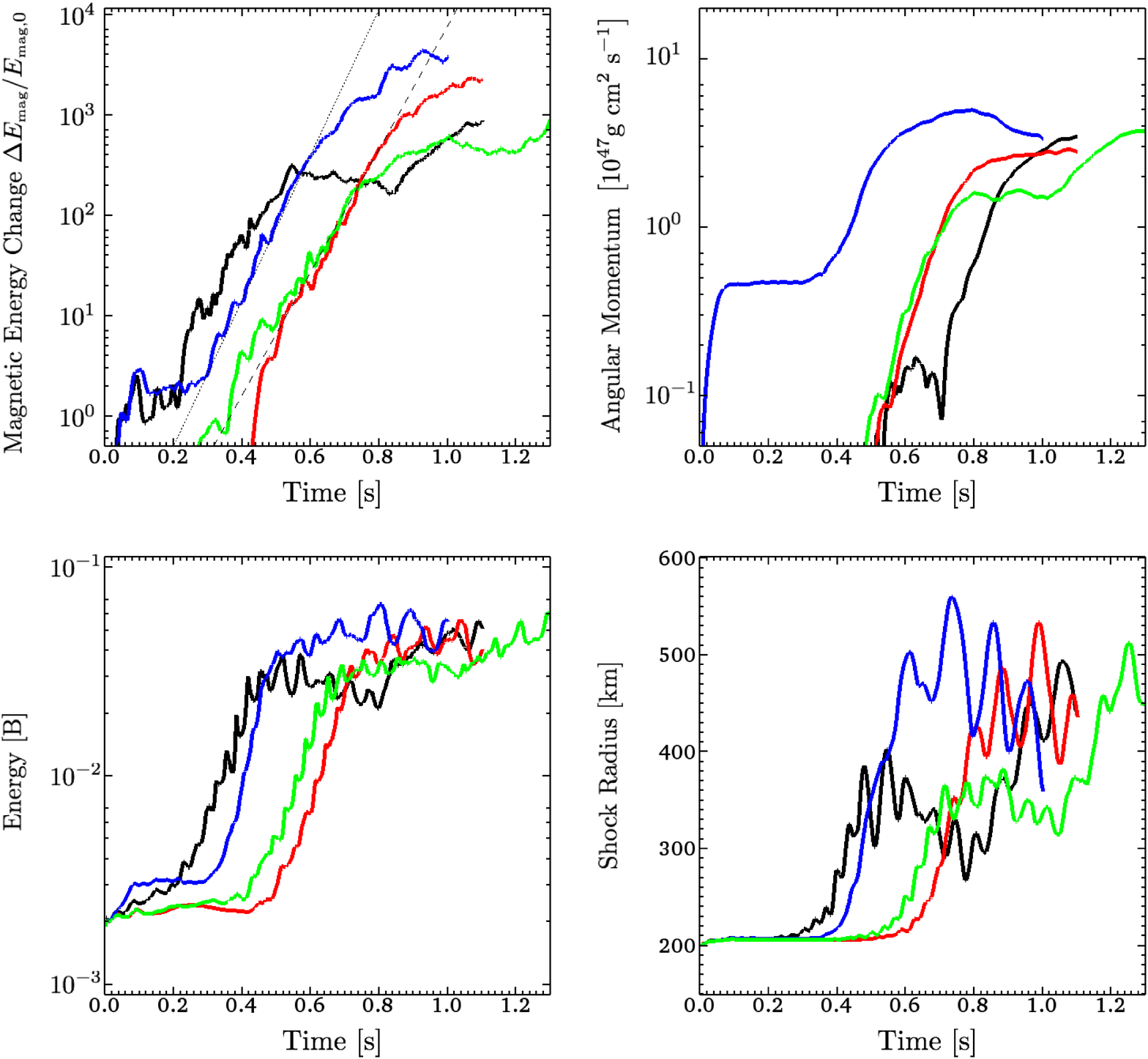}
  \caption{Overview of 3D models:  Relative change in total magnetic energy between the PNS and the shock (upper left panel), total angular momentum of flow between the PNS and the shock (upper right panel), kinetic energy of the flow between the PNS and the shock (lower left panel), and the average shock radius $\bar{R}_{\mbox{\tiny Sh}}=(3V_{\mbox{\tiny Sh}}/4\pi)^{1/3}$. Results are plotted for the reference model 3DB12Am (black), model 3DB12Rm (red), model 3DB12$\Omega$Rm (blue), and model 3DB10Rm (green).  (The initial magnetic energy is $2.3\times10^{-12}$~B and $2.3\times10^{-8}$~B for models with $B_{0}=10^{10}$~G and $B_{0}=10^{12}$~G, respectively.)  \label{fig:overview3DModels}}
\end{figure}

A brief overview of the outcome of these numerical experiments is given in Figure \ref{fig:overview3DModels} where we plot the relative change in total magnetic energy (upper left panel), the total angular momentum (upper right panel), and the total kinetic energy (lower left panel). All quantities are integrated over the volume bounded by the cutout boundary and the accretion shock surface.  We also plot the average shock radius $\bar{R}_{\mbox{\tiny Sh}}$ (lower right panel).  Here we define the average shock radius as the radius of a sphere whose volume is equal to that encompassed by the shock.  Models 3DB12Rm, 3DB10Rm, and 3DB12$\Omega$Rm are represented by the red, green, and blue, lines, respectively.  For easy comparison we also plot the results from the reference model (3DB12Am, black line).  

Although the models differ in the details, there are several common features.  All the models experience a period where the magnetic energy undergoes exponential growth, followed by a period of less vigorous growth.  (The dashed and dotted reference lines in the upper left panel are proportional to exponential functions with e-folding times of 71~ms and 60~ms, respectively.)  The magnetic energy eventually gets boosted by a factor of one to a few thousand.  The kinetic energy beneath the shock reaches the same level in all models.  Concerning the hydrodynamics in our suite of 3D simulations we can draw conclusions similar to \citetalias{blondin_mezzacappa_2007}:  the late-time evolution of the 3D SASI is dominated by the spiral mode.  The angular momentum reaches similar values in all models (a few $\times10^{47}$ g cm$^{2}$ s$^{-1}$), which is sufficient to impact the rotation rate of the underlying PNS.  Furthermore, magnetic energy growth also responds to the operation of the SASI spiral mode.  

The initiation of exponential magnetic energy growth coincides with the onset of the nonlinear SASI, when the kinetic energy beneath the shock starts to grow rapidly, but the magnetic energy continues to grow even when the kinetic energy begins to level off:  there is a clear change in the kinetic energy growth rate around $t=500$~ms for model 3DB12$\Omega$Rm (blue line), while the magnetic energy continues to grow at a nearly unchanged rate beyond this point.  A similar trend is also seen in model 3DB12Rm (red line) where the kinetic energy starts to level off around 700~ms while the magnetic energy growth rate is nearly unchanged beyond 800~ms.  However, the magnetic energy levels off shortly thereafter.  The magnetic energy is still growing in most models near the end of the simulation, but at a slower pace.  From these calculations it does not seem plausible that the magnetic energy will attain significantly higher levels, at least on a time scale that is relevant for core-collapse supernovae.  

The non-rotating models with the random perturbation, as opposed to the reference model, show signs of both the sloshing and the spiral mode early on and the linear evolution resembles a superposition of these modes with comparable amplitudes, but the spiral mode dominates later in the transition to the nonlinear stage.   \citet{blondin_shaw_2007} pointed out a close relationship between these modes.  We also observe that the rotating model (3DB12$\Omega$Rm) enters the nonlinear stage sooner, and then its magnetic energy grows somewhat faster than in the non-rotating models. The rotating model also develops more directly into one that exhibits a clear spiral mode pattern.  During the simulation the rotating model's angular momentum stays roughly aligned with the original rotation axis (to within an angle of $\arctan \left( \sqrt{L_x^2 + L_y^2} / L_z\right) \lesssim 0.2$ rad), while the non-rotating models have angular momentum vectors that change direction in a seemingly random fashion.  

The fact that the model with initial field of $10^{10}$~G develops in a manner similar to the models with $B_{0}=10^{12}$~G (i.e., attains a similar relative boost in magnetic energy) is an indication that the magnetic field has little effect on the nonlinear evolution of the SASI. It should be pointed out that model 3DB12Rm seems to go nonlinear somewhat later than model 3DB10Rm, which may indicate that the initial field strength has a minor effect on the linear development of the SASI.  

\subsection{Mechanisms for Magnetic Field Amplification}

We have found through our numerical experiments that SASI-induced flows are capable of amplifying the magnetic field beneath the shock.  In axisymmetric 2D models the sloshing mode results in a somewhat artificial accumulation of magnetic field along the symmetry axis.  The spiral mode eventually dominates the fluid flows in 3D models and drives the amplification of the magnetic field.   The kinetic energy of the flow, the immediate source of magnetic energy, becomes quite significant when the SASI is in the nonlinear stage.  From Figures \ref{fig:magneticAndKineticEnergy_medRes_2D_B0_1e10_l0_0e00_perturbed} and \ref{fig:magneticAndKineticEnergy_medRes_3D_B0_1e12_l0_0e00_axiSymmetricPerturbation} we can make the following preliminary observation:  the kinetic energy reservoir beneath the shock is a few times $10^{-2}$~B---several orders of magnitude above the levels reached by the magnetic energy at the end of the simulations.  We now turn to explain how magnetic field amplification is realized in our models.  

The basic mechanism responsible for magnetic field growth near the symmetry axis, seen in the axisymmetric 2D models and (initially) in the 3D model with the axisymmetric perturbation, can be summarized in simple physical terms:  there is a tendency for more magnetic field to be advected towards the axis than away from it, resulting in an overall accumulation of magnetic field in the polar regions. SASI-induced lateral flows towards the symmetry axis carry the radial component of the magnetic field along with them, and the associated magnetic flux conservation results in a stronger field as the fluid is compressed against the symmetry axis.  Under the constraint of axial symmetry, without a (at least momentarily) symmetry-breaking initiation of a toroidal flow, the fluid arriving at the axis is forced to turn parallel to it; therefore the field component parallel to the axis gets `left behind' at this point, deposited near the axis, because only the field component perpendicular to the flow suffers advection. Upon encountering either the neutron star below or continuing infall from above, the fluid flows parallel to the axis must once again become lateral, now advecting magnetic field away from the axis.  But to the extent flows impinging on the axis are preferentially diverted away from rather than towards the neutron star, the field advected away is of lesser strength than the field left behind.  

The phenomenon of near-axis field amplification is also easily understood mathematically in terms of Faraday's law of induction, expressed in Eq. (\ref{eq:induction}) for a perfectly conducting medium with electric field $\vect{E}=-\vect{u}\times\vect{B}$.  The stationary unperturbed initial condition has vanishing electric field, but perturbations give rise to lateral components of both $\vect{u}$ and $\vect{B}$, and while $\vect{u}$ and $\vect{B}$ remain poloidal, they are no longer strictly parallel (or antiparallel). This results in a purely toroidal $\vect{E}$, whose nonvanishing curl around the symmetry axis implies changes to $\vect{B}$.  In cylindrical coordinates $(r_\perp, \phi, z)$,
\begin{equation}
  \pderiv{B_{z}}{t} = - \f{E_{\phi}}{r_{\perp}} - \pderiv{E_{\phi}}{r_{\perp}}, \label{eq:dbzdt}
\end{equation}
where 
\begin{equation}
  E_{\phi} = u_{r_{\perp}} B_{z} - u_{z} B_{r_{\perp}}.  \label{eq:ephi}
\end{equation}
Near the axis $\partial E_{\phi} / \partial r_{\perp} \rightarrow 0$, and while $u_{r_{\perp}} \rightarrow 0$ as well, $u_{r_{\perp}} / {r_{\perp}}$ remains finite. Therefore, to the extent the first term of Eq. (\ref{eq:ephi}) dominates in Eq. (\ref{eq:dbzdt}), $B_z$ is susceptible to episodes of exponential growth---or exponential decline---near the symmetry axis.  

Magnetic field amplification in non-axisymmetric flows are not as straightforwardly explained.  In the following analysis we find it natural to consider the evolution of the (scalar) magnetic energy density, rather than focusing on the vector magnetic field.  From the magnetic induction equation (cf. Eq. (\ref{eq:induction})) it is easily found that the magnetic energy density evolves according to
\begin{eqnarray}
  \pderiv{e_{\mbox{\tiny mag}}}{t} 
  &=& \f{1}{\mu_{0}}
  \vect{B}\cdot
  \left[\left(\vect{B}\cdot\nabla\right)\vect{u}
     -\left(\vect{u}\cdot\nabla\right)\vect{B}
     +\vect{u}\divergence\vect{B} \right. \nonumber \\
     & & \hspace{.45in}
     \left.
     -\vect{B}\divergence\vect{u} 
     -\nabla\times\left(\eta\vect{J}\right)\right], 
  \label{eq:magneticEnergyEquation}
\end{eqnarray}
where the first, second, and fourth terms on the right-hand side are, conventionally, said to represent magnetic field evolution due to stretching, advection, and compression, respectively.  In Eq. (\ref{eq:magneticEnergyEquation}) we have retained the term due to magnetic monopoles which vanishes analytically (third term on the right-hand side).  We will use terms from Eq. (\ref{eq:magneticEnergyEquation}) in the analysis to help discern the mechanisms driving magnetic field amplification in our models.  Note that we have also added the dissipative term (last term on the right-hand side of Eq. (\ref{eq:magneticEnergyEquation})) containing the (scalar) resistivity $\eta$.  The dissipative term appears when the (non-ideal) electric field $ - \mathbf{u} \times \mathbf{B} + \eta \mathbf{J}$ is used in the magnetic induction equation. 
The current density is obtained from Amp{\`e}res law $\vect{J}= (\curl\vect{B})/\mu_0$.  

Another convenient way of stating the evolution of the magnetic energy density is through the MHD Poynting theorem, which is easily obtained by rewriting Eq. (\ref{eq:magneticEnergyEquation})
\begin{equation}
\pderiv{e_{\mbox{\tiny mag}}}{t}
  + \divergence\left[ \vect{P} + \eta \vect{J}\times\vect{B}\right]
  = - \vect{u} \cdot \left(\vect{J}\times\vect{B}\right)
     - \eta\vect{J}\cdot\vect{J}, 
  \label{eq:poyntingTheorem}
\end{equation}
where the Poynting vector is $\vect{P}=(\vect{E}\times\vect{B}) / \mu_0 = \left[ \vect{u}\left( \vect{B}\cdot\vect{B}\right)-\vect{B}\left(\vect{B}\cdot\vect{u}\right)\right] /\mu_0$.  Modulo any losses or gains through the boundaries of the computational domain, the magnetic energy may increase through work done against the Lorentz force (first term on the right-hand side), provided this term overcomes any losses due to Joule dissipation, i.e. resistivity (second term on the right-hand side).  The Lorentz work term, $W_{\mbox{\tiny L}}=-\vect{u}\cdot(\vect{J}\times\vect{B})$, can be both positive and negative, while the dissipative term, $Q_{\mbox{\tiny J}}=\eta\vect{J}\cdot\vect{J}$, only acts to decrease the magnetic energy.  

We are concerned with ideal MHD is this paper, but the numerical method for solving the magnetic induction equation contains dissipative terms that amount to numerical resistivity in regions where the magnetic field varies significantly over a few computational grid cells (e.g., current sheets).  Physically, the magnetic Reynolds number (obtained from the ratio $R_{m}=|W_{\mbox{\tiny L}}|/|Q_{\mbox{\tiny J}}|$) is extremely large in the supernova environment (the time scale for magnetic energy dissipation, $\tau_{\mbox{\tiny d}}\sim\lambda_{B}^{2}/(2\eta)$, is very long\footnote{With the value for resistivity $\eta$ listed in Table 1 in \citet{thompson_duncan_1993} the dissipation time for a magnetic field varying on a spatial scale of say $\lambda_{B}\sim1$~m (i.e., much smaller than any scale resolved by our simulations) exceeds $10^{7}$~s.})  and Joule dissipation may not be important on the explosion time scale.  Nevertheless, realistic magnetic Reynolds numbers are computationally prohibitive and not possible to achieve in numerical simulations of the type presented in this study.  Numerical Joule dissipation, however, may become significant in simulations, and it is important to also consider the effects of numerical resistivity in the analysis of magnetic field evolution.  We will do this, to some extent, later in this study when we present results from simulations that have been performed with different spatial resolutions.  

\begin{figure}
  \epsscale{1.0}
  \plotone{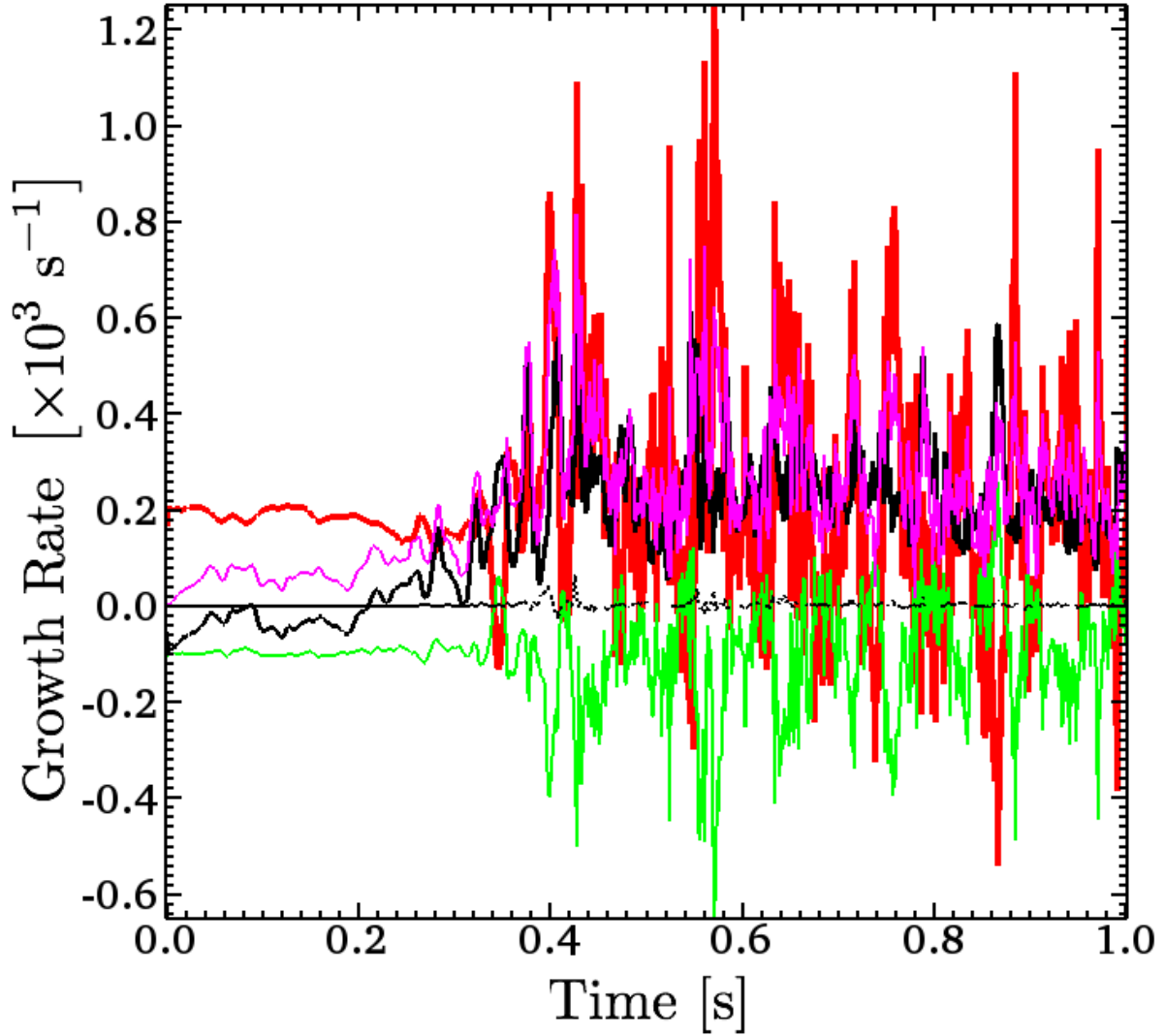}
  \caption{Magnetic energy growth rates extracted from an axisymmetric model with an initial field $B_{0}=10^{12}$~G (model 2DB12Am):  Compression $\sigma_{\divergence\vect{u}}$ (red curve), stretching $\sigma_{\nabla\vect{u}}$ (black solid curve), advection $\sigma_{\vect{u}\cdot\nabla}$ (green curve), and $\sigma_{\vect{J}\times\vect{B}}$ (magenta curve), respectively.  We have also plotted the corresponding rate due to magnetic monopoles $\sigma_{\divergence\vect{B}}$ (thin black dotted curve), which remains small throughout the simulation.  \label{fig:magneticEnergyGrowth_1e12G_2D}}
\end{figure}

\begin{figure}
  \epsscale{1.0}
  \plotone{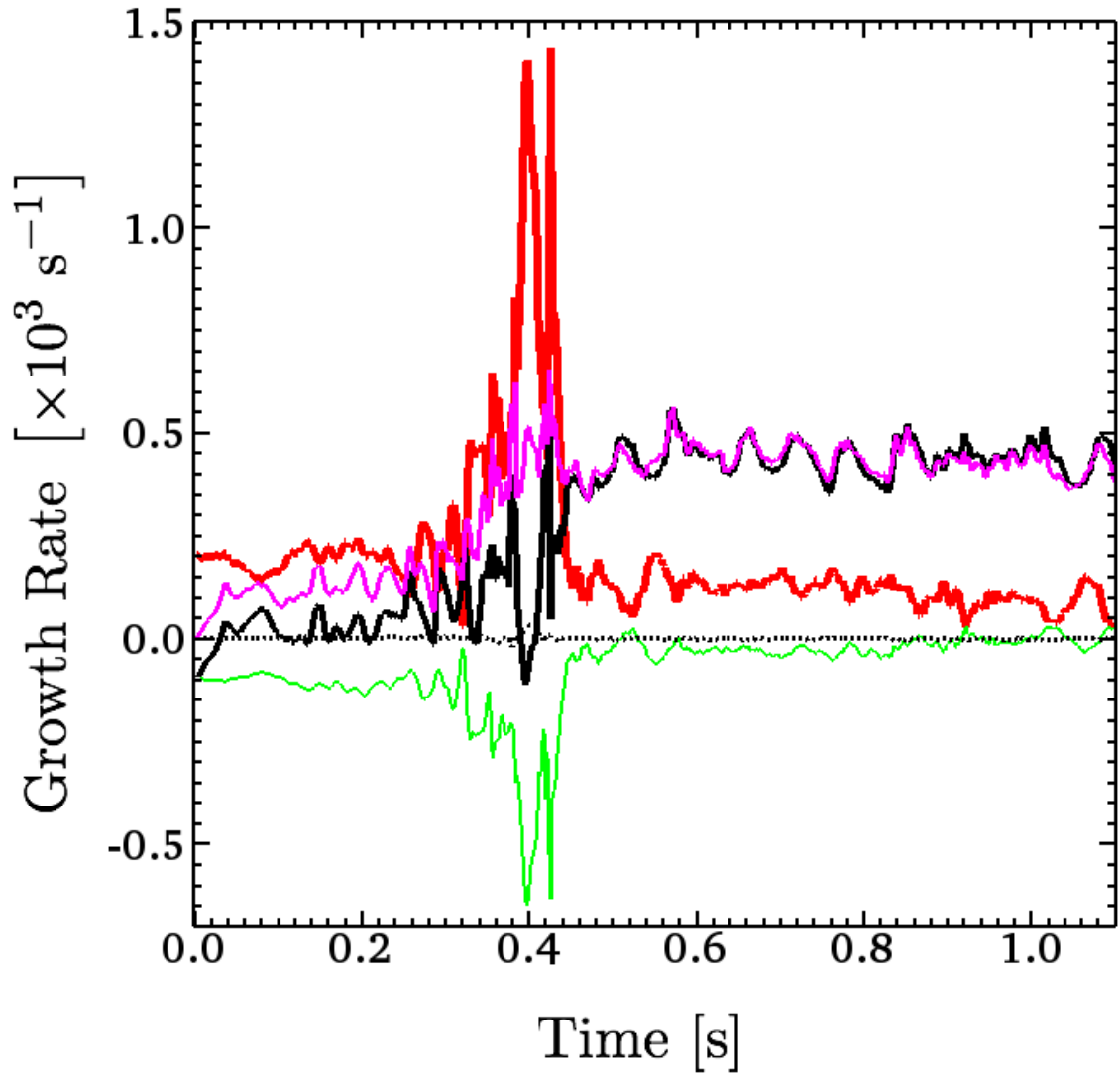}
  \caption{Same plot as Figure \ref{fig:magneticEnergyGrowth_1e12G_2D}, but for the 3D model with the axisymmetric perturbation (model 3DB12Am).  \label{fig:magneticEnergyGrowth_1e12G_3D_axisymmetric}}
\end{figure}

\begin{figure}
  \epsscale{1.0}
  \plotone{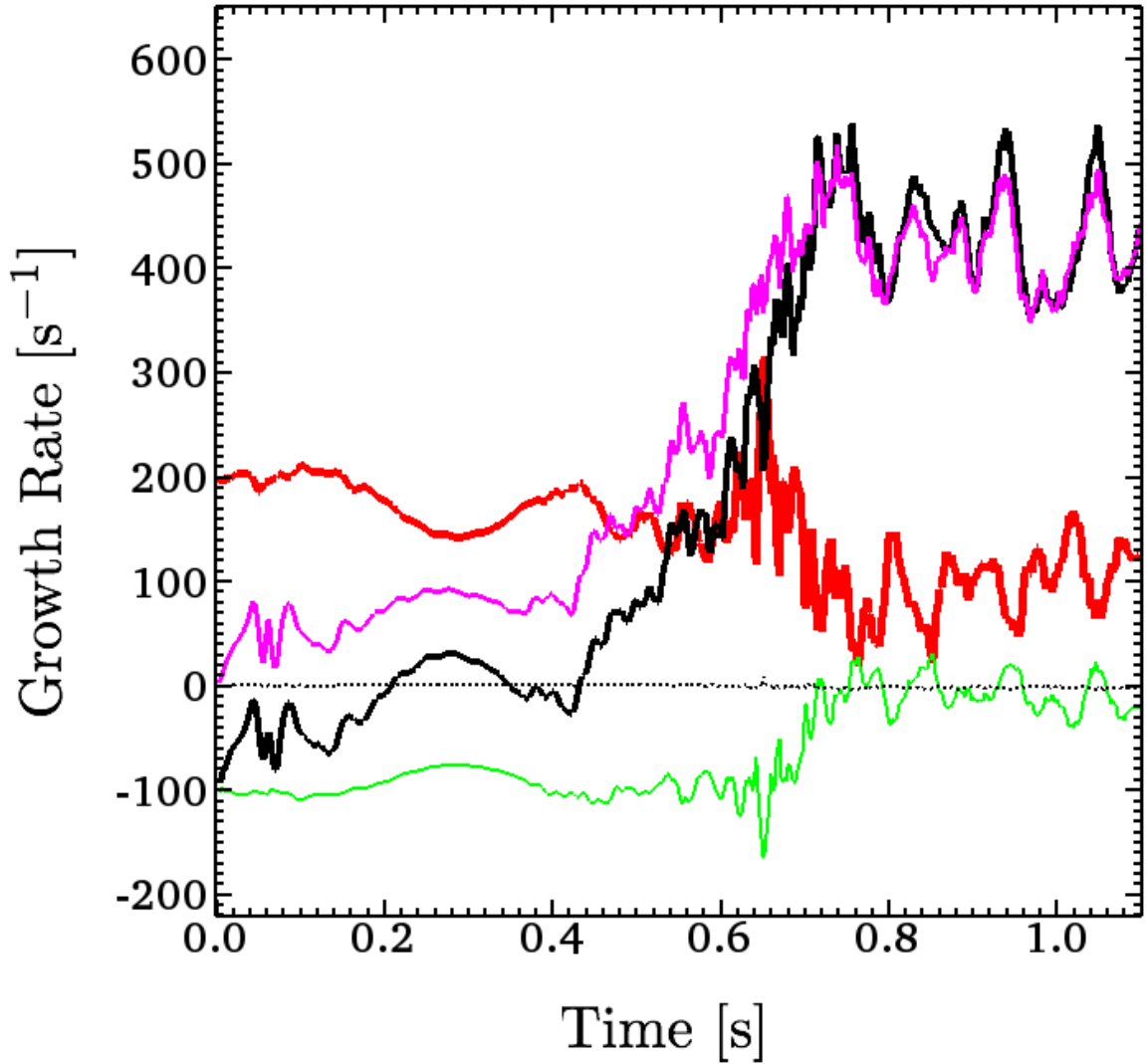}
  \caption{Same plot as Figure \ref{fig:magneticEnergyGrowth_1e12G_2D}, but for the 3D model with the random perturbation (model 3DB12Rm).  \label{fig:magneticEnergyGrowth_1e12G_3D_random}}
\end{figure}

\begin{figure}
  \epsscale{1.0}
  \plotone{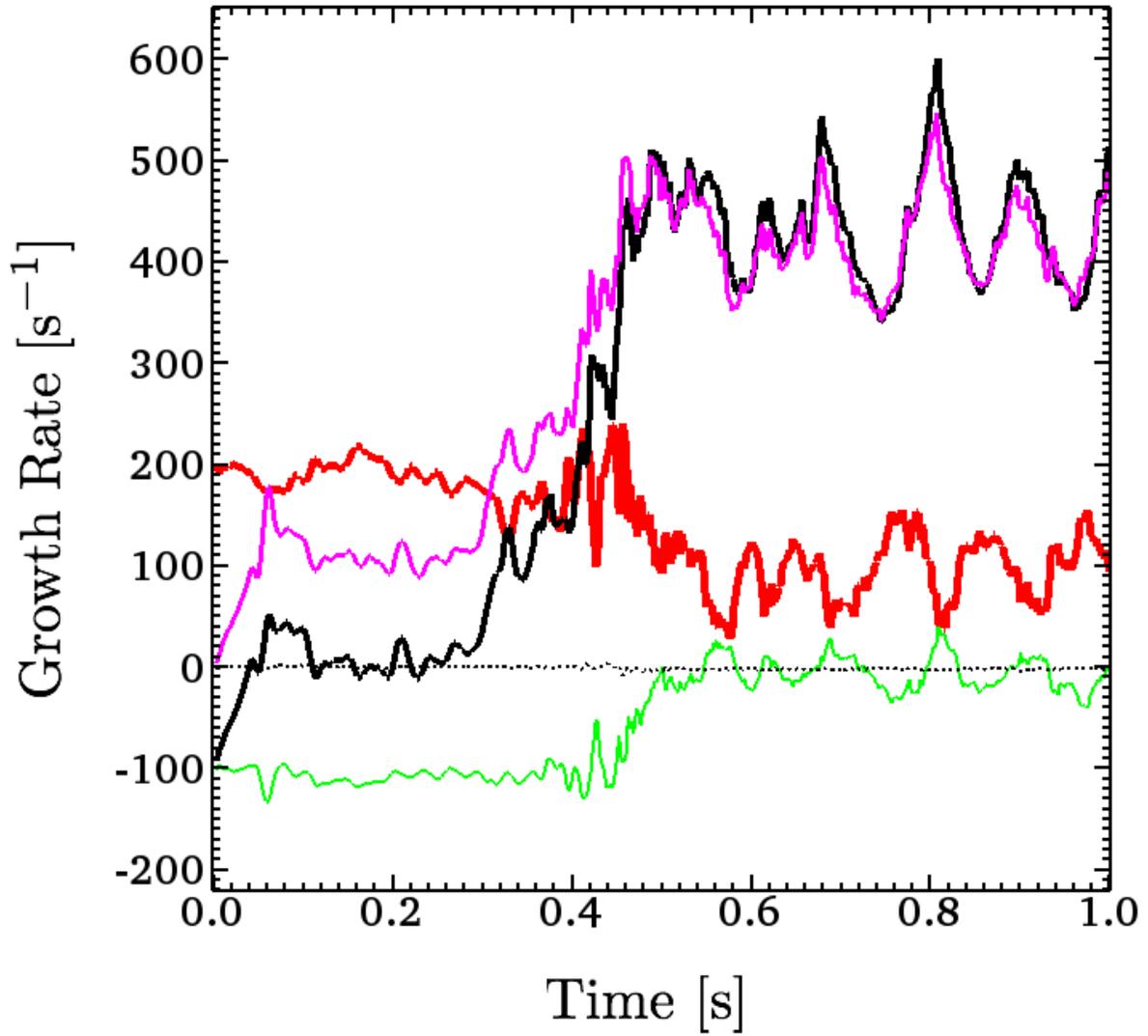}
  \caption{Same plot as Figure \ref{fig:magneticEnergyGrowth_1e12G_2D}, but for the rotating 3D model with the random perturbation (model 3DB12$\Omega$Rm).  \label{fig:magneticEnergyGrowth_1e12G_3D_random_rotating}}
\end{figure}

Considering ideal MHD ($\eta\to0$) for the moment, we now try to identify the mechanism(s) responsible for magnetic energy growth in our simulations by comparing the individual terms on the right-hand side of Eq. (\ref{eq:magneticEnergyEquation}) through magnetic energy growth rates due to stretching, advection, and compression, defined respectively as
\begin{eqnarray}
  \rateStretching
  &=& 
  \f{2\langle\vect{B}\cdot\left[\left(\vect{B}\cdot\nabla\right)\vect{u}\right]\rangle}
    {\langle\vect{B}\cdot\vect{B}\rangle}, \label{eq:stretchingRate} \\
  \rateAdvection
  &=& 
  -\f{2\langle\vect{B}\cdot\left[\left(\vect{u}\cdot\nabla\right)\vect{B}\right]\rangle}
      {\langle\vect{B}\cdot\vect{B}\rangle}, \label{eq:advectionRate} \\
  \rateCompression
  &=& 
  -\f{2\langle\vect{B}\cdot\left[\vect{B}\divergence\vect{u}\right]\rangle}
      {\langle\vect{B}\cdot\vect{B}\rangle} \label{eq:compressionRate}.  
\end{eqnarray}
Angle brackets denote an average over the volume encompassed by the shock (excluding the PNS).  We define the growth rate due to magnetic monopoles as
\begin{equation}
  \rateMonopoles
  =
  \f{2\langle\vect{B}\cdot\left[\vect{u}\divergence\vect{B}\right]\rangle}
    {\langle\vect{B}\cdot\vect{B}\rangle} \label{eq:monopoleRate}, 
\end{equation}
which we also include in figures below as a consistency check and show that it remains small in our simulations.  In calculating the rate in Eq. (\ref{eq:monopoleRate}) the divergence of the magnetic field was, for consistency with the computation of the other rates, computed from limited central differences of the zone centered magnetic field rather than with the face centered magnetic field components evolved by the constrained transport algorithm.  This results in a magnetic field-divergence that is not preserved to numerical round-off error, but still stays relatively small as can be seen from the dotted line in Figures \ref{fig:magneticEnergyGrowth_1e12G_2D}-\ref{fig:magneticEnergyGrowth_1e12G_3D_random_rotating}.  

For ideal MHD the magnetic energy growth rate $\sigma_{e_{\mbox{\tiny mag}}}=\langle e_{\mbox{\tiny mag}}\rangle^{-1}\langle\partial e_{\mbox{\tiny mag}}/\partial t\rangle$ equals the sum of the rates in Eqs. (\ref{eq:stretchingRate})-(\ref{eq:monopoleRate}).  Note that, except for the vanishing $\sigma_{\divergence\vect{B}}$, these terms are all non-zero for the unperturbed initial condition, but when added together result in a time-independent magnetic field.  

To supplement the analysis we also compute the magnetic energy growth rate due to the first term on the right-hand side of Eq. (\ref{eq:poyntingTheorem}), $W_{\mbox{\tiny L}}$.  SASI-driven flows producing net work done against the Lorentz force result in the conversion of kinetic energy into magnetic energy at a rate given by
\begin{equation}
  \rateLorentzWork
  =
  -\f{2\mu_{0}\langle\vect{u}\cdot\left(\vect{J}\times\vect{B}\right)\rangle}
     {\langle\vect{B}\cdot\vect{B}\rangle} \label{eq:lorentzRate}.  
\end{equation}

We compare a subset of our simulations by plotting the rates in Eqs. (\ref{eq:stretchingRate})-(\ref{eq:lorentzRate}) taken from one axisymmetric 2D model (2DB12Am), shown in Figure \ref{fig:magneticEnergyGrowth_1e12G_2D}, and three of our 3D models:  3DB12Am, 3DB12Rm, and 3DB12$\Omega$Rm, which are shown in Figures \ref{fig:magneticEnergyGrowth_1e12G_3D_axisymmetric}, \ref{fig:magneticEnergyGrowth_1e12G_3D_random}, and \ref{fig:magneticEnergyGrowth_1e12G_3D_random_rotating}, respectively.  

Figure \ref{fig:magneticEnergyGrowth_1e12G_2D} illustrates that compression (red curve) and stretching (black curve) contribute to magnetic field amplification in the 2D axisymmetric calculation (model 2DB12Am) during the epoch of most rapid growth, lasting from 300~ms to 400~ms, with contributions from compression reaching larger peak values.  Advection (green curve) tends to reduce the magnetic energy between the PNS and the shock.  This role of advection is expected since magnetic energy is continuously flowing through the inner cutout boundary.  In the saturated state ($t>400$~ms) we find that compression and stretching contribute roughly equally, but with large spikes appearing in the compression rate (and also in the advection rate) when a plunging stream from the shock triple point hits the PNS and the symmetry axis.  Decompression also results in weakening of the magnetic field:  the compression rate $\rateCompression$ is negative in several intervals, while the stretching rate $\rateStretching>0$ for $t>400$~ms.  We find that the rate $\rateLorentzWork$ (magenta curve) is similar in magnitude to both compression and stretching (though always positive) in this stage of the run.  We have calculated the time-averaged\footnote{The temporal average $\langle f\rangle=\f{1}{T}\int_{t_{1}}^{t_{2}}f \, dt$ of a variable $f$ over an interval $T=t_{2}-t_{1}$  is computed numerically using the trapezoid rule.} rates over the interval from 400~ms to 1~s and find $\langle\rateStretching\rangle\approx 242$~s$^{-1}$, $\langle\rateAdvection\rangle\approx -103$~s$^{-1}$, $\langle\rateCompression\rangle\approx 207$~s$^{-1}$, and $\langle\rateLorentzWork\rangle\approx 282$~s$^{-1}$.  

The evolution of model 3DB12Am is very similar to the corresponding axisymmetric 2D model (2DB12Am) up to about $t=430$~ms, with significant contributions from compression.  This is a result of fluid flows driven by the SASI sloshing mode converging at the (temporary) symmetry axis.  The similarity with the 2D model disappears abruptly when non-axisymmetric modes become significant.  For $t>460$~ms the rate due to stretching dominates, with $\rateLorentzWork\approx\rateStretching$, while contributions from compression ($>0$) are subdominant, and advection (mostly $<0$) appears to play an even smaller role.  The time-averaged rates for this run, over the interval from 460~ms to 1.1~s, are  $\langle\rateStretching\rangle\approx 435$~s$^{-1}$, $\langle\rateAdvection\rangle\approx -19$~s$^{-1}$, $\langle\rateCompression\rangle\approx 112$~s$^{-1}$, and $\langle\rateLorentzWork\rangle\approx 434$~s$^{-1}$.  The time interval over which the time-averages are computed have been chosen individually for each model to cover most of the nonlinear stage, but excluding the initial ramp-up of the SASI.  In the 3D models the bulk of magnetic energy is, in fact, generated during the highly nonlinear stage covered by these intervals.  

Both the non-rotating and the rotating models with the random pressure perturbation (3DB12Rm and 3DB12$\Omega$Rm) are very similar to model 3DB12Am in the nonlinear regime, being dominated by the stretching rate with $\rateLorentzWork\approx\rateStretching$.  As a point of comparison we also list the computed time-averaged rates for these models:  in the interval from 700~ms to 1.1~s for model 3DB12Rm we found $\langle\rateStretching\rangle\approx 435$~s$^{-1}$, $\langle\rateAdvection\rangle\approx -9$~s$^{-1}$, $\langle\rateCompression\rangle\approx 97$~s$^{-1}$, and $\langle\rateLorentzWork\rangle\approx 422$~s$^{-1}$.  For model 3DB12$\Omega$Rm the time averaged rates are $\langle\rateStretching\rangle\approx 434$~s$^{-1}$, $\langle\rateAdvection\rangle\approx -6$~s$^{-1}$, $\langle\rateCompression\rangle\approx 97$~s$^{-1}$, and $\langle\rateLorentzWork\rangle\approx 420$~s$^{-1}$ over the interval from 500~ms to 1~s.  

The randomly perturbed 3D models do not exhibit the large spikes in $\rateCompression$ that result from the sloshing mode induced by the axisymmetric pertubation early on in model 3DB12Am.  In the randomly perturbed models there is a transient period, lasting for about 100~ms, during the early nonlinear development where $\rateCompression\approx\rateStretching$, but compression plays a subdominant role for magnetic field amplification.  The spiral mode dominates at late times in all the 3D models presented here.  

The stretching rate stays at a relatively constant level in the nonlinear stage in all the 3D models.  Despite variations on a $\sim100$~ms time scale, there does not seem to be any noticeable rising or declining trend in $\rateStretching$.  We also note that a small amount of rotation in the infalling matter has no noticeable effect on the stretching rate.  

We note that these rates indicate that the magnetic energy grows on a millisecond time scale, while the magnetic energy beneath the shock grows on a time scale of tens of milliseconds; i.e., the sum of the individual rates ($\rateStretching+\rateAdvection+\rateMonopoles+\rateCompression$) does not match up exactly with the total rate $\sigma_{e_{\mbox{\tiny mag}}}$ (not shown).  Several factors contribute to this:  (1) we compute the individual rates using limited second order centered finite differences (cf. Eq. (\ref{eq:slopeLimiter})), even where they are not well defined (e.g., in shocks and discontinuities, which are abundant in the nonlinear stage); (2) the time derivative in $\sigma_{e_{\mbox{\tiny mag}}}$ extends over several milliseconds (output files are written every 2~ms) and is compared with terms computed with spatial operators acting on an instant of the solution; (3) we have not taken into account  the effects of numerical diffusion due to $Q_{\mbox{\tiny J}}$, which likely plays a significant role when the magnetic field is concentrated in thin flux tubes; (4) the rates are computed from terms in the equation for the magnetic energy density, the square of the magnetic field, while in the code we evolve the individual components of the magnetic field; and (5) the volume average used to compute the rates is essentially an arithmetic mean, which is greatly influenced by outliers.  We do find that the match improves with increased spatial resolution, in particular during the initial linear development.  Nevertheless, we believe that the individual rates as we compute them in this paper are helpful in identifying the physical mechanisms responsible for magnetic field amplification in our simulations, as well as in the comparison of 2D and 3D simulations.  

The magnetic energy growth seen in our 3D models is driven by the fluid motions from the SASI resulting in a net work done against the Lorentz force.  This is caused by stretching of the magnetic field:  In all our 3D models we find that $\rateLorentzWork\approx\rateStretching$ at late times during the operation of the SASI spiral mode.  The spiral mode results in a persistent shear flow inside the supernova shock which extends from the shock triple point down towards the PNS.  The shearing region generates fluid vorticity, triggers secondary fluid instabilities (e.g., the Kelvin-Helmholtz instability), and results in a turbulent flow filling a significant fraction of the volume beneath the shock.  (We loosely use the term `turbulent flow' when we refer to the swirling nonlinear flows generated by the SASI.)  The separation between two (initially adjacent) fluid elements grows exponentially with time in a turbulent flow.  If the fluid elements are connected by a weak magnetic field the frozen-in condition of ideal MHD results in stretching, and thereby strengthening, of the magnetic field and an increase in the magnetic energy \citep[e.g.,][]{ott_1998}:  in the context of a `flux rope' topology (see more below), the decreased cross-sectional area of a longitudinally stretched (and axially twisted) flux tube implies, under flux conservation, an increased field strength.

\begin{figure}
  \epsscale{1.0}
  \plotone{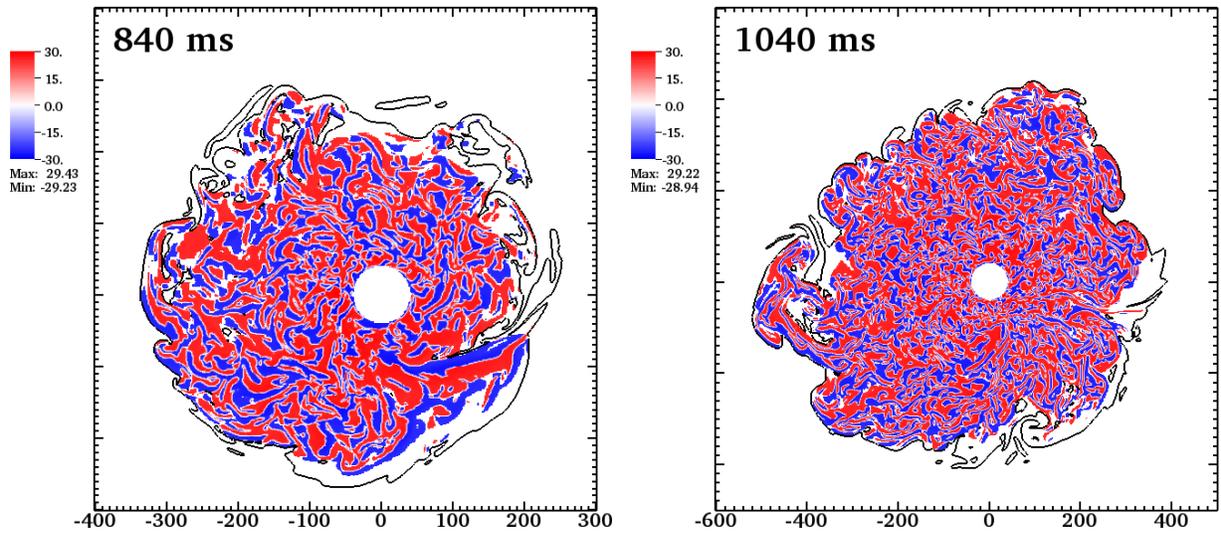}
  \caption{Color plot showing the distribution of the magnetic energy generation rate $W_{\mbox{\tiny L}}=-\vect{u}\cdot(\vect{J}\times\vect{B})$.  These panels correspond to the two lower panels of the magnetic field in Figure \ref{fig:magneticFieldOverview_medRes_3D_B0_1e12_l0_0e00_perturbed}.  Black contours are drawn where the magnitude of the magnetic field equals $6\times10^{10}$~G (left panel) and $4\times10^{10}$~G (right panel).  \label{fig:sasi_MHD_3D_medRes_B0_1e12_logLorentzWork_twoPanel_spiralMode}}
\end{figure}

\begin{figure}
  \epsscale{1.0}
  \plotone{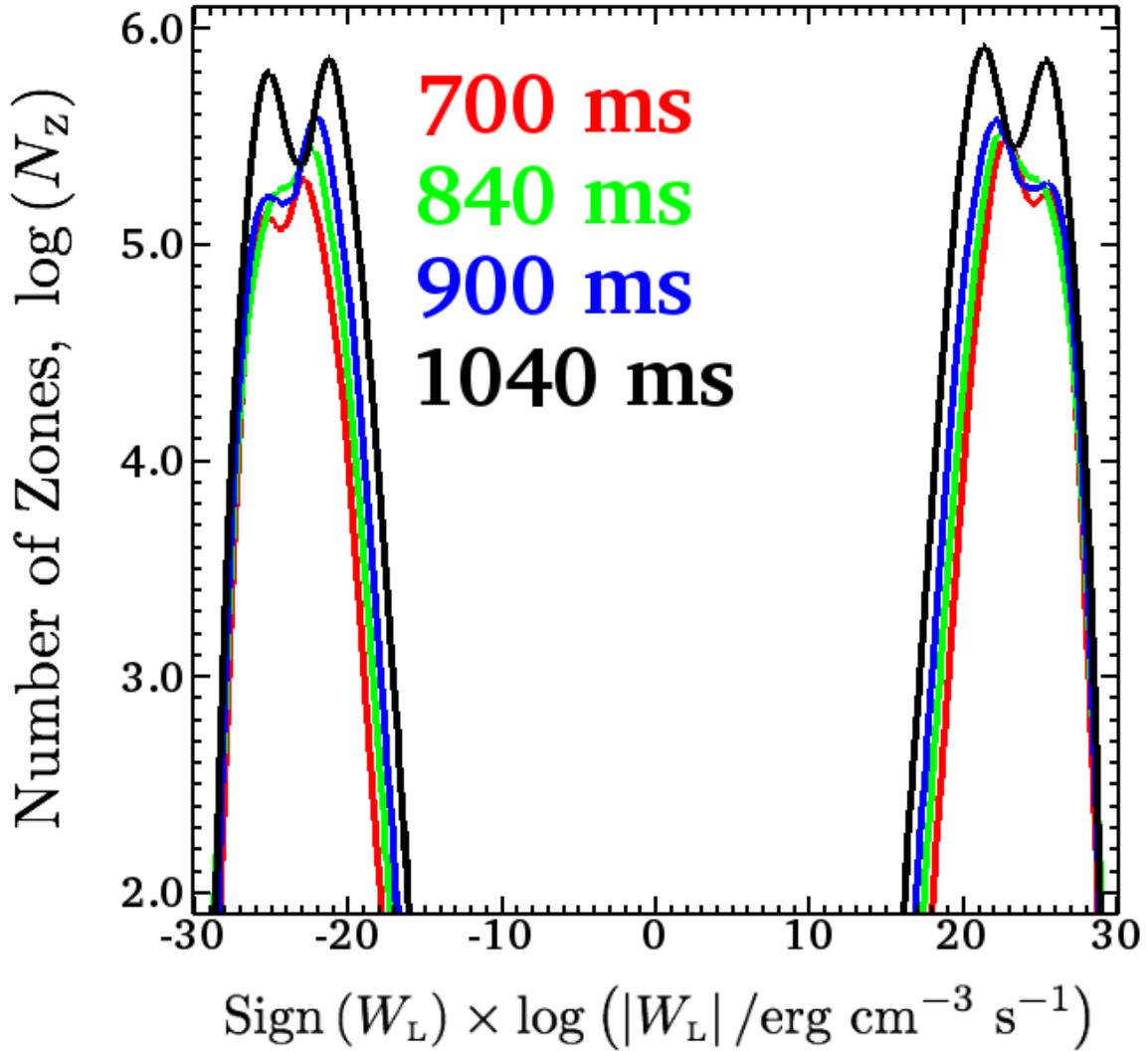}
  \caption{Histogram showing the distribution of $W_{\mbox{\tiny L}}$ beneath the shock for model 3DB12Am at selected times during the nonlinear evolution of the SASI:  700~ms (red line), 840~ms (green line), 900~ms (blue line), and 1040~ms (black line).  \label{fig:sasi_MHD_3D_medRes_B0_1e12_lorentzWorkHistogram}}
\end{figure}

The spatial distribution of the magnetic energy generation (and destruction) rate $W_{\mbox{\tiny L}}$ is plotted in Figure \ref{fig:sasi_MHD_3D_medRes_B0_1e12_logLorentzWork_twoPanel_spiralMode}, and in Figure \ref{fig:sasi_MHD_3D_medRes_B0_1e12_lorentzWorkHistogram} a histogram showing the distribution of $W_{\mbox{\tiny L}}$ in zones inside the accretion shock is plotted.  The bulk of magnetic energy does not seem to be generated in the shear layer itself.  It is the hydrodynamical consequences of the presence of the shear layer that drives the amplification of the magnetic field.  The magnetic energy generation (and destruction) rate $W_{\mbox{\tiny L}}$ is isotropically distributed inside the accretion shock with positive {\it and} negative values of similar magnitude, which overall results in a net generation of magnetic energy.  Notice the nearly mirror-symmetric shape of the distribution of positive and negative values of $W_{\mbox{\tiny L}}$ at the selected times shown in Figure \ref{fig:sasi_MHD_3D_medRes_B0_1e12_lorentzWorkHistogram}.  Figure \ref{fig:sasi_MHD_3D_medRes_B0_1e12_logLorentzWork_twoPanel_spiralMode} shows the distribution of magnetic energy generation and destruction scattered over a large volume; $W_{\mbox{\tiny L}}$ tends to be largest in magnitude around the PNS, near the termination of the supersonic stream plunging down from the shock triple point.  Also note that the distribution of $W_{\mbox{\tiny L}}$ is confined to roughly the same region as the fluid vorticity in Figure \ref{fig:sasi_MHD_3D_medRes_B0_1e12_vorticityMagnitude_twoPanel_spiralMode}.  

We have seen that the magnetic field evolves into a complicated flux rope structure.  The total magnetic energy levels off in the nonlinear stage at a level that is well below the kinetic energy of the highly turbulent flow beneath the shock, even as the magnetic energy growth rate $\sigma_{\vect{J}\times\vect{B}}$ (dominated by stretching) remains at a relatively constant level ($\approx 435$ s$^{-1}$).  Also, with the exception of transient periods in a few localized regions for models with $B_{0}=10^{12}$~G, the magnetic energy density remains locally well below the internal and kinetic energy densities of the flow.  Our calculations therefore do not suggest that magnetic energy growth is quenched because the magnetic field establishes dynamical equipartition with the fluid on any spatial scale.  On the other hand, the bulk of the magnetic energy is concentrated on spatial scales where numerical resistivity inevitably plays a role.  We will now proceed to investigate the effect of varying the spatial resolution in our simulations.  

\subsection{Varying the Spatial Resolution}

To further elucidate the evolution of the magnetic field in the SASI we have computed 2D and 3D models with different spatial resolutions.  We present results from three axisymmetric 2D models with $B_{0}=10^{10}$~G (cf. Table \ref{tbl:2Dmodels}): 2DB10Al, 2DB10Ah, and 2DB10Ah$^{+}$, respectively.  We also compute three 3D models with $B_{0}=10^{12}$~G (cf. Table \ref{tbl:3Dmodels}): 3DB12Al, 3DB12Am, and 3DB12Ah.  The difference in initial field strength in the 2D and 3D models is of less significance.  Note that there is a factor of four difference in spatial resolution between the lowest and the highest resolution 2D model, while the spatial resolution in the lowest and the highest resolution 3D model differ only by a factor of two.  The 2D models are of course much less computionally intensive than the 3D models.  While this allows us to explore a large range in spatial resolution, the obvious limitations of imposing axial symmetry makes it less interesting to pursue this path.  We will focus mostly on the 3D models in this section.  All the models presented in this section are initiated from non-rotating initial conditions and perturbed with the axisymmetric perturbation.  

\subsubsection{Two-dimensional models}
\label{sec:varyingResolution2D}

\begin{figure}
  \epsscale{1.0}
  \plotone{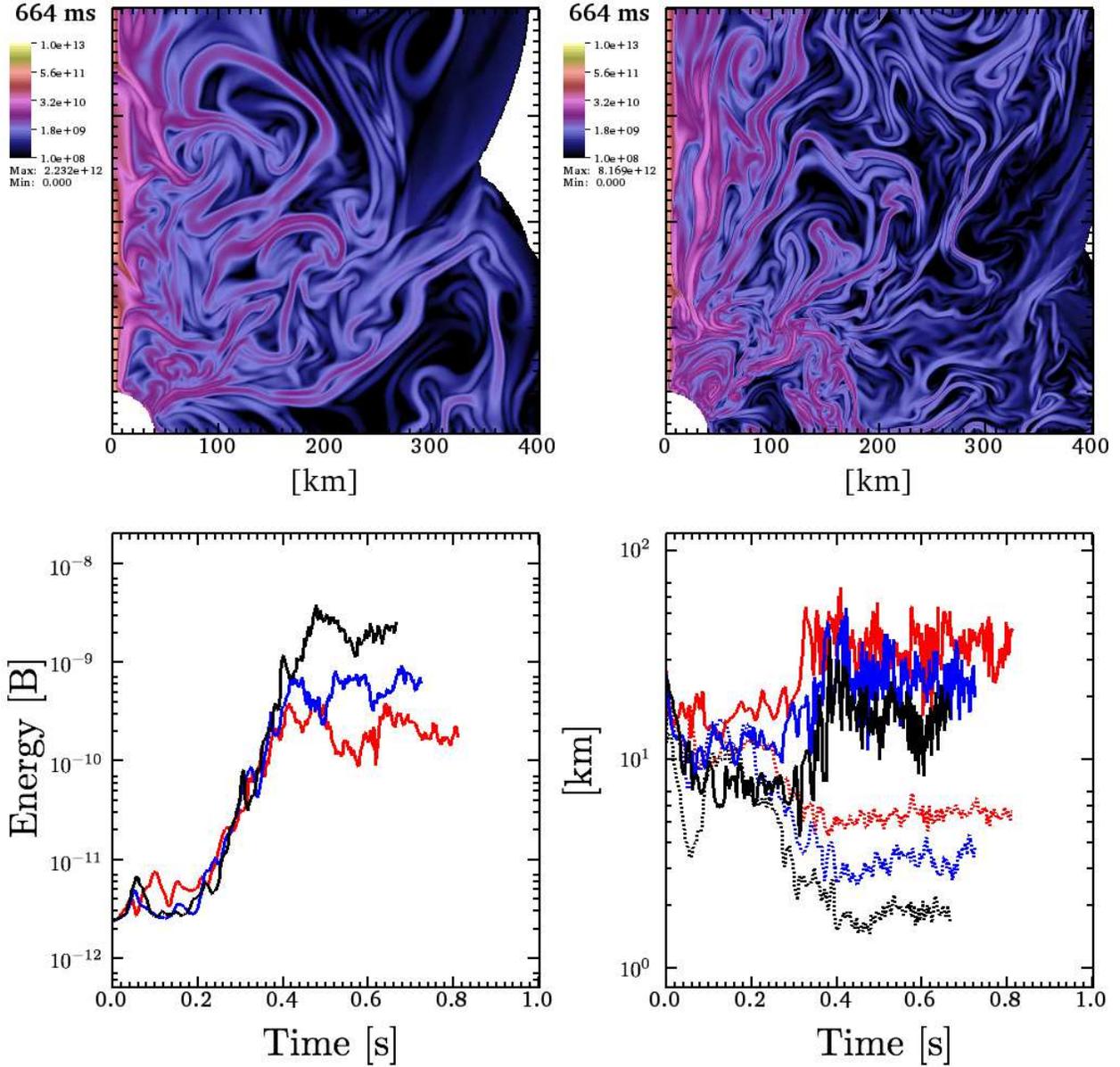}
  \caption{Results from axisymmetric calculations where the spatial resolution has been varied:  Color plot of the magnitude of the magnetic field for a model with $\Delta l\approx 1.17$~km (2DB10Ah; upper left panel) and a model with $\Delta l\approx 0.59$~km (2DB10Ah$^{+}$; upper right panel).  In the lower two panels we plot results from models where the zone width $\Delta l$ has been set to 2.34~km, 1.17~km, and 0.59~km (red, blue, and black lines, respectively).  In the lower left panel we plot the total magnetic energy between the shock and the PNS.  In the lower right panel we plot the magnetic curvature radius $\lambda_{\mbox{\tiny c}}$ (solid lines) and the magnetic rms scale $\lambda_{\mbox{\tiny rms}}$ (dotted lines).   \label{fig:varyingResolution2D}}
\end{figure}

Figure \ref{fig:varyingResolution2D} provides an overview of the effects increased spatial resolution has on the structure of the magnetic field, and the evolution of the magnetic energy between the PNS and the shock in axisymmetric calculations (2DB10Al, 2DB10Ah, and 2DB10Ah$^{+}$).  Obviously, higher spatial resolution allows smaller structures to be captured by the numerical simulations.  This can be seen in the two upper panels of Figure \ref{fig:varyingResolution2D} where a color plot of the magnitude of the magnetic field reveals a highly intermittent `flux tube'\footnote{The imposed symmetry in the 2D calculations prevents the formation of flux tubes.  Instead, flux sheets are formed.  For conformity with the description of the 3D results below, we also refer to these structures as flux tubes.  } structure.  

The energy stored in the magnetic field beneath the shock increases with spatial resolution (cf. lower left panel in Figure \ref{fig:varyingResolution2D}).  The magnetic energy follows the same evolutionary path in all models, with nearly identical growth rates, during the ramp-up of the SASI.  All models have reached a saturated state around $t=500$~ms, but the models with higher spatial resolution saturate at a somewhat higher magnetic energy.  (The total magnetic energy is dominated by the field around the symmetry axis in these models, but a similar dependence on spatial resolution is also seen when the magnetic field around the symmetry axis is excluded.)  The magnetic energy in model 2DB10Ah$^{+}$ is about an order of magnitude higher than in model 2DB10Al for $t\gtrsim500$~ms.  Furthermore, our results have not converged in the range of resolutions shown in Figure \ref{fig:varyingResolution2D}.  The model with the highest spatial resolution (2DB10Ah$^{+}$, $\Delta l\approx0.59$~km) is computed on a $r_{\perp}\times z$ grid consisting of $2048\times4096$ zones.  

We further quantify the structure of the magnetic field by plotting two useful characteristic scales, namely the magnetic curvature radius
\begin{equation}
  \lambda_{\mbox{\tiny c}}
  =\sqrt{\langle B^{4} \rangle/\langle \left|\left(\vect{B}\cdot\nabla\right)\vect{B}\right|^{2} \rangle}
\end{equation}
and the so-called magnetic rms scale
\begin{equation}
  \lambda_{\mbox{\tiny rms}}
  =\sqrt{\langle B^{2} \rangle/\langle \left|\nabla B\right|^{2} \rangle}.  
\end{equation}
Again, angle brackets denote a volume average over the volume bounded by the surface of the PNS and the shock.  The magnetic curvature radius measures how sharply the magnetic field is bent \citep[e.g.,][]{ryu_etal_2000}, while the magnetic rms scale \citep[cf.][]{brandenburgSubramanian_2005} provides a measure on the thickness of magnetic flux tubes when the magnetic field has evolved into a highly intermittent flux tube structure.  

Indeed, for $t>500$~ms both the magnetic curvature radius and the magnetic rms scale decrease as the spatial resolution in increased (cf. lower right panel in Figure \ref{fig:varyingResolution2D}).  This trend is consistent with the two upper panels of Figure \ref{fig:varyingResolution2D}:  higher spatial resolution resolves smaller eddies in the turbulent flow beneath the shock, which again is reflected in how sharply the magnetic flux tubes are bent.  It is very interesting to note that the total magnetic energy is more or less saturated when $\lambda_{\mbox{\tiny rms}}\approx\mbox{a few}\times\Delta l$.  This is consistent with stretching being a contributing mechanism for magnetic field amplification.  Magnetic field amplification due to stretching can proceed until the magnetic field becomes strong enough to impact the fluid flow at the appropriate spatial scale, or until the smallest spatial dimension of the magnetic field (i.e., the flux tube thickness) is comparable to a few computational grid cells, at which point numerical diffusion inevitably becomes significant.  In the latter case, the magnetic energy generation rate is nearly balanced by (numerical) Joule dissipation with $W_{\mbox{\tiny L}}\lesssim Q_{\mbox{\tiny J}}$, and this seems to be what occurs in the simulations presented here.  

Cowling's anti-dynamo theorem, which requires that the magnetic field eventually decay to zero in an axisymmetric system \citep[see e.g., ][]{nunez_1996}, is not in conflict with our axisymmetric simulations. Our stationary unperturbed model system is not isolated, but features a continuous inflow of magnetized fluid (though in a physical system the accretion would not last indefinitely). More importantly, the theorem does not forbid transient magnetic field amplification, and the duration of our simulations is much shorter than the decay timescale implied by the magnitude of numerical resistivity. (This is not to say numerical diffusivity is unimportant in our simulations; as described above, it is responsible for setting a minimum flux tube width, and it therefore limits magnetic field amplification.)

\subsubsection{Three-dimensional models}
\label{sec:varyingResolution3D}

The 3D models discussed in this subsection are initiated from the same initial condition and are perturbed in the same manner---the axisymmetric perturbation.  This places models to be compared (both 2D and 3D) on a fairly equal footing, and facilitates a somewhat deterministic evolution early on.

As far as the hydrodynamics is concerned, these models evolve in a qualitatively similar manner, sharing the same gross features:  the initial ramp-up of the SASI is driven by the $\ell=1$ sloshing mode, leading to a buildup (and partial destruction)  of magnetic field along the temporary symmetry axis.  The emergence of non-axisymmetric modes occurs around $t=400$~ms in all three models, and is followed by a rearrangement of the flow which eventually leads to a prominent spiral mode and a post shock flow with significant angular momentum about the PNS.  The integrated kinetic energy and total angular momentum between the PNS and the shock evolve similarly with time in all three models and have, at the end of the simulation, reached values of about $E_{\mbox{\tiny kin}}\approx 3-4\times 10^{-2}$~B and $|\vect{L}|\approx 3-4\times 10^{47}$ cm$^{2}$ s$^{-1}$, respectively.  The kinetic energy, the source of magnetic energy, is virtually the same in all models.  The internal energy evolves in a very similar manner in all models as well.  The evolution of the magnetic field remains sensitive to spatial resolution, however.  

\begin{figure}
  \epsscale{1.0}
  \plotone{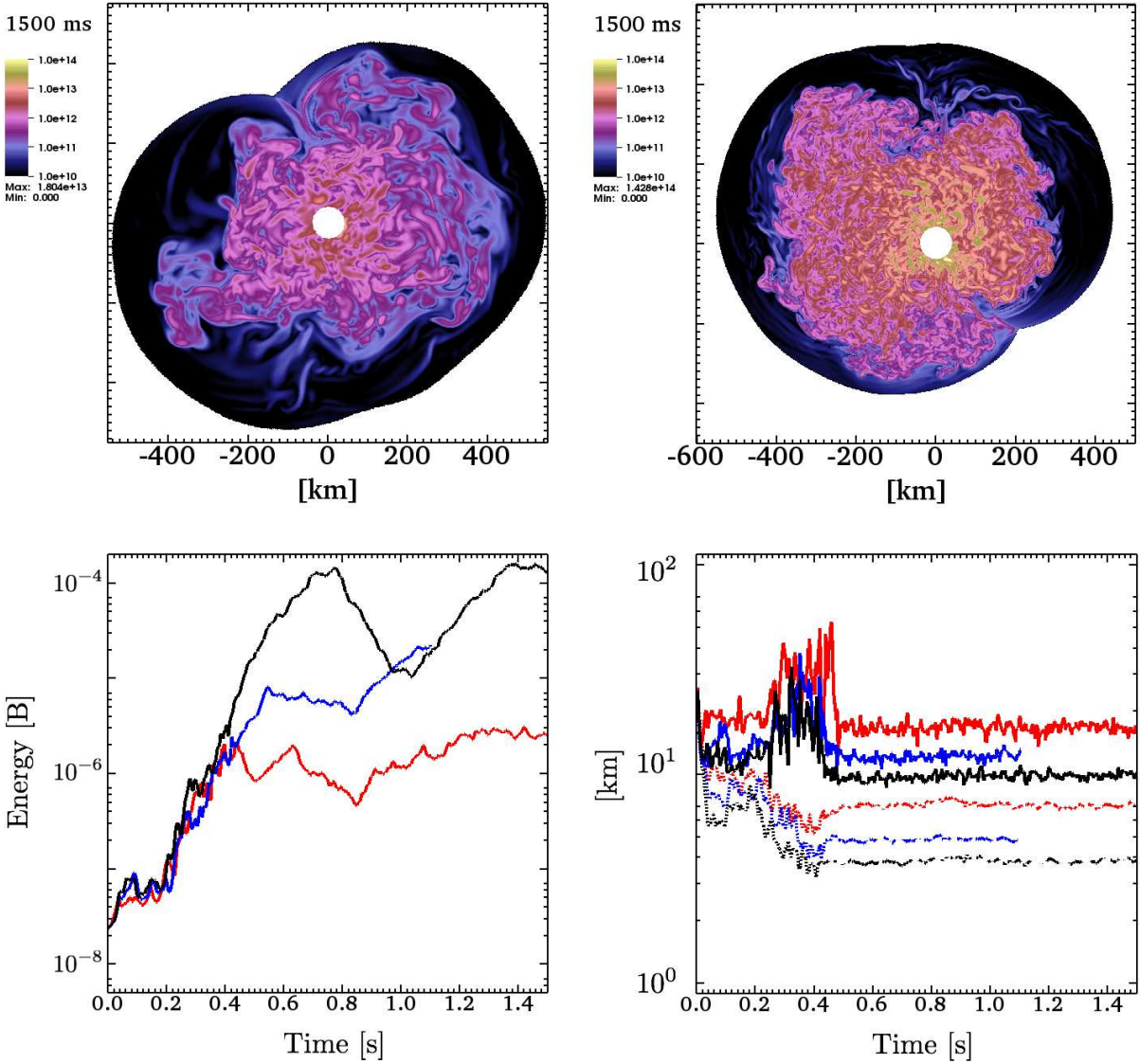}
  \caption{Similar to Figure \ref{fig:varyingResolution2D}, but from 3D calculations where the spatial resolution has been varied:  Snapshot of the magnitude of the magnetic field at $t=1500$~ms for models with $\Delta l\approx 2.34$~km (3DB12Al; upper left panel) and $\Delta l\approx 1.17$~km (3DB12Ah; upper right panel).  The orientation of the plots is chosen so that the normal vector of the slicing plane is parallel to the total angular momentum vector of the flow between the PNS and the shock surface.  The two lower panels show (versus time) results from models where the zone width $\Delta l$ has been set to 2.34~km, 1.56~km, and 1.17~km (red, blue, and black lines, respectively):  The left panel shows the total magnetic energy between the shock and the PNS, and the right panel shows the magnetic curvature radius $\lambda_{\mbox{\tiny c}}$ (upper three solid lines) and the magnetic rms scale $\lambda_{\mbox{\tiny rms}}$ (lower three dotted lines).   \label{fig:varyingResolution3D_noPoyntingFlux}}
\end{figure}

The overview of the 3D models given in Figure \ref{fig:varyingResolution3D_noPoyntingFlux} illustrates the remarkable sensitivity of the magnetic field to the spatial resolution.  The two upper panels show the distribution of the magnitude of the magnetic field at the end of the simulations, $t=1500$~ms, in a slice through the origin of the computational box.  The upper right panel, showing the highest resolution model (3DB12Ah), reveals a highly intermittent magnetic field structure with large regions occupied by magnetic fields that are significantly stronger than in the lowest resolution model (3DB12Al) shown in the upper left panel.  

The total magnetic energy integrated over the volume between the surface of the PNS and the shock is shown in the lower left panel of Figure \ref{fig:varyingResolution3D_noPoyntingFlux}.  The magnetic energy is basically insensitive to resolution during the `axisymmetric' stage ($t\lesssim 400$~ms), but the magnetic energy curves begin to diverge severely as the flow beneath the shock becomes more turbulent ($t\gtrsim 400$~ms).  During this stage of the runs the magnetic energy in the highest resolution model reaches values which are up to two orders of magnitude higher than in the lowest resolution model, despite only a factor of two difference in spatial resolution.  This can be contrasted with the results from the 2D calculations above.  

We find, as expected, that stretching is the dominant magnetic field amplification mechanism at late times in all the 3D models presented in this section, with $\rateLorentzWork\approx\rateStretching$ for $t\gtrsim 500$~ms.  The stretching rate remains at a nearly constant level at late times in all three models, and increases somewhat with increasing spatial resolution (about 65\% from 3DB12Al to 3DB12Ah).  

The effect of the sloshing mode on the structure of the magnetic field is seen in the lower right panel in Figure \ref{fig:varyingResolution3D_noPoyntingFlux} where we plot the length scales $\lambdacurvature$ and $\lambdarms$, which up to $t\approx 420$~ms provide a rough measure of the two dimensions of the magnetic flux tubes (i.e., length and thickness) that develop in the polar regions:  the curvature radius reaches its largest values and the rms scale dips to a minimum during this epoch.  Later in the simulations ($t\gtrsim 500$~ms) both $\lambdacurvature$ and $\lambdarms$ settle down to nearly constant values with respect to time, with $\lambdacurvature>\lambdarms$.  Both scales decrease with increasing resolution, but the ratio $\lambdarms/\Delta l$ is virtually unchanged in all three models ($\sim 3.1$).  The ratio $\lambdacurvature/\Delta l$ increases slightly with increasing resolution (about 20\% from model 3DB12Al to model 3DB12Ah).  

It is apparent from these simulations (and also the 2D simulations) that the accretion-driven turbulent flow facilitated by the nonlinear SASI forces the magnetic field to develop on small scales, which become increasingly available through increased spatial resolution.  

\begin{figure}
  \epsscale{1.0}
  \plotone{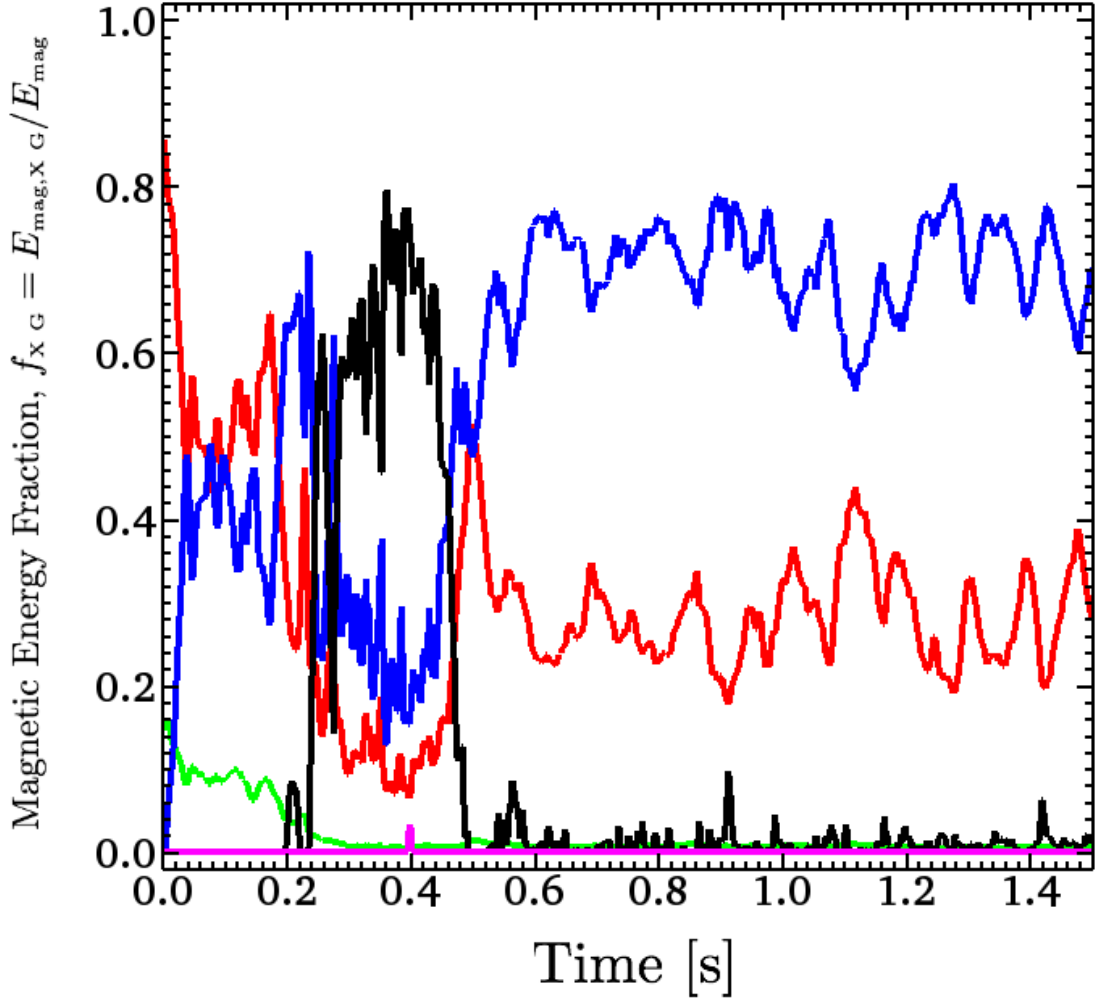}
  \caption{Fraction of the total magnetic energy between the PNS and the surface of the shock with magnetic field strength $|\vect{B}|\in[10^{10},10^{11})$~G;  $f_{\mbox{\tiny 10~G}}$ (green), $|\vect{B}|\in[10^{11},10^{12})$~G;  $f_{\mbox{\tiny 11~G}}$ (red), $|\vect{B}|\in[10^{12},10^{13})$~G;  $f_{\mbox{\tiny 12~G}}$ (blue line), $|\vect{B}|\in[10^{13},10^{14})$~G;  $f_{\mbox{\tiny 13~G}}$ (black), and $|\vect{B}|\in[10^{14},10^{15})$~G;  $f_{\mbox{\tiny 14~G}}$ (magenta), for the low resolution model (3DB12Al).  
  \label{fig:magneticEnergyRatios_B0_1e12_lowRes_axisymmetric}}
\end{figure}

\begin{figure}
  \epsscale{1.0}
  \plotone{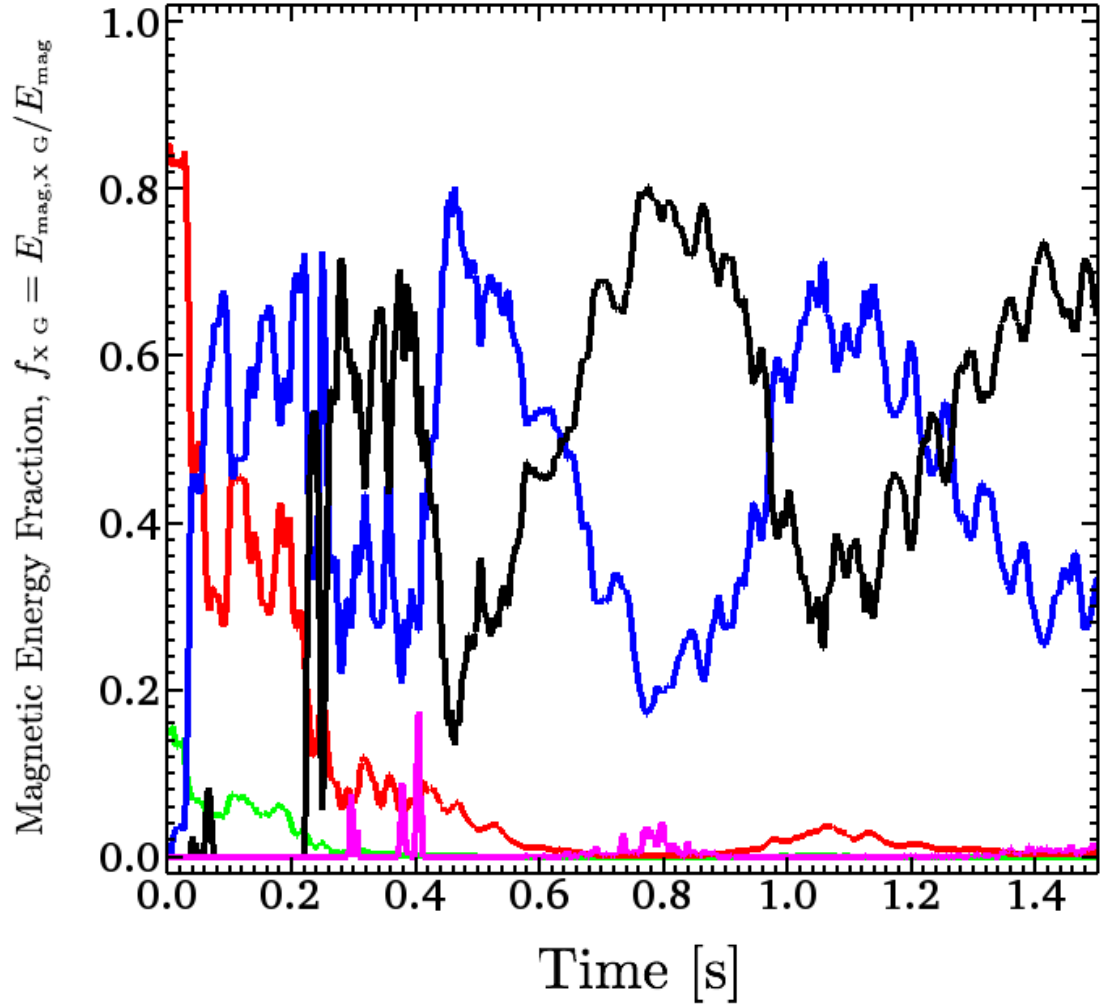}
  \caption{Same as Figure \ref{fig:magneticEnergyRatios_B0_1e12_lowRes_axisymmetric}, but for the high resolution model (3DB12Ah).  
  \label{fig:magneticEnergyRatios_B0_1e12_highRes_axisymmetric}}
\end{figure}

We further illustrate the effect of increasing the spatial resolution in Figures \ref{fig:magneticEnergyRatios_B0_1e12_lowRes_axisymmetric} and \ref{fig:magneticEnergyRatios_B0_1e12_highRes_axisymmetric} (for models 3DB12Al and 3DB12Ah, respectively) by plotting the fraction of total magnetic energy stored in magnetic fields with $|\vect{B}|\in\left[10^{\mbox{\tiny X}},10^{\mbox{\tiny X}+1}\right)$~G, $f_{\mbox{\tiny X G}}$, for $\mbox{X}=10,11,\ldots,14$.  Note that the sum $\sum_{\mbox{\tiny X}=10}^{14} f_{\mbox{\tiny X G}}\approx 1$, since none of the models generate magnetic fields exceeding $10^{15}$~G, and magnetic fields below $10^{10}$~G contributes negligibly to the total magnetic energy.  

We have already pointed out that the magnetic field evolution is similar up to $t\approx 500$~ms in all three models.  This is also evident in the magnetic field distributions in Figures \ref{fig:magneticEnergyRatios_B0_1e12_lowRes_axisymmetric} and \ref{fig:magneticEnergyRatios_B0_1e12_highRes_axisymmetric}:  during the time period bracketed by $t=200$~ms and $t=500$~ms $\fEmag{13}$ reaches (or exceeds) $0.7$ in all three models.  (Some intermittent episodes with spikes in $\fEmag{14}$ are also present.)  This occurs during the operation of the sloshing mode when magnetic field is advected towards and compressed against the $z$-axis (cf. upper left panel in Figure \ref{fig:magneticFieldOverview_medRes_3D_B0_1e12_l0_0e00_perturbed}).  The strong magnetic field disappears quickly once axial symmetry is broken; it is dispersed by the flow, and also advected through the inner cutout boundary.  

For $t\gtrsim500$~ms model 3DB12Al is dominated by magnetic fields of order $10^{12}$~G, but a significant fraction of the magnetic energy is in magnetic fields of order $10^{11}$~G.  A negligible fraction of the magnetic field is in the $10^{13}-10^{14}$~G range (cf. black curve in Figure \ref{fig:magneticEnergyRatios_B0_1e12_lowRes_axisymmetric}).  In the interval from 500~ms to 1500~ms we find time-averaged values $\langle\fEmag{11}\rangle\approx0.28$, $\langle\fEmag{12}\rangle\approx0.7$, and $\langle\fEmag{13}\rangle\approx0.09$ for the lowest resolution model.  

The evolution of the magnetic energy distributions in model 3DB12Am is very similar to 3DB12Al during operation of the sloshing mode and the spiral mode.  For $t\gtrsim500$~ms the relative position of the curves is shifted towards a stronger magnetic field when the resolution increases by $50\%$, but the time evolution is uneventful.  Averaged over the interval from $500$~ms to $1100$~ms we find $\langle\fEmag{11}\rangle\approx0.07$, $\langle\fEmag{12}\rangle\approx0.81$, and $\langle\fEmag{13}\rangle\approx0.12$, respectively.  There is a shallow slope in the curves for $\fEmag{12}$ (decreasing) and $\fEmag{13}$ (increasing), while $\fEmag{11}$ remains at a nearly constant level.  

Model 3DB12Ah (cf. Figure \ref{fig:magneticEnergyRatios_B0_1e12_highRes_axisymmetric}) displays a very different evolution than models 3DB12Al and 3DB12Am beyond $500$~ms.  We find that $\fEmag{12}+\fEmag{13}$ stays above $0.95$ for $t>500$~ms, but also that there are large time intervals where $\fEmag{13}$ stays significantly above $\fEmag{12}$.  This is not seen in the two lower resolution models.  In particular, $\fEmag{12}$ declines steadily at the expense of an increase in $\fEmag{13}$ from about $t=460$~ms to $780$~ms.  A few percent of the total magnetic energy is also due to magnetic fields exceeding $10^{14}$~G near the end of this epoch.  At $t=780$~ms we find $\fEmag{12}\approx0.17$ and $\fEmag{13}\approx0.8$.  Then there is a period lasting until $t=1040$~ms where $\fEmag{13}$ drops down to about $0.26$.  The time interval where $t\in[780,1040]$ coincides with the order of magnitude drop in magnetic energy (cf. lower left panel of Figure \ref{fig:varyingResolution3D_noPoyntingFlux}) when the post shock flow rearranges itself after the breakup of axial symmetry.  The shock volume shrinks dramatically when this occurs, and the average shock radius dips down to a minimum of about $240$~km for $t\approx900$~ms.  The drop in the total magnetic energy between the shock and the PNS is balanced by an increase in the Poynting flux through the surface enclosing the PNS.  A clear spiral mode appears (and persists until the end of the calculation) when $\fEmag{13}$ begins to grow again.  At the end of the calculation most of the magnetic energy consists of magnetic fields with $|\vect{B}|>10^{13}$~G ($\fEmag{13}\approx 0.7$).  Only a small fraction ($\sim 1\%$) of the magnetic energy is due to magnetic fields exceeding $10^{14}$~G.  Note that this `pausing phase' seems to be an artifact of the applied perturbation, and results in a significant delay in magnetic energy growth.  

\begin{figure}
  \epsscale{1.0}
  \plotone{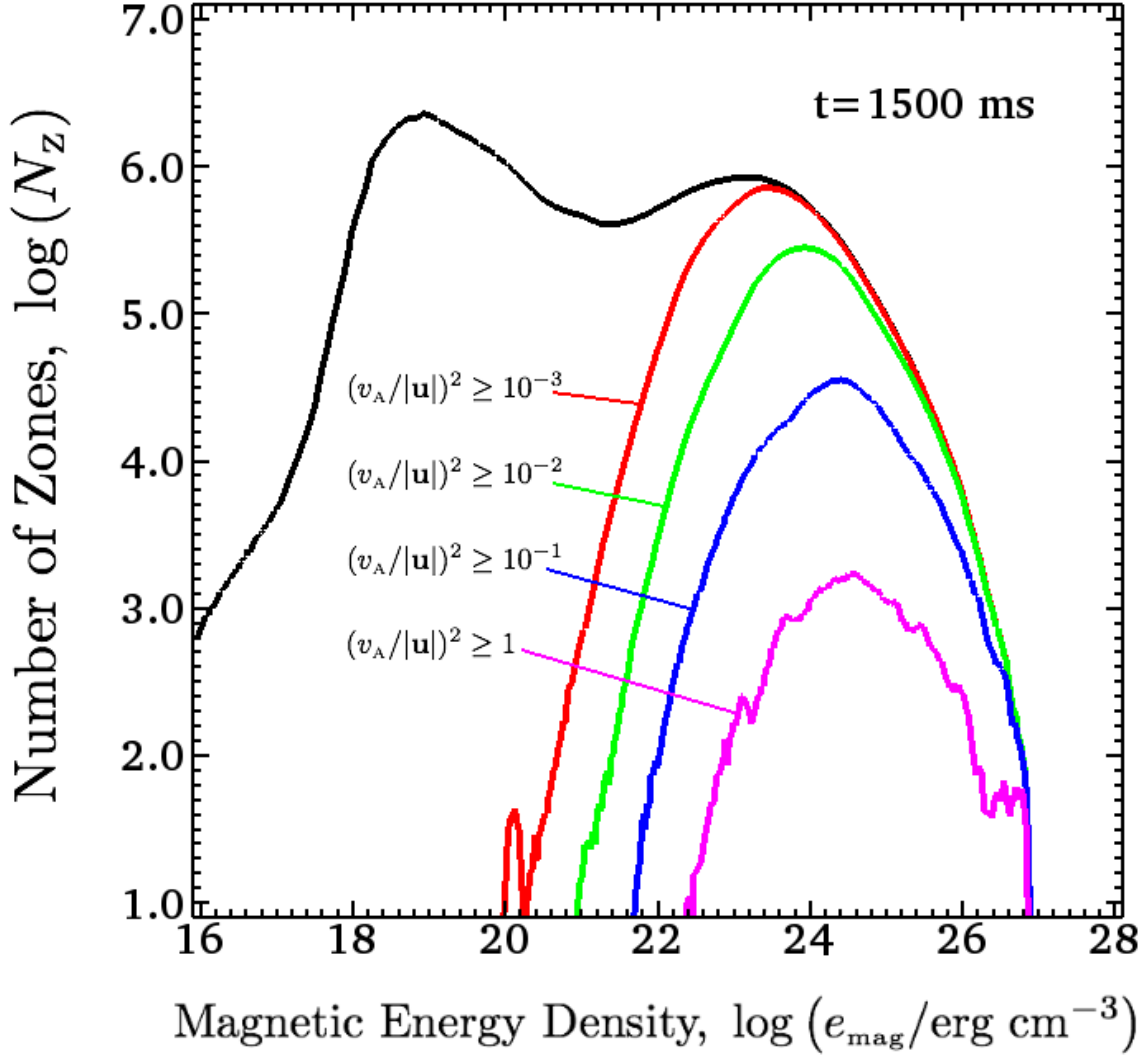}
  \caption{Distribution of the magnetic energy density in zones beneath the shock (black curve) for model 3DB12Ah.  The red, green, blue, and magenta curves represent subsets of zones where the ratio of magnetic to kinetic energy ($e_{\mbox{\tiny mag}}/e_{\mbox{\tiny kin}}=v_{\mbox{\tiny A}}^{2}/|\vect{u}|^{2}$) is greater than or equal to $10^{-3}$, $10^{-2}$, $10^{-1}$, and $1$, respectively.  
  \label{fig:magneticEnergyDistribution_3D_highRes}}
\end{figure}

Despite the dramatic increase in magnetic energy with increasing resolution, the global characteristics of the flow interior to the shock remain unaffected by the magnetic field.  The strength of the magnetic field generated by the SASI does not carry much physical meaning by itself; in order to determine any influence of the magnetic field on the fluid, the magnetic energy must be compared locally with the kinetic energy and/or internal energy.  The magnetic energy grows at the expense of the kinetic energy in the flow, and it is expected (assuming that realistic magnetic Reynolds numbers can be attained) that the magnetic energy can grow until a dynamical equilibrium (at least on some spatial scale) is established.  We investigate if this has occurred in Figure \ref{fig:magneticEnergyDistribution_3D_highRes} in which we plot the distribution of the magnetic energy density in the volume between the PNS and the shock (black line) at the end of model 3DB12Ah.  We compare the magnetic energy to the kinetic energy by plotting the distribution of zones where the ratio of magnetic to kinetic energy $\beta_{\mbox{\tiny kin}}^{-1}=e_{\mbox{\tiny mag}}/e_{\mbox{\tiny kin}}$ $(=v_{\mbox{\tiny A}}^{2}/|\vect{u}|^{2})$ is grater than or equal to $10^{-3}$ (red), $10^{-2}$ (green), $10^{-1}$ (blue), and $1$ (magenta), respectively.  At the end of this simulation we find that most of the magnetic energy ($97\%$) is stored in zones where $\beta_{\mbox{\tiny kin}}^{-1}\ge 10^{-3}$, while a progressively smaller fraction of the total magnetic energy is in zones where $\beta_{\mbox{\tiny kin}}^{-1}\ge 10^{-2}$ and $\beta_{\mbox{\tiny kin}}^{-1}\ge 10^{-1}$ ($0.73\%$ and $21\%$, respectively), and only about $2\%$ of the total magnetic energy is in zones where $\beta_{\mbox{\tiny kin}}^{-1}\ge 1$.  The ratio of magnetic to internal energy remains even lower:  less than $10\%$ of the magnetic energy is in zones where $\beta^{-1}\ge 0.01$, while there are no zones with $\beta^{-1}\ge 0.1$ for $t=1500$~ms.  Thus, it does not seem plausible that magnetic energy growth is curbed by an established dynamical equilibrium.  The bulk of the magnetic energy quickly becomes concentrated on the smallest available spatial scale (determined by the spatial resolution), and the magnetic field cannot be further strengthened through stretching.  Numerical diffusion is therefore, in effect, limiting further growth of the magnetic energy.  

The growth of the magnetic energy has not converged in the range of spatial resolutions covered by the 3D simulations presented in this section.  However, it is evident that simulations with high spatial resolution (even higher than we can afford in this study) are necessary in order to evolve the magnetic field as it is stretched to smaller and smaller scales in the turbulent flow beneath the shock.  For a flux tube with thickness $d$ we expect from flux conservation during stretching that $B\times (r_{\perp}d)\approx \mbox{constant}$ for the 2D models, and $B \times d^{2}\approx\mbox{constant}$ for the 3D models.  This expectation agrees well with Figure \ref{fig:rmsFieldVSrmsScale} in which we plot the time-averaged rms magnetic field $\langle B_{\mbox{\tiny rms}}\rangle$ versus the time-averaged magnetic rms scale $\langle\lambdarms\rangle$ (i.e., flux tube thickness) for the 2D and the 3D models presented in this section.  Moreover, the limitations of axisymmetric calculations become evident:  the steeper slope displayed by the 3D results underscores the importance of evolving the supernova magnetic field with 3D simulations.  According to this stretching relationship, an average flux rope thickness of $\langle\lambdarms\rangle\approx0.5$~km implies $\langle B_{\mbox{\tiny rms}}\rangle\approx 10^{14}$~G, for models with $B_{0}=10^{12}$~G, while a similar rms magnetic field for models with $B_{0}=10^{10}$~G requires $\langle\lambdarms\rangle\approx50$~m!  Our results suggest that the strength of the magnetic fields generated by the SASI in core-collapse supernovae can attain levels similar to (or even significantly higher than) those attained in the simulations presented here; although the exact values depend on the initial (progenitor) magnetic field, and the spatial scale at which a presumed dynamical equilibrium between the fluid and the magnetic field is established.  On the other hand, the kinetic energy available from the SASI-driven turbulent flow, and hence the magnetic energy that can be generated by the SASI is likely too low for SASI-generated magnetic fields to have any significant impact on the global dynamics leading to the majority of core-collapse supernovae---i.e., those originating from non-rotating or slowly rotating progenitors \citep{heger_etal_2005}.  

\section{Neutron Star Magnetization}

The Poynting flux through the inner boundary of our simulation implies neutron star magnetization.  The PNS is excised from our computational domain, but we can estimate the rms magnetic field in the volume it occupies---and therefore the magnetic field strength of the nascent neutron star due to the flux of electromagnetic energy through our inner (`cutout')  boundary.  

When the right-hand side of Eq. (\ref{eq:poyntingTheorem}) is neglected (and $\eta\to0$), an integral over the volume $V$ occupied by the PNS gives the rate of accumulation of magnetic energy $E_{\mbox{\tiny{mag}}}$ in terms of the Poynting flux through the PNS surface $\partial V$. In particular
\begin{eqnarray}
  B_{\mbox{\tiny rms}}
  &=&
  \left( \f{1}{V}\int_{V} \vect{B}\cdot\vect{B} \, dV \right)^{1/2} \nonumber \\
  &\approx& 
  3\times 10^{14} \mbox{ G} 
  \left( \f{E_{\mbox{\tiny{mag}}}}{10^{-3} \mbox{ B}}\right)^{1/2}
  \left( \f{\rPNS}{40 \mbox{ km}} \right)^{-3/2},  
\end{eqnarray}
where
\begin{equation}
  E_{\mbox{\tiny{mag}}} \left( t\right) \simeq - \int_{0}^{t} dt' \oint_{\partial V} \vect{P}\cdot d\vect{S}. 
  \label{eq:poyntingFlux}
\end{equation}
From our highest resolution 3D simulation (3DB12Ah) we find that $E_{\mbox{\tiny{mag}}}$ reaches almost $10^{-3}$~B by the end of the calculation. Thus the PNS is expected to be significantly magnetized ($B_{\mbox{\tiny rms}} \sim 10^{14}$~G) from the accreted field generated by the SASI model.  This expectation is of course sensitive to the initial magnetic field and the duration of SASI operation. Thus an observed pulsar magnetic field strength, in conjunction with knowledge of the delay to explosion (and therefore duration of the SASI) obtained from neutrino or gravitational wave signals, might even provide information on the initial magnetic field strength. 

This estimate directs attention to the fact that excision of the PNS prohibits treatment of some issues that may be worthy of further study.  For instance, our neglect of the Maxwell stress in estimating the neutron star magnetization $B_{\mbox{\tiny rms}}$ corresponds to the fact that excision of the PNS prohibits meaningful treatment of the `back-reaction' of PNS magnetization on fluid flows and magnetic field evolution in the computational domain.  

In addition the SASI-generated magnetic field may also provide a significant input field to be further amplified by the convective dynamo that potentially operates inside the PNS \citep[e.g.,][]{thompson_duncan_1993}.  

\section{Summary, Discussion, and Conclusions}

We have presented 2D (axisymmetric) and 3D simulations of an idealized MHD model of a stalled supernova shock, which demonstrate that SASI-driven flows---initiated from non-rotating or slowly rotating progenitors---are able to significantly amplify the magnetic field beneath the shock.  An almost four orders of magnitude increase in magnetic energy is observed in the highest resolution 3D model.  The amplification does not result in magnetic fields that have any direct impact on global features of the shock evolution in our models, but the final magnetic energy and field strength remain sensitive to the spatial resolution in our numerical simulations.  The magnetic field evolves in a turbulent flow, and develops into a highly intermittent `flux rope' structure.  The average `flux rope' thickness is limited by finite spatial resolution, being resolved by only a few computational cells, and numerical dissipation suppresses further growth of the magnetic energy (with $Q_{\mbox{\tiny J}}\lesssim W_{\mbox{\tiny L}}$; cf. Eq. (\ref{eq:poyntingTheorem})).  

The early evolution in axisymmetric models, and in axisymmetrically perturbed 3D models, is characterized by continued focusing of fluid flows toward the polar axes due to the SASI sloshing mode.  As flows directed toward the axis are forced to turn parallel to it, the magnetic field is advected into the polar regions, `combed' parallel to the polar axes by the fluid flow, and left behind.  Relatively modest initial oscillations in the magnetic field strength are followed by an epoch of exponential field growth with the emergence of the SASI-induced internal shock and its associated streams (or sheets, in axisymmetry) plunging toward the PNS and the symmetry axis.  The 2D models evolve to an end state where the bulk of the magnetic energy is concentrated around the imposed axis of symmetry.  The amplification is due to compression {\it and} stretching in 2D models.  

\begin{figure}
  \epsscale{1.0}
  \plotone{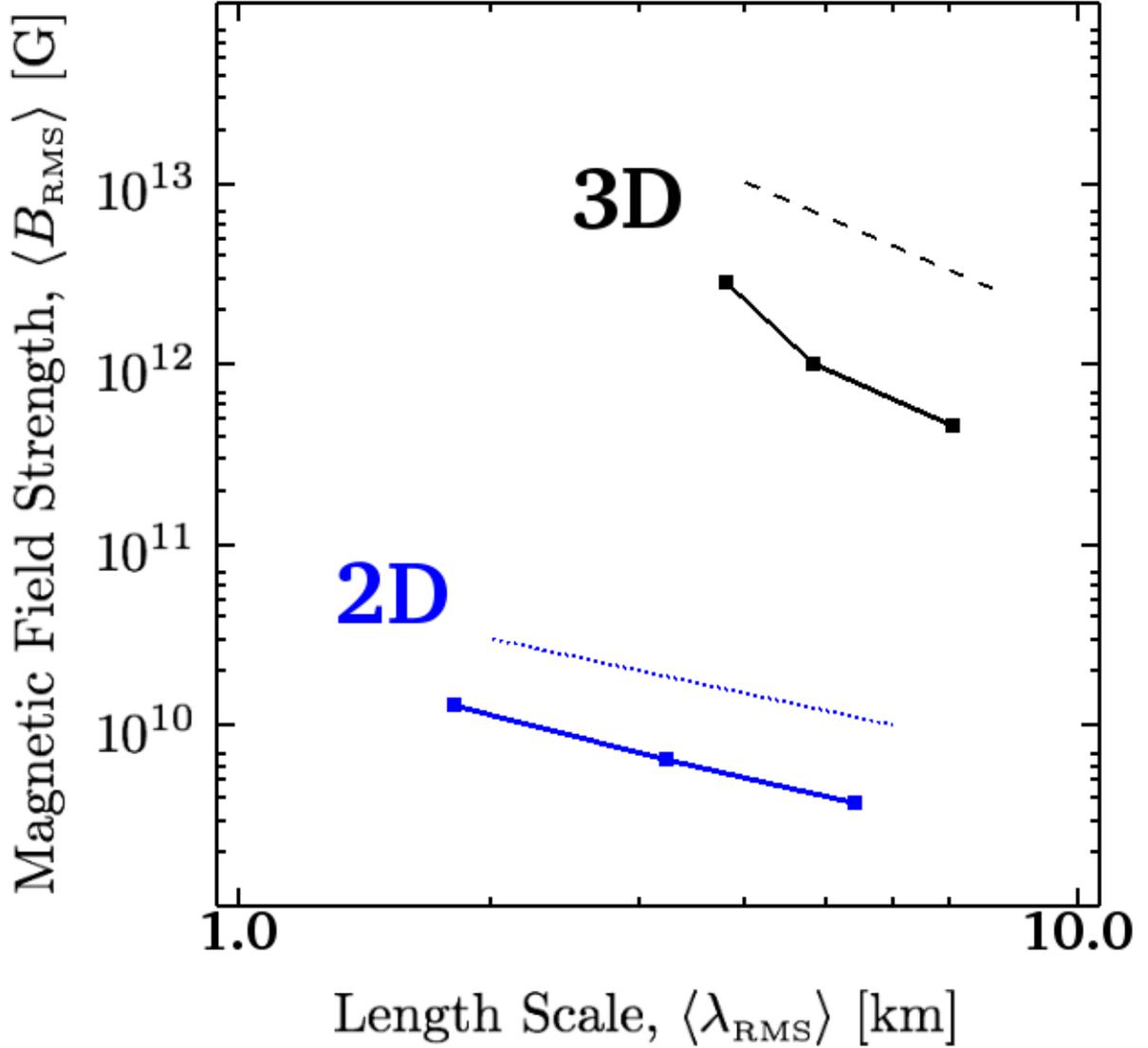}
  \caption{Plot showing time-averaged rms magnetic field strength $\langle B_{\mbox{\tiny rms}}\rangle$,versus time-averaged magnetic rms scale $\langle\lambda_{\mbox{\tiny rms}}\rangle$, in the `saturated' nonlinear stage.  Results are shown (in blue) for the axisymmetric 2D models (2DB10Al, 2DB10Ah, and 2DB10Ah$^{+}$) and (in black) for the 3D models (3DB12Al, 3DB12Am, and 3DB12Ah).  (The time-average extends from $t=500$~ms to the end of each model.)  The dotted blue and dashed black reference lines are proportional to $\langle\lambda_{\mbox{\tiny rms}}\rangle$ and $\langle\lambda_{\mbox{\tiny rms}}\rangle^{2}$, respectively.  
  \label{fig:rmsFieldVSrmsScale}}
\end{figure}

Non-axisymmetric flows (i.e., the SASI spiral mode) emerge and become dominant in all the 3D models.  These flows quickly destabilize and disperse the magnetic structure generated by the axisymmetric SASI (cf. model 3DB12Am), demonstrating the need for 3D simulations to avoid spurious relics of axisymmetry on the magnetic field evolution.  The importance of doing 3D calculations is further underscored by the dominance of the inherently non-axisymmetric SASI spiral mode, and by the obvious geometrical limitation on magnetic field amplification by stretching when axisymmetry is imposed (stretching of sheets vs. stretching of tubes; cf. Figure \ref{fig:rmsFieldVSrmsScale}).

Magnetic field stretching, facilitated by the turbulent flows driven by the SASI spiral mode, plays the dominant role in amplifying the magnetic field in the 3D models.  The development of turbulence beneath the shock is associated with the plunging supersonic stream emanating from the shock triple point.  These nonlinear flows develop from the accompanying shear and stagnation regions, and spread during the operation of the SASI spiral mode to fill a large volume beneath the shock.  The magnetic field amplification is not driven directly by this shear flow, but rather indirectly by the hydrodynamical consequences of the presence of the plunging stream, which drives a turbulent flow beneath the shock.  There is, of course, some magnetic field amplification occurring in the shear layer as well, but the magnetized fluid elements spend a relatively short time in this region compared to the time they spend `churning' in the vigorous turbulence driven by the SASI.  

The growth of magnetic energy in our numerical simulations eventually becomes limited by numerical resistivity because of the magnetic field's tendency to evolve into thin flux ropes.  (Numerical resistivity inevitably becomes important when the magnetic field varies significantly over only a few grid zones.)  This is, of course, a nonphysical effect in our simulations since the fluid between the PNS and the shock is expected to behave as a nearly perfect electrical conductor.  We are therefore unable to make exact predictions about the magnetic field in a presumed saturated state, but it is reasonable to expect that the magnetic field may eventually become dynamically significant, at least on relatively small spatial scales, where the drag of the fluid on the flux ropes is balanced by the tension of the flux ropes \citep[cf.][]{thompson_duncan_1993}.  The turbulent kinetic energy available on the scales where the magnetic field may become dynamically important will determine its final state and ultimate impact on the dynamics.  In fact, a sizable fraction of the kinetic energy beneath the shock is due to chaotic motions, but we do not expect this to be sufficient for SASI-generated magnetic fields to become important to global dynamics of core-collapse supernovae emanating from non-rotating or slowly rotating progenitors.  A more detailed analysis is required to investigate the issues related to magnetic field saturation in further detail, including a Fourier spectral decomposition of the magnetic and kinetic energies.  Such an analysis is beyond the scope of this initial paper, but we plan to include it in a forthcoming study.  

Given infinite spatial resolution, the magnetic energy in our models is still limited by the kinetic energy of the SASI-driven post-shock flows, which is a few $\times10^{-2}$~B.  The rotational energy associated with the spin-up of the PNS (cf. Eq (\ref{eq:omegaDot})) is of similar magnitude, and not sufficient to produce MHD-driven outflows.  A larger energy reservoir---e.g., a rotational energy reservoir provided by relic angular momentum from progenitor rotation---appears to be needed for MHD effects to play a principal role in driving the explosion.  The simulations presented in this study do not produce sufficient rotational energies.  The rotational energy, $E_{\mbox{\tiny rot}}$, of a nascent neutron star with a rotation period of a few milliseconds is in the range of $1-10$~B \citep[e.g.,][]{ott_etal_2006}, and the free energy available due to differential rotation is about 10\% of $E_{\mbox{\tiny rot}}$ \citep[][]{shibata_etal_2006}.  \citet{burrows_etal_2007} estimate free energies exceeding 3~B for their fastest spinning models, which are capable of driving outflows.  However, such rapidly rotating PNSs are not predicted by current stellar evolution theory \citep[e.g.,][]{heger_etal_2005}, and MHD-driven explosions are therefore likely to be limited to a only small subset of progenitor stars.  

Indeed, past studies have concluded that rotation is needed both as the energy reservoir {\em and} for amplification of the post-bounce magnetic field.  In particular, the conventional wisdom has been that rapid rotation is needed to amplify the magnetic field to dynamical significance {\em before} any rotational energy can be magnetically tapped to drive outflows.  Both field line wrapping and the MRI require differential rotation and result in field amplification on the rotation time scale.  The MRI, hailed as the most promising amplification mechanism, may lead to exponential field growth in the differentially rotating fluid between the PNS and the supernova shock, and perhaps result in magnetorotationally-driven outflows \citep[e.g.,][]{akiyama_etal_2003, burrows_etal_2007}.  The lack of sufficient rotation (and also spatial resolution) in our models excludes the development of the MRI, but an interesting question for future work is whether the SASI may (perhaps in conjunction with the MRI) generate the magnetic fields required to extract rotational energy from the PNS and result in important dynamical effects---in particular, for moderately-rotating progenitor stars.  Moderate rotation alone may not be able to rapidly generate a strong magnetic field; but if a sufficiently strong magnetic field can develop through the mechanisms discussed in this paper, the (`modest') rotational energy available to be harnessed by that field still might be significant enough to impact pre-explosion dynamics.  In other words, the SASI may be able to extend the range of progenitors---that is, range of rotation rates---for which magnetic fields play a role in their explosion dynamics.  Even if these magnetic fields turn out not to be important to the explosion, then perhaps they may impact the proto-neutron star wind that follows \citep[e.g.,][]{thompson_2003}.  

The impact of magnetic fields in core collapse supernovae must ultimately be investigated in sophisticated supernova models with neutrino transport and a nuclear equation of state, across the range of progenitors predicted by stellar evolution theory, but our models offer proof of principle that magnetic fields may be significantly amplified post bounce without initial core rotation, contrary to conclusions reached in past core collapse supernova simulations that included MHD \citep[e.g.,][]{leblanc_wilson_1970, symbalisty_1984}.  

Although our calculations result in magnetic fields that may not play a direct dynamical role in the explosion itself, they do suggest interesting implications for neutron star magnetization.  The rotational energy available from the spin-up of the PNS might account for a strongly magnetized neutron star; but the turbulent kinetic energy spawned by the SASI is of comparable magnitude, and our simulations suggest that it may be tapped to create strongly magnetized neutron stars even from {\em non-rotating} progenitors.  The SASI generates magnetic fields in a large volume outside the PNS that is accreted as long as the SASI persists. Because the strength of the generated field also depends on the initial field strength, an observed pulsar magnetic field strength, in conjunction with knowledge of the delay to explosion (and therefore duration of the SASI) obtained from neutrino or gravitational wave signals, might even provide information on the initial (progenitor) magnetic field strength.  

In addition, further amplification may be attained through field line wrapping, and perhaps the MRI, in the differentially rotating region extending from the PNS, and also due to the convective dynamo that potentially operates inside the PNS.  Our model excludes the region occupied by the PNS, where these processes take place, but it is entirely plausible that the magnetic field generated in our models can be further amplified due to activities occurring closer to, or inside, the PNS.  

Our calculations also demonstrate the computational difficulty of capturing MHD effects in the large scale simulations that are needed to ultimately assess the role of magnetic fields in core collapse supernovae, namely the need for high spatial resolution in localized regions.  In the context of the MRI the wavelength of the fastest-growing unstable mode (proportional to the magnetic field strength) becomes prohibitively short when considering the field strengths predicted by stellar evolution calculations.  SASI-induced magnetic fields are also limited by spatial resolution because the thickness of magnetic flux ropes, which harbor the strongest magnetic fields, are currently impossible to resolve properly in global multiphysics calculations.  Furthermore, the ubiquitous development (and perhaps importance) of turbulent flows poses additional challenges by itself:  shock-capturing methods (similar to the method employed in our calculations to evolve the MHD equations) are inherently diffusive for turbulence, while methods commonly used in the study of turbulence are less effective in capturing discontinuities in the flow.  However, resolution of these issues is beyond the scope of this study.  

\acknowledgments

This work was supported by the U.S. Department of Energy Office of Nuclear Physics and Office of Advanced Scientific Computing Research, and by grants from the NASA Astrophysics Theory and Fundamental Physics Program and the NSF PetaApps Program.  The simulations were performed on the National Center for Computational Sciences (NCCS) Leadership Computing Facility under the Innovative and Novel Computational Impact on Theory and Experiment (INCITE) Program.  We thank an anonymous referee for discerning comments that led to significant improvements of the manuscript.  We acknowledge support from members of the NCCS, especially Bronson Messer and Ross Toedte.

\clearpage

\clearpage

\clearpage

\clearpage

\clearpage

\clearpage

\clearpage

\clearpage

\clearpage

\clearpage

\clearpage

\clearpage

\clearpage

\clearpage

\clearpage

\clearpage

\clearpage

\clearpage

\clearpage

\clearpage

\clearpage

\clearpage

\clearpage

\clearpage

\clearpage

\begin{table}
\begin{center}
\caption{Tabular overview of two-dimensional (axisymmetric) calculations.  \label{tbl:2Dmodels}}
\begin{tabular}{lcccc}
\tableline\tableline
Model Name & $B_{0}$ [G] & $l$ [cm$^{2}$ s$^{-1}$] & Perturbation & Spatial Resolution [km] \\
\tableline
2DB10Nm & $1\times10^{10}$ & 0 & none & $300/192$ \\
2DB10$\Omega$Nm & $1\times10^{10}$ & $1.5\times10^{15}$ & none & $300/192$ \\
2DB10Al & $1\times10^{10}$ & 0 & axisymmetric & $300/128$  \\
2DB10Am & $1\times10^{10}$ & 0 & axisymmetric & $300/192$ \\
2DB10Ah & $1\times10^{10}$ & 0 & axisymmetric & $300/256$ \\
2DB10Ah$^{+}$ & $1\times10^{10}$ & 0 & axisymmetric & $300/512$ \\
2DB12Am & $1\times10^{12}$ & 0 & axisymmetric & $300/192$ \\
2DB13Am & $1\times10^{13}$ & 0 & axisymmetric & $300/192$ \\
2DB14Am & $1\times10^{14}$ & 0 & axisymmetric & $300/192$ \\
\tableline
\end{tabular}
\end{center}
\end{table}

\clearpage

\begin{table}
\begin{center}
\caption{Tabular overview of three-dimensional calculations.  \label{tbl:3Dmodels}}
\begin{tabular}{lcccc}
\tableline\tableline
Model Name & $B_{0}$ [G] & $l$ [cm$^{2}$ s$^{-1}$] & Perturbation & Spatial Resolution [km] \\
\tableline
3DB12Al & $1\times10^{12}$ & 0 & axisymmetric & $300/128$ \\
3DB12Am & $1\times10^{12}$ & 0 & axisymmetric & $300/192$ \\
3DB12Ah & $1\times10^{12}$ & 0 & axisymmetric & $300/256$ \\
3DB12Rm & $1\times10^{12}$ & 0 & random & $300/192$ \\
3DB12$\Omega$Rm & $1\times10^{12}$ & $1.5\times10^{15}$ & random & $300/192$ \\
3DB10Rm & $1\times10^{10}$ & 0 & random & $300/192$ \\
\tableline
\end{tabular}
\end{center}
\end{table}

\end{document}